# Crystal chemistry criteria of the existence of spin liquids on the kagome lattice

**L M Volkova and D V Marinin**

Institute of Chemistry, Far Eastern Branch, Russian Academy of Sciences, 690022 Vladivostok, Russia

E-mail: volkova@ich.dvo.ru

**Abstract**
The structural-magnetic models of 25 antiferromagnetic kagome cuprates similar to herbertsmithite ($ZnCu_3(OH)_6Cl_2$) – a perspective spin liquid – have been calculated and analyzed. Main correlations between the structure and magnetic properties of these compounds were revealed. It has been demonstrated that, in all AFM kagome cuprates, including herbertsmithite, there exists the competition between the exchange interaction and the antisymmetric anisotropic exchange one (the Dzyaloshinskii-Moriya interaction), as magnetic ions are not linked to the center of inversion in the kagome lattice. This competition is strengthened in all the kagome AFM, except herbertsmithite, by one more type of the anisotropy (duality) of the third in length $J3$ magnetic couplings (strong $J3(J1_2)$ next-to-nearest-neighbor couplings in linear chains along the triangle edges and very weak FM or AFM $J3(J_d)$ couplings along the hexagon diagonals). The above couplings are crystallographically identical, but are divided to two types of different in strength magnetic interactions. The existence of duality of $J3$ couplings originated from the structure of the kagome lattice itself. Only combined contributions of dual $J3$ couplings with anisotropic Dzyaloshinskii-Moriya interactions are capable to suppress frustration of kagome antiferromagnetics. It has been demonstrated that the possibility of elimination of such a duality in herbertsmithite, which made it a spin liquid, constitutes a rare lucky event in the kagome system. Three crystal chemistry criteria of the existence of spin liquids on the kagome lattice have been identified: first, the presence of frustrated kagome lattices with strong dominant antiferromagnetic nearest-neighbor $J1$ couplings competing only with each other in small triangles; second, magnetic isolation of these frustrated kagome lattices; and third, the absence of duality of the third in length $J3$ magnetic couplings.

___________________________________________________________________________

## 1. Introduction

Spin liquid is a rare quantum body state at low temperature. It emerges only in some dielectric crystals with antiferromagnetic (AFM) exchange interaction, when spins are not ordered even at approaching the absolute zero. Frustration of magnetic interactions serves as a main quality characterizing candidates for quantum spin liquid (QSL), as it disrupts the spin long-range order. Instead, they form collective complex states. Such crystals were named collective paramagnetics or spin liquids [1 - 8]. The search of potential materials with QSL realization constitutes a very important task in spintronics.

Kagome antiferromagnetics are considered as the most promising quantum spin liquids. They have been intensively studied in the recent years. These compounds include herbertsmithite ($ZnCu_3(OH)_6Cl_2$) [9, 10] characterized with an ideal kagome geometry and considered as the most perfect spin liquid. Due to geometric frustration of spin interactions, the magnetic ordering in it is suppressed down to zero temperature.

We built up the structural-magnetic model of herbertsmithite using the earlier developed [11-13] crystal chemistry method of calculation of the sign (type of magnetic moment orientation) and strength of magnetic interactions ($J_{ij}$) on the basis of structural data. We revealed the main correlations between the structure and magnetic properties of herbertsmithite and determined crystal chemistry criteria of targeted search of new perspective spin liquids. In the inorganic crystal structure database (ICSD) and literature, we identified 25 AFM cuprate kagome similar to herbertsmithite. For each of these compounds, we calculated the sign and strength of magnetic interactions not only between nearest neighbours, but also for longer-range neighbours as in the kagome plane as between the planes. A majority of these compounds was investigated repeatedly, as theoretically as experimentally, and the same result was obtained in all cases. The only fact was of virtually no doubt that herbertsmithite was a spin liquid, whereas the main reason of ordering in other similar compounds upon the temperature decrease was the emergence of the antisymmetric anisotropic exchange interaction - the Dzyaloshinskii-Moriya interaction [14, 15]. That is why the main objective of our study was a search of reasons of blocking the DM interaction in herbertsmithite and, on the contrary, facilitating its emergence in other similar antiferromagnetics with cuprate kagome lattice. To put it more correctly, we had to identify specific magnetic couplings in the crystal lattice resulting in the emergence of such an interaction. After an extensive search, we found a solution, which, as usual, was on the surface.

We have analyzed the obtained results of study of 25 AFM kagome of cuprates and established that the strong bond model that takes into consideration only the nearest-neighbor bond cannot explain the emergence of the magnetic ordering under effect of DM interaction in compounds with magnetic kagome lattices. In spite of the fact that the space groups (except N9 and N36), in which the compounds we have studied are crystallized (including herbertsmithite ($ZnCu_3(OH)_6Cl_2$), are centrosymmetric, copper ions occupy noncentrosymmetric equivalent positions in them. In this case, the antisymmetric anisotropic exchange interaction (the Dzyaloshinskii-Moriya interaction) contributes to total magnetic exchange interaction between two nearest magnetic spins in the kagome lattice (angle Cu-O-Cu = 101–141º; parameter of the Dzyaloshinskii-Moriya interaction D≠0). In magnetic crystals without a center of inversion, the competition between the exchange interaction and the Dzyaloshinskii-Moriya interaction caused by the spin-orbital coupling induces spin canting and, thus, serves as a source of a weak ferromagnetic behavior in an antiferromagnetic. However, such a competition takes place in all the compounds, including herbertsmithite, which, unlike others, resists ordering upon the temperature decrease under effect of this DM interaction. This means that, in all the kagome AFM, except herbertsmithite, there exists an additional anisotropic interaction, which, in combination with the DM interaction, suppresses frustration and orders these 24 frustrated antiferromagnetics. We have identified this second type of anisotropic interactions in the kagome plane among long-range ones beyond the nearest neighbors. They emerge in the case of existence of magnetic nonequivalence of crystallographically equivalent interactions. Although the bonds between similar ions located at equal distances and occupying equivalent positions in the space group are crystallographically equivalent, the parameters of magnetic couplings between these ions may differ. In other words, when one of the types of crystallographically equivalent interactions in the crystal structure is divided into two types of magnetic interactions in the magnetic structure differing in sign and/or strength. We observed such a phenomenon in the $Cr_{1/3}NbS_2$ intercalate, where a dramatic nonequivalence of the strengths of crystallographically equivalent AFM $J6$ and $J6'$ magnetic interactions ($J6'/J6 = 0.007$) facilitated the emergence of the DM interaction, whose role in $Cr_{1/3}NbS_2$ consists in final ordering and stabilization of chiral spin helices into a chiral magnetic soliton lattice [16].

We will demonstrate that the duality – nonequivalence of crystallographically identical third in length $J3$ magnetic couplings ($J3(J_d)$) couplings along the hexagon diagonals and $J3(J1_2)$ next-to-nearest-neighbour couplings in linear chains along the triangles sides) – originates from the crystal structure of the kagome lattice itself. Only the elimination of such a duality preserve the frustration of magnetic inetractions and, therefore, the spin liquid state. We will then consider the conditions, under which such a duality can be eliminated.

**2. Method of Calculation**

The structural-magnetic models are based on crystal chemistry parameters (crystal structure and ions charge and size). These models characteristics include: (1) sign and strength of magnetic interactions $J_{ig}$; (2) dimensionality of magnetic structures (not always coinciding with that of those of crystal structures) (3) presence of magnetic frustrations on specific geometric configurations; (4) possibility of the transition of the antiferromagnetic (AFM) – ferromagnetic (FM) type. Structural-magnetic models enable one to reveal main correlations between the compounds structure and magnetic properties and determine, on their basis, the crystal chemistry criteria of targeted search of novel functional magnetics, also in the inorganic crystal structure database (ICSD).

To determine the characteristics of magnetic interactions (type of the magnetic moments ordering and strength of magnetic coupling) in materials, we used the earlier developed phenomenological method (named the "crystal chemistry method") and the "MagInter" program created on its basis [11-13]. In this method, three well-known concepts about the nature of magnetic interactions are used. First, it was the Kramers's idea [17], according to which, in exchange couplings between magnetic ions separated by one or several diamagnetic groups, the electrons of nonmagnetic ions play a considerable role. Second, we used the Goodenough–Kanamori–Anderson's model [18-20], in which the crystal chemical aspect points clearly to the dependence of strength interaction and the type of orientation of spins of magnetic ions on the arrangement of intermediate anions. Third, we used the polar Shubin–Vonsovsky's model [21]: by consideration of magnetic interactions, we took into account not only anions, which are valence bound to the magnetic ions, but also all the intermediate negatively or positively ionized atoms, except cations of metals without unpaired electrons.

The method enables one to determine the sign (type) and strength of magnetic couplings on the basis of structural data. According to this method, a coupling between the magnetic ions $M_i$ and $M_j$ emerges in the moment of crossing the boundary between them by an intermediate ion $A_n$, with the overlapping value of ~0.1 Å. The area of the limited space (local space) between the $M_i$ and $M_j$ ions along the bond line is defined as a cylinder, whose radius is equal to these ions radii. The strength of magnetic couplings and the type of magnetic moments ordering in insulators are determined mainly by the geometrical position and the size of intermediate $A_n$ ions in the local space between two magnetic ions ($M_i$ and $M_j$) The positions of intermediate $A_n$ ions in the local space are determined by the $h(A_n)$ distance from the center of the $A_n$ ion up to the $M_i$-$M_j$ bond line and the degree of the ion displacement to one of the magnetic ions expressed as a ratio ($l_n'/l_n$) of the lengths $l_n$ and $l_n'$ ($l_n \leq l_n'$; $l_n' = d(M_i - M_j) - l_n$) produced by the bond line $M_i$-$M_j$ division by a perpendicular made from the ion center (Supplementary Note1: figure 1, (https://stacks.iop.org/JPCM/33/415801/mmedia)).).

The intermediate $A_n$ ions will tend to orient magnetic moments of $M_i$ and $M_j$ ions and make their contributions $j_n$ into the emergence of antiferromagnetic (AFM) or ferromagnetic (FM) components of the magnetic interaction in dependence on the degree of overlapping of the local space between magnetic ions ($\Delta h(A_n)$), the asymmetry ($l_n'/l_n$) of position relatively to the middle of the $M_i$-$M_j$ bond line, and the distance between magnetic ions ($M_i$-$M_j$). Among the above parameters, only the degree of space overlapping between the magnetic ions $M_i$ and $M_j$ ($\Delta h(A_n) = h(A_n) - r_{A_n}$) equal to the difference between the distance $h(A_n)$ from the center of the $A_n$ ion up to the bond line $M_i$-$M_j$ and the radius ($r_{A_n}$) of the $A_n$ ion determined the sign of magnetic interaction. If $\Delta h(A_n) < 0$, the $A_n$ ion overlaps (by $|\Delta h|$) the bond line $M_i$–$M_j$ and initiates the emerging contribution into the AFM-component of the magnetic interaction. If $\Delta h(A_n) > 0$, there remains a gap (the gap width $\Delta h$) between the bond line and the $A_n$ ion, and this ion initiates a contribution to the FM-component of the magnetic interaction. The distance from the boundary of the local space between the magnetic ions $M_i$ and $M_j$ to the surface of the intermediate ion $A_n$ ($D = r_M - (h(A_n) - r_{An})$ equal to the difference between the radius $r_M$ and the distance $h(A_n)$ from the center of the $A_n$ ion up to the bond line $M_i$-$M_j$ and the radius $r_{An}$.

The sign and strength of the magnetic coupling $J_{ij}$ are determined by the sum of the above contributions:

$$J_{ij} = \sum_n j_n$$

The $J_{ij}$ value is expressed in Å$^{-1}$ units. If $J_{ij} < 0$, the type of $M_i$ and $M_j$ ions magnetic ordering is AFM and, in opposite, if $J_{ij} > 0$, the ordering type is FM.

To translate the $J_n$ value in per angstrom (Å$^{-1}$) into the energy units degree Kelvin (K) more conventional for experimenters, it is necessary to select a magnetic fragment similar in crystal structure and chemical composition, which was studied experimentally, to calculate the parameters of magnetic couplings by the crystal chemistry method based on the structural data, and to determine the coefficients ($K$) of the relationship between theoretical and experimental data for individual coupling. For instance, the measurements of the magnetic susceptibility [22-24] $ZnCu_3(OH)_6Cl_3$ show antiferromagnetic couplings of the order $J1^{Zn} \approx 190$ K, whereas the *ab initio*-based analysis of the Cu-Cu exchange coupling constants yields a smaller value $J1^{Zn} = 182$ K [25]. In $YCu_3(OH)_6Cl_3$, the nearest-neighbor Heisenberg exchange $J1 = 82(2)$ K [26] is in good agreement with the numerical modeling and complementary ESR measurements and magnetic susceptibility. In spite of the similarity of structural data for these compounds, the values of the strengths of dominant nearest-neighbor exchange $J1$ interactions differ in 2.2-2.3 times, whereas, in accordance with our calculations based on the structural data, the values AFM $J1^{Zn} = -0.0670$ Å$^{-1}$ and AFM $J1^Y = -0.0649$ Å$^{-1}$ (AFM) are virtually equal. The latter could indicate on the presence, in case of $YCu_3(OH)_6Cl_3$, of additional (both structural and "non-structural") interactions inducing a weakening of the AFM nearest-neighbor $J1$ couplings. Possibly, they may include the competition between the AFM $J1$ and $J2$ couplings ($J2/J1 = 0.25$) and a strengthening of the DM interaction (see paragraphs *3.3.1* and *3.2.2.1*). The coefficients values appeared to be as follows: $K^{Zn}_{exp} = -2835.8$, $K^{Zn}_{calculated} = -2716.4$ and $K^Y = -1263.5$. One should emphasize that, unlike the experiment, the crystal chemistry method enables one to calculate the magnetic parameters of individual couplings as ideal ones. This method does not take into account possible effects of other interactions and forces on these couplings.

The format of the initial data for the "MagInter" program (crystallographic parameters, atom coordinates) complies with the cif-file in the Inorganic Crystal Structure Database (ICSD) (FIZ Karlsruhe, Germany). The ionic radii of Shannon [27] ($r(^{IV}Cu^{2+}) = 0.57$ Å, $r(^{VI}O^{2-}) = 1.40$ Å, $r(^{VI}F^{1-}) = 1.331$, $r(^{VI}Cl^{1-}) = 1.81$, $r(^{VI}S^{6+}) = 0.12$ Å et al.) were used for calculations.

The method of selection of the material for our study was rather simple. Using the ICSD data, we identified 52 magnetic compounds with the kagome plane. Among them, only 32 compounds were based on copper. For these compounds, we calculated the parameters of magnetic couplings inside and between kagome planes. As a result of calculations, we excluded 11 compounds due to one or several reasons: (1) boucle kagome plane; (2) not all three interactions in small triangles in the kagome plane are antiferromagnetic; (3) presence of additional strong AFM $J2$ interactions in the kagome plane; (4) presence of strong interplane AFM or FM interactions, also due to location of additional magnetic and non-magnetic ions between kagome planes, which could participate in the coupling as intermediate ions; (5) strong non-stoichiometry in positions of as copper as intermediate ions.

Finally, we obtained just 21 magnetic compounds of copper, whose characteristics were more or less similar to those in the structural-magnetic model of herbertsmithite ($ZnCu_3(OH)_6Cl_2$). We also found 4 more compounds in the literature that are not available in the ICSD. All these compounds are characterized with strong AFM $J$1 couplings only between nearest neighbors and weak additional $J$2 and $J$3($J_d$) couplings. Magnetic couplings are very weak or virtually absent between kagome lattices. As was shown by the literature search, almost all of them had been studied repeatedly as experimentally, as theoretically. The latter provides substantial opportunities to reveal the role of structural factors in the formation of spin liquid, since, in spite of extensive studies of this material, there are still many ambiguities on the most basic problems. In the following section, we will demonstrate the necessity of taking into account the Jahn–Teller effect at calculations of the parameters of magnetic interactions between $Cu^{2+}$ ions by the crystal chemistry method.

## *2.1. Role of the Jahn–Teller effect in determination of the parameters of magnetic interactions in copper kagome compounds*

In 1934, when talking to E. Teller, L.D. Landau for the first time put forward the idea that, in the presence of electronic degeneracy (two or more orbital electronic states of equal energies), the nuclear configuration could turn out to be unstable and spontaneously deform. This idea verified later by Jahn and Teller [28] for all types of nonlinear molecules served as a basis for the Jahn–Teller theorem (JT), whose implications are now extensively used in analysis of the structure and properties of multiatom systems [29-34]. According to the Jahn–Teller theorem, if the system ground state is characterized by several nonequivalent degenerated energy levels, then the system distortion must split the degeneracy and lower one of the system energy levels.

The JT effect is the most explicitly manifested in $Cu^{2+}$ compounds. Depending on the concrete orbital the electron is located on, the geometric effect from its location is expressed in elongation or compaction of octahedra. Usually, the octahedron elongation corresponds to elongation of two bonds along both sides of the equatorial plane – (4+2)–coordination. Elongation is believed to be a regular phenomenon, whereas compaction in the form of a compressed octahedron is observed rather rarely. It is believed that the emergence of the (2+4) configuration as a compressed octahedron is related to two-dimensional distortion as a result of the dynamic JT effect – through mutual exchange of long and short bonds in the equatorial plane, which yields, by averaging over time, the average value of equatorial bonds lengths exceeding those of axial bonds. Bersucker described [31] the JT effect as a vibron one and considered this phenomenon from the point of d-d transitions. He does not exclude the situation when the conditions in a crystal stabilize not one, but several configurations of the coordination sphere close in energy that differ in this sphere spatial structure, creating isomers. These isomers differ in the fact that, at the same composition and ligand surrounding, the copper–ligand distances differ in different isomers. Besides, there are possible transitions from one isomer to another under effect of pressure, temperature, or storage time. In this case, not only two main isomers, but also a group of the so-called intermediate samples are possible. Here, they could differ not only in the synthesis method, but also in the appearance, crystal shape, chemical behavior, solubility, spectroscopic and magnetic properties etc.

The latter conclusion was corroborated by the emergence, due to the JT effect, of three isomers of volborthite ($Cu_3(V_2O_7)(OH)_2(H_2O)_2$) [34-36], which will be considered below. Two of them differ in the type of octahedron, whereas the third one contains copper ions in two coordinations: elongated and compacted octahedra.

We calculate the parameters of magnetic couplings based on the structural data, also in isomers – compounds that are identical in atomic composition, but differ in atoms spatial locations and, therefore, in magnetic interactions parameters. All the compounds we consider belong to a specific class of substances, whose crystal structure and, consequently, magnetic properties are largely determined by the presence of Jahn–Teller $Cu^{2+}$ ions with orbital degeneracy [29-33]. Earlier, at studies of magnetic structures of $KCuF_3$ [37-43], kamchatkite ($KCu_3OCl(SO_4)_2$) [44], averievite $Cu_5O_2(VO_4)_2(Cu+Cl)$ [45], ilinskite $NaCu_5O_2(SeO_3)_2Cl_3$ [45], and avdoninite $K_2Cu_5Cl_8(OH)_4 \cdot 2H_2O$ [45], we demonstrated that intermediate X ions, whose bond with copper has a JT elongation, do not contribute to the magnetic coupling. That is why, at calculations of magnetic couplings parameters $J_n$ by the crystal chemistry method, the $j(X^{ax})$ contribution to the magnetic coupling from intermediate X ions located at elongated positions with at least one of two $Cu^{2+}$ ions participating in the interaction must be neglected.

## 3. Results and discussion

In this chapter, we will calculate and discuss the structural-magnetic model of herbertsmithite ($ZnCu_3(OH)_6Cl_2$) and determine, on its basis, the crystal chemistry criteria of the existence of a spin liquid in kagome antiferromagnetics. We will calculate magnetic couplings parameters in 25 kagome cuprates and show how in strongly frustrated AFM systems even small factors could disrupt so fragile liquid state and result in ordering or other magnetic state.

### *3.1. Ways of formation of dominant AFM nearest-neighbor couplings on the kagome lattice*

There exist two main ways of formation of dominant AFM nearest-neighbor couplings on the kagome lattice in dependence on the number of copper ions per one intermediate X ion making the main contribution to the AFM component of this coupling (figure 1). The first type – $XCu_2$, when each of three edges of a small triangle in the kagome plane is centered by the X ion (figures 1(a) and (b)). These X ions are located approximately in the middle of its edges, but outside the planes of triangles above or below as, for example, in $ZnCu_3(OH)_6Cl_2$ [9], so that the AFM couplings along each triangle edge are formed due to contributions of individual intermediate X ions.

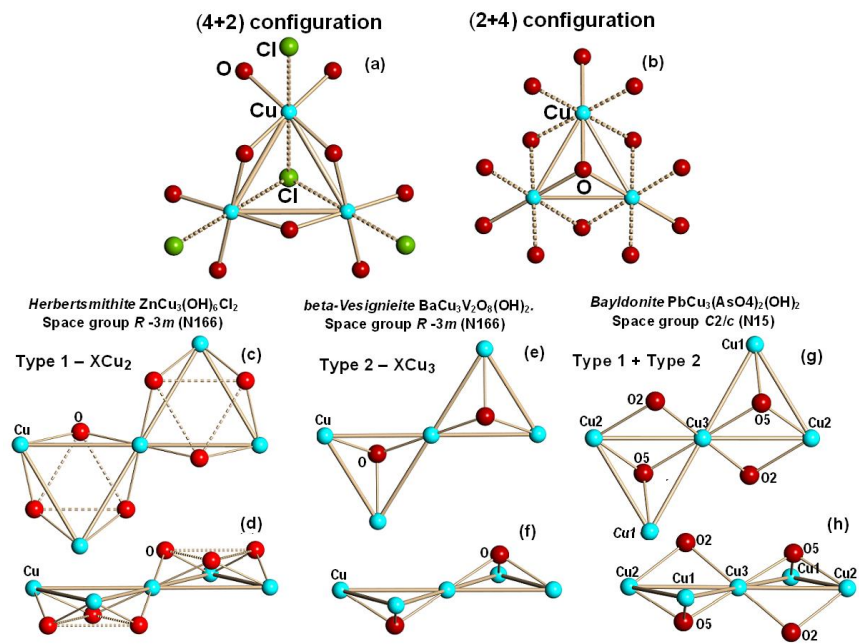

**Figure 1.** The configurations of (4+2) and (2+4) JT distortions of $Cu^{2+}$ octahedron (a) and(b). Two types of locations of intermediate oxygen ions forming magnetic couplings in the kagome plane. Projections of the fragments of crystal structures on the kagome plane (c, e, g) and perpendicularly to the plane (d, f, h) are shown.

The second type – $XCu_3$, when each small triangle in the kagome plane is centered by the X ion (figures 1(c) and (d)). This configuration can be considered as an anion-centered triangle ($XCu_3$) similarly to the oxy-centered tetrahedron ($OCu_4$). In this case, couplings along three triangle sides are formed due to one intermediate X ion located in the triangle center above or below its plane with some specific order, for example, in *beta-vesignieite* $BaCu_3V_2O_8(OH)_2$ [46].

There are cases, when the second type of formation of magnetic couplings is supplemented by the contribution of one more intermediate ion according to the type 1, for example, the O2 ion, into the coupling along the Cu2-Cu3 edge in *bayldonite* $Cu_3Pb(AsO_4)_2(OH)$ [47] (figures 1(e) and (f)). However, the O2 ion makes a small FM contribution and decreases insignificantly the strength of the AFM coupling between Cu2 and Cu3 ions. Note that $ZnCu_3(OH)_6Cl_2$ and *beta-vesignieite* $BaCu_3V_2O_8(OH)_2$ crystallize in the same space group, but the type of JT distortion is different for them. Nevertheless, we have not managed to find a simple relationship between the ways of formation of magnetic couplings in the kagome plane and types of configurations of (4+2) and (2+4) JT distortions or the number of anions per $Cu^{2+}$ ion in 25 compounds under consideration.

Hereinafter, discussing concrete materials, we will focus on the ways of formation of magnetic couplings and types of configurations if JT distortions.

### *3.2. Structural-magnetic model of herbertsmithite γ-$ZnCu_3(OH)_6Cl_2$ and crystal chemistry criteria of the existence of spin liquids on the kagome lattice*

Herbersmithite (γ-$ZnCu_3(OH)_6Cl_2$) [9] comprises the structurally perfect spin-1/2 kagome antiferromagnet [3] and a likely candidate of a quantum spin liquid [10]. It crystallizes in the centrosymmetric trigonal space group *R*-3*m* (N166). Magnetic $Cu^{2+}$ ions occupy 1 crystallographically independent site and have a JT distorted coordination polyhedron $CuO_4Cl_2$ in the form of an octahedron elongated along the axial direction Cl1-Cu1-Cl1 (type 4+2), where d(Cu1-O1) = 1.984Åx4 and d(Cu1-Cl1) = 2.765Åx2. These octahedra are linked through oxygen ions located in corners of their equatorial square planes and form kagome lattices of copper atoms located in oxygen surrounding (figure 2(a)). According to the ratio of the number of oxygen atoms to that of copper atoms equal to 2, each of the three *J*1 couplings along Cu1-Cu1 sides of small triangles in the kagome lattice corresponds to one intermediate oxygen O1 ion, located outside the triangle, in particular, from opposite sides in neighboring triangles. As a result, in a half of small triangles, oxygen ions are located above the plane, while in another half they are, in opposite, located below it. Between these copper-oxygen layers forming magnetic couplings, diamagnetic $Zn^{2+}$ ions not participating in magnetic couplings and $Cl^-$ ions occupying axial positions in octahedra are located. The minimal Cu1-Cu1 distance between layers is relatively small and equal to just 5.090 Å.

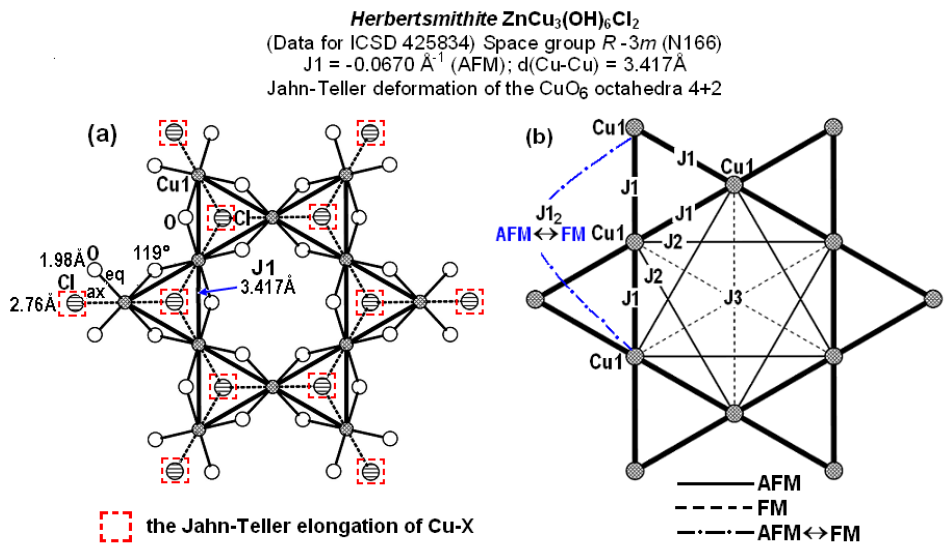

**Figure 2.** Kagome layer$CuO_4Cl_2$ octahedra (a) and Jn couplings in the kagome lattice (b) in herbertsmithite $ZnCu_3(OH)_6Cl_2$ (projected perpendicular to the c axis). In this and other figures, the thickness of lines shows the strength of Jn couplings. AFM and FM couplings are indicated by solid and dashed lines, respectively. The possible FM → AFM transitions are shown by the stroke in dashed lines.

Let us now determine, using the crystal chemistry method, what characteristics of magnetic couplings would describe herbertsmithite ($ZnCu_3(OH)_6Cl_2$), if their formation was determined exclusively by the crystal structure. The calculation of the sign (type of orientation of magnetic moments) and strength of magnetic inetractions ($J_{ij}$) was performed on the structural data for $ZnCu_3(OH)_6Cl_2$ obtained in Braithwaite et al. [9].

According to our calculations, the AFM nearest-neighbor $J1$ couplings ($J1$ = -0.0670 Å$^{-1}$, d(Cu1-Cu1) = 3.417 Å) between Cu1 ions in the kagome lattice are dominant and compete to each other in small triangles (figures 2(a) and (b) Supplementary table 1). These $J1$ couplings are formed under effect of O1 ions forming Cu1-O1-Cu1 angles equal to 118.9° with Cu1 ions from small triangles. Along the sides of two large triangles fit into honeycombs, the AFM $J2$ couplings ($J2$ = -0.0108 Å$^{-1}$, d(Cu1-Cu1) = 5.918 Å, $J2/J1$ = 0.16) are 6.2-fold weaker than the AFM $J1$ couplings.

The role of the $J2/J1$ ratio in the emergence of the spin liquid has been studied extensively in the literature, albeit without obtaining unambiguous results. For example, as was shown by studies of the phase diagram of the $J1$–$J2$ Heisenberg model on the kagome lattice by Kolley *et al* [48], the magnetic order in the range -0.1≤$J2/J1$≤0.2 was absent in a narrow interval around $J2$ ~ 0, which was compatible with the spin-liquid behavior. As was shown in [49], a substantial spin-liquid phase was centered near $J2/J1$ = 0.05-0.15, while in [50] the limits of the existence of spin liquid were expanded to $J_2/J_1$≤0.3. In the case of kagome, the narrowest range of stability for the existence of the gapless spin-liquid ground state (-0.03 ≤$J_2/J_1$≤ 0.045) is presented in [51]. Besides, the system of $J1$ and $J2$ couplings can be considered as AFM zigzag spin ½ chains with nearest- and next-nearest neighbor interactions $J1$ and $J2$, где $J1$>>$J2$. These chains are elongated in the directions of $J2$ couplings in the kagome lattice. Possible quantum states on the spin chain at different $J2/J1$ rations were studied in [52-55]. The value for the latter critical frustration of $J2$ = 0.24$J1$ [52]

The crystallographic equivalents of $J1$ and $J2$ couplings in the kagome lattice of $Cu^{2+}$ ions are magnetic equivalents as well.

*3.2.1. Role of J3 couplings in ordering of kagome antiferromagnets.* A special place in the kagome lattice belongs to the third in length (d(Cu1-Cu1) = 6.834 Å) $J3$ coupling (figure 3, Supplementary table 1). It is dual, consisting of two crystallographically identical parts of the same sixfold $J3$: $J3(J_d)$ couplings along hexagon diagonals and $J3(J1_2)$ next-to-nearest-neighbour couplings in linear chains along the triangles sides (figure 3(a) and (b)). The crystallographically equivalent $J3(J_d)$ and $J3(J1_2)$ couplings are not magnetic equivalents and decrease the symmetry of the magnetic lattice relatively to the symmetry of its crystal structure. Nonequivalence of the parameters of the $J3(J_d)$ and $J3(J1_2)$ magnetic couplings originates from the kagome lattice structure itself, namely, the presence of an empty hexagonal space in the triangular kagome lattice. In spite of the fact that the space groups N12, N15, N148, N164, N166 and N19, in which the compounds we examine crystallize, are centrosymmetric, copper ions occupy in them noncentrosymmetirc equivalent positions that facilitates the emergence of DM interaction in kagome lattices. In two other centrosymmetric space groups (N11 and N14), copper ions occupy as noncentrosymmetric as centrosymmetric equivalent positions. Just two of the examined compounds crystallize in the noncentrosymmetric space groups (N9 and N36), in which copper ions occupy noncentrosymmetrcic equivalent positions. The kagome lattice can be obtained from the triangular lattice with shared corners and edges, if one removes from it a part of ions, so that the triangular lattice would be just corner shared (figure 2 (b) and 3 (b)). In view of this, let us consider in detail the structural and magnetic nonequivalence in the kagome lattice of $ZnCu_3(OH)_6Cl_2$ (figures 3(a) and (b)) and its role in the emergence of magnetic ordering.

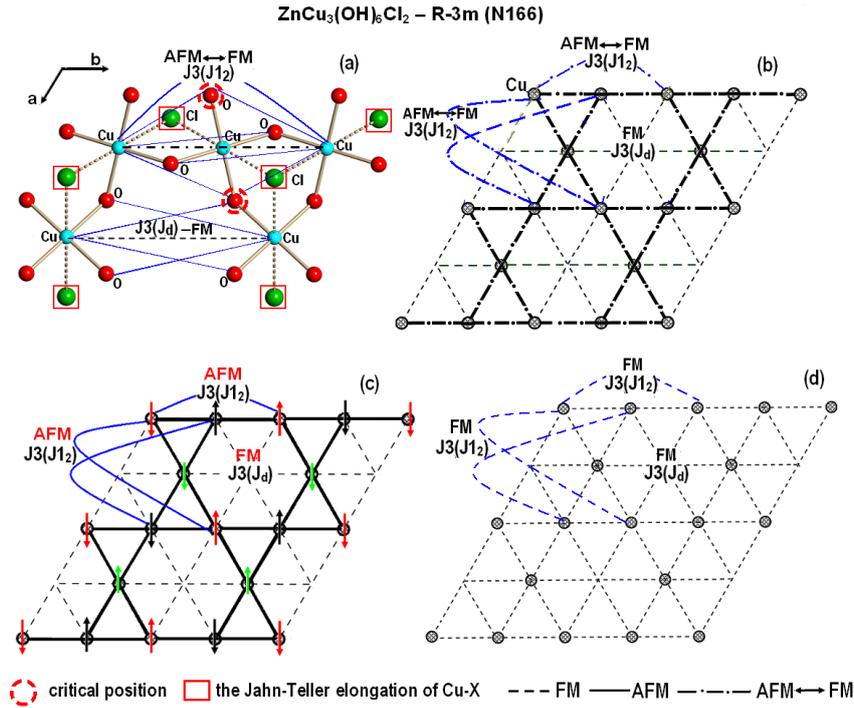

**Figure 3.** The arrangement of intermediate ions in the local spaces of $J3(J_d)$ and $J3(J1_2)$ couplings in the kagome lattice of $ZnCu_3(OH)_6Cl_2$ (a); nonequivalence of magnetic parameters of the crystallographically equivalent $J3(J_d)$ and $J3(J1_2)$ couplings (b); scheme of possible ordering in the kagome lattice under effect of the strong $J3(J1_2)$ couplings (c); absence of duality of the third in length $J3(J1_2)$ and $J3(J_d)$ magnetic couplings (d). Canting of the atomic moments is not shown.

According to our calculations, the $J3(J_d)$ coupling ($J3 = 0.0018$ Å$^{-1}$, d(Cu1-Cu1) = 6.834 Å, $J3(J_d)/J1$ = -0.03) is ferromagnetic and very weak, since its local space is virtually free from intermediate ions. Just four oxygen atoms make significant contributions ($j(O1) = 0.00045 \times 4$) to the ferromagnetic component of this coupling.

The things are different with its crystallographically equivalent coupling. The AFM $J3(J1_2)$ coupling ($J3(J1_2) = -0.0300$ Å$^{-1}$, d(Cu1-Cu1) = 6.834 Å) is formed in the chain along the sides of small triangles mainly under effect of the intermediate $Cu^{2+}$ ion (figure 3(a)). This intermediate $Cu^{2+}$ ion is locetd in the middle of the Cu-Cu bond line and makes a substantial AFM contribution ($j(Cu1)$: -0.0244 Å$^{-1}$). Also, two oxygen atoms add small contributions (-0.0028 Å$^{-1}$) into the AFM component of the $J3(J1_2)$ coupling. As a result, the next-nearest AFM $J3(J1_2)$ coupling is sufficiently strong to compete with the nearest AFM $J1$ coupling ($J3(J1_2)/J1 = 0.45$). Moreover, these couplings could suppress the frustration of AFM $J1$ couplings in triangles. Figure 3(c) shows the scheme of one of possible variants of such an ordering.

However, the AFM $J3(J1_2)$ couplings are unstable. The point is, in herbertsmithite, the local space of the $J3(J1_2)$ coupling is crossed by two more oxygen ions at the distance $\Delta a = 0.026$ Å (figure 3(a)) that is smaller than the critical value $\Delta a \sim 0.1$ Å ($\Delta a = (r_M + r_{An}) - h_{An}$) [11-13], supplementary figure 1), so that they cannot initiate the emerging of magnetic interaction ($j(O) = 0$). In herbertsmithite, at the temperature decrease, these oxygen ions could move slightly deeper inside this local space. Then, there will emerge a significant FM contribution from these ion, which will partially or completely suppress the strength of the AFM component of the $J3(J1_2)$ coupling until reorientation of magnetic moments (AFM → FM transition), also creating the main effect – equalization of the parameters of the $J3(J_d)$ and $J3(J1_2)$ magnetic couplings. The mechanism of AFM to FM transition in $J3(J1_2^n)$ couplings on the kagome lattice is characteristic for all the compounds we studied.

For example, without changing the space group and cell parameters, let us shift two additional oxygen ions deeper (by 0.161 Å) into the local space of $J3(J1_2)$ until $\Delta a = 0.187$ Å through replacement of the initial coordinates O1 (x = 0.2056, y = -0.2056, z = 0.0612) by new ones (x = 0.2150, y = -0.2150, z = 0.0410). In this case, the FM contribution from this two ions increases up to 0.0342 Å$^{-1}$ and suppresses the strength of the AFM component (-0.0342Å$^{-1}$) of this coupling. Besides, the following changes would occur as result of this shift: the d(Cu1-O1) distance would decrease from 1.984 down to 1.850 Å; the AFM $J1$ coupling would increase (1.76-fold) to $J1 = -0.1181$ Å$^{-1}$; and the AFM $J2$ one would increase to a noticeably smaller degree (1.18-fold) to $J2 = -0.0128$ Å$^{-1}$ ($J2/J1 = 0.11$). To sum up, we have demonstrated that the shift of two oxygen ions located in the critical position "$a$" ($\Delta a \sim 0.1$) deeper into the local space of $J3(J1_2)$ could result in equalization of magnetic parameters of two crystallographically equivalent $J3(J_d)$ and $J3(J_d)$ couplings and, therefore, hamper ordering of the frustrated magnetic structure.

There exist many theoretical studies [56 - 59] outlining the frames, which must include the parameters of $J1$, $J2$, and $J3(J_d)$ magnetic couplings (figure 2) in the kagome plane determining the possibilities of the formation of the quantum spin-liquid state (QSL). However, there is no unambiguous conclusion on the matter. According to [60], the strength of the interplane $J_\perp$ bond must not exceed 15 % of that of the intraplane $J1$ bond to preserve the spin-liquid state.

According to our calculations, relative strengths of the couplings in the kagome plane of $ZnCu_3(OH)_6Cl_2$ would be located in the ranges $-0.03 \leq J3(J_d)/J1 \leq J2/J1 \leq 0.16 \leq J3(J1_2)/J1 \leq 0.45$ and $J2/J1 \leq 0.11$ ($J3(J1_2)/J1 = J3(J_d)/J1=0$) before and after the oxygen ion shift inside the local space of the $J3(J1_2)$ coupling, respectively. The strength of couplings ($J_\perp$ - AFM $J4$, $J5$', $J6$, and FM $J5$) at distances d(Cu-Cu) = 5.09 Å – 7.018 Å is insignificant and constitutes just from 2 to 5 % of that of intraplane $J1$ couplings (figure 4, supplementary table 1). The strengths of two (FM $J7$ and AFM $J8$) couplings located at a distance larger than ~8 Å attain 12 and 9 % of the $J1$ coupling strength, respectively. However, the strengths of these remote couplings could be exaggerated, since the decrease of the coupling strength accelerates along with the distance increase and must be inversely proportional not to the distance square, as accepted in our method, but to its cube. Unfortunately, the available literature does not contain a sufficient bulk of reliable data to take this effect into account in our method.

Many works [3-6, 22, 24, 61, 62] are devoted to revealing the role of different types of defects, structural distortions, and impurities in as destruction as stabilization of the spin-liquid ground state. The point is, different experiments revealed 4–7 % of substitution of $Cu^{2+}$ ions by $Zn^{2+}$ ions in $ZnCu_3(OH)_6Cl_2$.

Not denying a substantial effect of impurities on the substance crystal structure and, therefore, its magnetic properties, we believe that the emergence and destruction of spin liquid in compounds is controlled by nonequivalence of magnetic and structural subsystems. This nonequivalence is manifested in anisotropy (duality) of third in length $J3(J1_2)$ and $J3(J_d)$ magnetic couplings that are crystallographically identical.

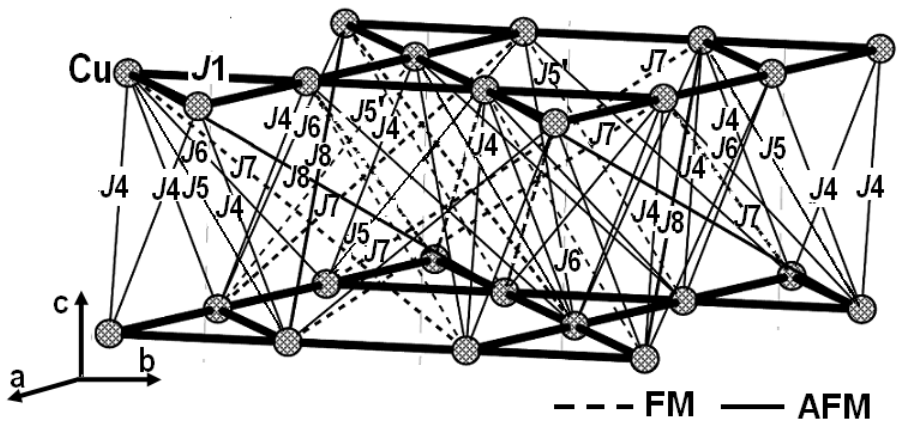

**Figure 4.** The Interplane magnetic couplings in $ZnCu_3(OH)_6Cl_2$.

However, insignificant shifts of intermediate oxygen ions in herbertsmithite could attenuate or even eliminate this anisotropy. It is possible in the kagome system only in the case when the $J3(J1_2)$ value becomes as weak as $J3(J_d)$, so that the system remains frustrated at the temperature decrease down to 0 K. The latter phenomenon that is extremely rare for kagome systems was found absolutely by accident in $ZnCu_3(OH)_6Cl_3$.

To sum up, elimination of the duality in the magnetic subsystem will and support for emergence of the DM ordering. From the point of crystal chemistry, our conclusions are in agreement with those put forward by Cépas et al. [63] that the kagome compound $ZnCu_3(OH)_6Cl_3$ may be in a quantum critical region controlled by Dzyaloshinskii-Moriya coupling.

*3.2.2. Crystal chemistry criteria of the existence of spin liquids on the kagome lattice.* In the course of buildup of the structural-magnetic model of herbertsmithite $ZnCu_3(OH)_6Cl_3$, we established three main characteristics required for the existence of spin liquids on the kagome lattice. They include:

(a) presence of frustrated kagome lattices with strong dominant AFM nearest neighbor $J1$ couplings competing only with each other in small triangles;

(b) magnetic isolation of these frustrated lattices;

(c) absence of duality (anisotropy) of third in length $J3(J1_2)$ and $J3(J_d)$ magnetic couplings that are crystallographically identical.

As was shown above, the presence of duality of this $J3$ in the magnetic structure originates from the crystal structure of the kagome lattice itself. Elimination of such a duality - nonequivalence of magnetic parameters of two crystallographically identical $J3(J_d)$ couplings along the hexagon diagonals and $J3(J1_2)$ next-to-nearest-neighbour couplings in linear chains along triangles sides could occur through insignificant shifts of intermediate ions in local spaces of these couplings. The latter means that the additional anisotropic interaction, which suppresses frustration in combination with the DM one, disappears, and the probability of the existence of a spin liquid increases.

The first two criteria are evident and accepted by a majority of researchers, whereas feasibility of the third one will be grounded below at discussions of other spin-1/2 kagome antiferromagnets.

### *3.3. Kagome antiferromagnets with (4+2)-elongated octahedral $Cu^{2+}$ – analogs of herbertsmithite*

*3.3.1. Isostructural analog of herbertsmithite – $\gamma$-$MgCu_3(OH)_6Cl_2$.* Mg-herbertsmithite, $\gamma$-$MgCu_3(OH)_6Cl_2$, [64] is isostructural to Zn-herbertsmithite, $\gamma$-$ZnCu_3(OH)_6Cl_2$, and, therefore, their structures belong to the same space group, in which atoms occupy identical regular point systems, whereas the atoms coordinations in points of occupied regular systems are identical (supplementary table 1). Moreover, due to similarity of $Zn^{2+}$ and $Mg^{2+}$ radii, geometric parameters of copper-oxygen layers in these isostructural systems differ insignificantly. Since we calculated the sign and strength of magnetic interactions using these very geometric parameters, the result was a virtually unambiguous correlation of the above magnetic parameters in Zn- and Mg-systems (supplementary table 1). Mg-herbertsmithite, just like Zn-herbertsmithite, is a structurally-perfect kagome antiferromagnet, in which there exists the possibility to eliminate anisotropy of crystallographically equivalent $J3(J_d)$ and $J3(J1_2)$ couplings through shifts of two intermediate oxygen ions located in the critical position "a" ($\Delta a$ = 0.017 Å), slightly deeper inside the local space of $J3(J1_2)$.

According to Colman et al. [64], the Mg-herbertsmithite phases show no evidence of magnetic transitions down to 1.8 K, and low-temperature magnetometer measurements suggest the formation of a quantum spin liquid (QSL). The spin-liquid behavior in AFM Mg-herbertsmithite is clearly established by the muon spin resonance ($\mu$SR) experiments [65].

*3.3.2. Non-isostructural analogs of herbertsmithite.* Let us consider 7 structurally-perfect kagome antiferromagnets (supplementary table 1), in which ideal kagome lattices are realized, but these compounds are not isostructural to γ-ZnCu$_3$(OH)$_6$Cl$_2$. They crystallize in two different trigonal/rhombohedral centrosymmetric space groups. Four compounds – YCu$_3$(OH)$_6$Cl$_2$ [66], *haydeeite* MgCu$_3$(OH)$_6$Cl$_2$ [67], MgCu$_3$(OH)$_6$Br$_2$ [68], and CdCu$_3$(OH)$_6$(NO$_3$)$_2$H$_2$O [69] – crystallize in the space group *P*-3*m*1 (N164), whereas three fluorides – Cs$_2$SnCu$_3$F$_{12}$ [70], Cs$_2$ZrCu$_3$F$_{12}$ [71], and Cs$_2$TiCu$_3$F$_{12}$ [72] – crystallize in the space group *R* -3*mH* N164, just like herbertsmithite. In these compounds, copper ions could be to an insignificant degree substituted by Mg, Cd, Ti, or Sn ions, but the X-ray diffraction analysis did not detect it. Since isomorphism of copper was not virtually observed with yttrium and zirconium, in YCu$_3$(OH)$_6$Cl$_2$ and Cs$_2$ZrCu$_3$F$_{12}$, kagome lattices cannot be distorted by substitution.

Our attempts to find, in the above compounds, some regular dependence of the Cu-Cu distances in the kagome plane and the distances from intermediate of oxygen (fluorine) to this plane on the charge value and diamagnetic ions metal ions radii were not successful. The latter is associated with different locations of diamagnetic ions, for example, between copper-oxygen layers, as in herbertsmithite or its polymorphous modification kapellasite, in which ions sit at the centers of the hexagons. Besides, introduction of additional ions to the space between copper-oxygen layers for charge compensation provide final complication of the problem of search of such regularity and force us to drop this idea. The reason here consists in simultaneous variation of these two values: Cu-Cu distances in the kagome lattice and $h(A_n)$ distances from intermediate oxygen (fluorine) ions to the kagome plane at substitution. One should emphasize that these very characteristics serve as a basis in formation of *J*1, *J*2, and *J*3 magnetic parameters of the kagome lattice, which makes it necessary to consider peculiar features of each compound individually.

*3.3.2.1. YCu$_3$(OH)$_6$Cl$_2$.* Let us calculate the parameters of magnetic couplings in YCu$_3$(OH)$_6$Cl$_2$ (figure 5, Supplementary table 1) and compare them with respective data in herbertsmithite ZnCu$_3$(OH)$_6$Cl$_2$. Upon substitution of Zn by Y, the parameters of dominant AFM nearest-neighbor $J1^Y$ couplings ($J1^Y$ = -0.0649 Å$^{-1}$, d(Cu1-Cu1)$^Y$ = 3.375 Å, $J1^Y/J1^{Zn}$ = 0.97, d(Cu-Cu)$^Y$/d(Cu-Cu)$^{Zn}$ = 0.99) in YCu$_3$(OH)$_6$Cl$_2$ remained close to respective values in herbertsmithite. It is natural that similar result was also obtained in the case of the next-nearest AFM $J3(J1_2)^Y$ couplings ($J3(J1_2)^Y$ = -0.0304AFM Å$^{-1}$, d(Cu1-Cu1)$^Y$ = 6.749 Å, $J3(J1_2)^Y/J3(J1_2)_2^{Zn}$ = 1.01, d(Cu-Cu)$^Y$/d(Cu-Cu)$^{Zn}$ = 0.99). AFM $J2^Y$ couplings ($J2^Y$ = -0.0160 Å$^{-1}$. d(Cu1-Cu1)$^Y$ = 5.845 Å, $J2^Y/J2^{Zn}$ = 1.48 and d$^Y$(Cu-Cu)/d(Cu-Cu)$^{Zn}$ = 0.99) along the sides of hexagons (in other words, along the sides of large triangles) underwent the most significant changes in the Y-system. The strength of $J2^Y$ couplings attained a one-fourth of that ($J2^Y/J1^Y$ = 0.25) of the nearest neighbor $J1^Y$ couplings, which determined the possibility of competition between them.

In opposite, couplings along diagonals of hexagons ($J3(J_d)$) became even weaker in the Y-system and changed the sign, thus transforming into AFM $J3(J_d)^Y$ couplings ($J3(J_d)^Y$ = -0.0012 Å$^{-1}$, d(Cu-Cu)$^Y$ = 6.749 Å, $J3(J_d)^Y/J3(J_d)^{Zn}$ = 0.67, d$^Y$(Cu-Cu)/d$^{Zn}$(Cu-Cu) = 0.99). The probability of attaining, in some point, an equal strength of crystallographically identical, but magnetically nonequivalent $J3(J_d)^Y$ and $J3(J1_2)^Y$ ($J3(J1_2)^Y/J3(J_d)^Y$ = 25.3) couplings, due to the shift deep inside the local space $J3(J1_2)^Y$ of two intermediate oxygen ions located in the critical position "*a*" (Δ*a* = 0.007 Å), is very low. This would require significant changes in Cu-O distances upon the temperature decrease.

The ranges of parameters of $J1^Y$, $J2^Y$, and $J3^Y$ couplings of YCu$_3$(OH)$_6$Cl$_2$ (figure 5(b)) became somewhat expanded (0.02 ≤ $J3^Y/J1^Y$ < $J2^Y/J1^Y$ ≤ 0.25) in comparison with ZnCu$_3$(OH)$_6$Cl$_2$ because of the increase of the strength of the AFM *J*2 coupling.

The Y-system does not virtually include interplane magnetic couplings at nearest distances ($J4^Y$ = 0, d(Cu1-Cu1)$^Y$ = 5.625 Å), $J5^Y$ ($J5^Y/J1^Y$ = 0.01, d(Cu1-Cu1)$^Y$ = 6.559 Å). However, at long distances, interplane magnetic couplings are slightly stronger ($J6(J6')^Y/J1^Y$ = 0.04(-0.02), d(Cu1-Cu1)$^Y$ = 8.111; $J7(J7')^Y/J1^Y$ = 0.02(0.06), d(Cu1-Cu1)$^Y$ = 8.785). Besides, unlike ZnCu$_3$(OH)$_6$Cl$_2$, the kagome planes in YCu$_3$(OH)$_6$Cl$_3$ contain the Cl2 ions that are not included into the copper coordination, but mainly serve to create electroneutrality at substitution of Zn$^{2+}$ ions by Y$^{3+}$ ions. However, these very Cl2 ions enter the local space of the $J7$ couplings (Supplementary figure 5(c)). If one formally takes into account the AFM contribution that these ions could make into such couplings, their strength could increase substantially.

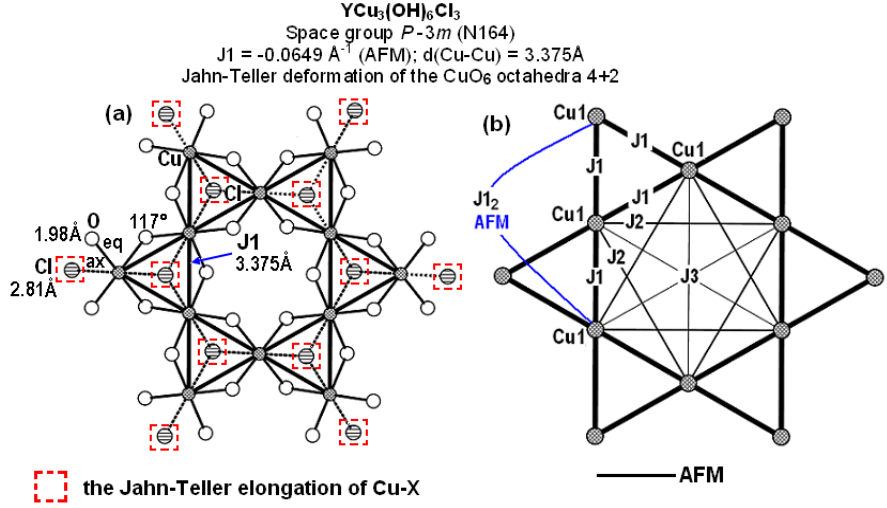

**Figure 5.** Kagome layer CuO$_4$Cl$_2$ octahedra (a) and $J_n$ couplings in the kagome lattice (b) in YCu$_3$(OH)$_6$Cl$_2$ (projected perpendicular to the $c$ axis)

The quantum kagome antiferromagnet YCu$_3$(OH)$_6$Cl$_3$ attracted a substantial attention of researches as a possible holder of the spin-liquid state. However, Barthélemy et al. [73] demonstrated on the basis of low-temperature neutron diffraction data that a disordered static magnetic is under development in the compound YCu$_3$(OH)$_6$Cl$_3$. Zorko et al. [74, 75] revealed experimentally the magnetic ordering emerging in this material at $T_N$ = 15K. The authors see one of tentative reasons for ordering in the existence of further-neighbor exchange interactions, which could have a decisive role in destabilizing the quantum spin-liquid state in YCu$_3$(OH)$_6$Cl$_3$. Based on the results of our calculations, we can suggest the AFM $J2^Y$ coupling in the kagome plane (figure 5(b)) for the role of the above interactions. This coupling is just fourfold weaker than the dominant AFM nearest neighbor coupling $J1^Y$, so that it could affect the spin orientation in small triangles.

However, later [46], these authors corrected the magnetic parameters in YCu$_3$(OH)$_6$Cl$_3$ and concluded that the nearest-neighbor exchange interaction $J1$ = 82(2) K and exchange interactions beyond the nearest neighbors are limited by 5 % of $J1$ at the maximum. Moreover, they presented extra arguments in favor of the dominating role of the DM in the origin of magnetic ordering in a structurally perfect quantum kagome antiferromagnet). From our side, we affirm this conclusion and demonstrate that the main reason of the emergence of the DM interaction in YCu$_3$(OH)$_6$Cl$_3$ could consist in the duality – nonequivalence of crystallographically identical $J3^Y$ ($J3(J_d)^Y/J3(J1_2)^Y$ = 0.04) magnetic interactions in the kagome plane.

*3.3.2.2. Haydeeite α-MgCu$_3$(OH)$_6$Cl$_2$ and α-MgCu$_3$(OH)$_6$Br$_2$.* Our calculations demonstrate (supplementary table 1) that in haydeeite α-MgCu$_3$(OH)$_6$Cl$_2$ [67] the dominant AFM $J1$ ($J1$ = -0.0399 Å$^{-1}$, d(Cu1-Cu1) = 3.137 Å) nearest-neighbor couplings are virtually twofold weaker than in Zn- and Y-systems, since |Δh(O)| – the degree of overlapping of the local space between magnetic copper ions by the oxygen ion (Supplementary figure 1(a)) in the $J1$ coupling of this compound – is almost twofold smaller. The strength of the second in length AFM $J2$ coupling ($J2$ = -0.0162 Å$^{-1}$, d(Cu1-Cu1) = 5.433 Å) remains about the same as in YCu$_3$(OH)$_6$Cl$_3$, while the strengths ratio $J2/J1$ increases dramatically (up to $J2/J1$ = 0.41), and, therefore, the competition between AFM $J1$ and AFM $J2$ couplings becomes more active.

Aside from the above competition, which is undesirable for the emergence of the spin-liquid state, here, as in the Y-system, one observes the another remaining problem consisting in nonequivalence of the parameters of magnetic interactions of crystallographically equivalent $J3(J_d)$ and $J3(J1_2)$ third in length (d(Cu1-Cu1) = 6.273 Å) couplings. The AFM $J3(J_d)$ ($J3$ = -0.0028 Å$^{-1}$, $J3/J1$ = 0.07) couplings along the hexagon diagonals are weak, whereas the next-nearest AFM $J3(J1_2)$ couplings ($J3(J1_2)$ = -0.0324 Å$^{-1}$, $J3(J1_2)/J1$ = 0.81) in the chain along the sides of small triangles are strong and compete with the $J1$ couplings. However, unlike the case of herbertsmithite, it is hardly possible to exclude this nonequivalence ($J3(J1_2)/J3(J_d)$ = 11.6), since two additional oxygen ions capable to provide it do not even reach the border of the local space of the $J3(J1_2)$ coupling ($\Delta a$ = -0.005 Å).

We calculated the interplane couplings for the distances around d(Cu1-Cu1) ~8.51 Å. All of them except one AFM $J6'$ coupling ($J6'/J1$ = 0.18, d(Cu1-Cu1) = 7.909 Å) appeared to be weak or absent. According to [78, 79], below $T_c$ = 4.2 K, haydeeite α-MgCu$_3$(OH)$_6$Cl$_2$ manifests a long-range magnetic order.

Literature sources provide contradictory data on the parameters of magnetic couplings in haydeeite α-MgCu$_3$(OH)$_6$Cl$_2$. Based on electronic structure calculations, Janson et al. [78] demonstrated that haydeeite MgCu$_3$(OH)$_6$Cl$_2$ comprised a 2D magnet, with two relevant AFM exchanges: the NN exchange $J1$ and the exchange along ''diagonals'' of a kagome lattice $J3(J_d)$ and α ≡ $J3(J_d)/J1$ ≈1. The same conclusion was made by Colman et al. [76]. However, we revealed that the second in strength coupling was not the $J3(J_d)$ exchange along 'diagonals', but another one crystallographically identical to it AFM $J3(J1_2)$ coupling ($J3(J1_2)/J1$ = 0.81, d(Cu1-Cu1) = 6.273 Å) at the same distance (Supplementary table 1) This strong AFM $J3(J1_2)$ coupling is mainly formed under effect of the intermediate Cu1 ion (the angle Cu1-C1-Cu1 = 180°), whereas the central part of the local space of the $J3(J_d)$ coupling ($J3(J_d)/J1$ = 0.07, d(Cu1-Cu1) = 6.273 Å) is virtually void. Only from sides, there enter 4 oxygen ions making insignificant contributions ($j$(O1): -0.0007x4) to the coupling AFM component. According to the crystal chemistry method conditions, hydrogen ions are not taken into account.

Conclusions made by Boldrin et al. [77] based on the neutron inelastic scattering data differ from that made by Janson et al. [78] and Colman et al. [76] and from our results. In opposite to us, they believe that haydeeite, α-MgCu$_3$(OD)$_6$Cl$_2$, comprises a ferromagnetic with FM couplings between nearest neighbors $J1$ = -38 K and AFM couplings along diagonals $J3(J_d)$ = +11 K.

To sum up, the results of our calculations reflecting the contribution of structural factors in formation of the magnetic structure demonstrate that haydeeite, α-MgCu$_3$(OH)$_6$Cl$_2$, contains, in addition to the strong AFM nearest-neighbor $J1$ couplings along the sides of small triangles, two more comparatively strong AFM $J3(J1_2)$ ($J3(J1_2)/J1$ = 0.81) and $J2$ ($J2/J1$ = 0.41) couplings competing to each other. Such a competition could result, in the presence of anisotropy between the $J3(J1_2)$ and $J3(J_d)$ couplings eliminating centrosymmetry of the magnetic subsystem, in ordering under effect of the DM forces at the temperature decrease.

Substitution of chlorine ions by those of bromine in α-MgCu$_3$(OH)$_6$Cl$_2$ did not result in significant changes in the parameters of magnetic couplings in α-MgCu$_3$(OH)$_6$Br$_2$ (Supplementary table 1) we calculated using the structural data provided in [68]. The magnetic system orders antiferromagnetically at 5.4 K [68].

*3.3.2.3. CdCu$_3$(OH)$_6$(NO$_3$)$_2$H$_2$O.* The parameters of magnetic couplings in the kagome AFM CdCu$_3$(OH)$_6$(NO$_3$)$_2$H$_2$O compound [69] (supplementary table 1) differs insignificantly from those in the kagome AFM haydeeite, MgCu$_3$(OH)$_6$Cl$_2$, according to our calculations. In the kagome lattice, the AFM $J2$ couplings ($J2$ = -0.0130 Å$^{-1}$, d(Cu1-Cu1) = 5.648 Å) along the sides of long triangles can also compete with the dominant AFM nearest-neighbor couplings $J1$ ($J1$ = -0.0340 Å$^{-1}$, d(Cu1-Cu1) = 3.261 Å), since the strength of their couplings is just 2.6-fold smaller than that in the $J1$ couplings ($J2/J1$ = 0.38).

Our estimation of the type of orientation of magnetic moments and the relation of the strengths of the $J1$ and $J2$ magnetic couplings based on the structural data is in many aspects in agreement with the conclusions made by Okuma et al. [79] on the basis of magnetization, magnetic torque, and heat capacity measurements using single crystals. They established that CdCu$_3$(OH)$_6$(NO$_3$)$_2$H$_2$O was a kagome antiferromagnet with the AFM nearest-neighbor exchange interaction $J1$ = 45 K and suggested a spin order below $T_N$ ~ 4 K. Besides, according to their estimation, the antiferromagnetic $J2$ and ferromagnetic $J3(J_d)$ couplings are smaller than half of the dominant antiferromagnetic $J1$ ones. However, their estimation of parameters of the $J3(J_d)$ couplings comprising diagonals in the hexagon in the kagome plane does not agree with our data, according to which $J3(J_d)$ ($J3(J_d)$ = 0, d(Cu1-Cu1) = 6.522 Å). As was shown above (paragraph *3.2.1*), three $J3(J_d)$ couplings were just a ½ part of the sixfold $J3$ coupling. The second part of this coupling includes the crystallographically identical to them $J3(J1_2)$ next-to-nearest-neighbour couplings ($J3(J1_2)$ =-0.0296 Å$^{-1}$, d(Cu1-Cu1) = 6.522 Å) in linear chains along the triangles sides, whose strengths are just slightly ($J1/J3(J1_2)$ =1.15) smaller than those of the $J1$ couplings. Interplane couplings in this antiferromagnetic are weak.

To sum up, in four kagome antiferromagnets – YCu$_3$(OH)$_6$Cl$_2$, haydeeite MgCu$_3$(OH)$_6$Cl$_2$, MgCu$_3$(OH)$_6$Br$_2$ and CdCu$_3$(OH)$_6$(NO$_3$)$_2$H$_2$O – crystallizing in the same space group *P-3m*1 (N164), containing structurally perfect (ideal) kagome lattices, but not isostructural to herbertsmithite, two main criteria of the existence of spin liquid on the kagome lattice are invalid.

First, in addition to the dominant AFM nearest-neighbor $J1$ exchange interaction, we revealed comparatively strong long-range AFM $J2$ ($J2/J1$ = 0.25-0.41) interactions. The weakest AFM $J2$ coupling ($J2/J1$ = 0.25) is present in YCu$_3$(OH)$_6$Cl$_2$.

Second, there is a clearly expressed duality – the magnetic nonequivalence of crystallographically equivalent $J3(J1_2)$ ($J3(J1_2)/J1$ = 0.8 – 0.9) and $J3(J_d)$ ($J3(J_d)/J1$ = 0 – 0.09) couplings. Probably, it is possible to overcome such a nonequivalence for just one (YCu$_3$(OH)$_6$Cl$_2$) of four compounds examined in this paragraph.

*3.3.3. A family of mixed metal fluorides and structural and magnetic transitions with reduction of symmetry at low temperatures.* The Cs$_2$Cu$_3$BF$_{12}$ family (where B – Sn [70], Or [71], and Ti [72]) compounds crystallize until the structural phase transition in the same as herbertsmithite (γ-ZnCu$_3$(OH)$_6$Cl$_2$) rhombohedral crystal system with the space group *R-3m* (N166), in which Cu$^{2+}$ ions form an ideal kagome plane. Just like in herbertsmithite, magnetic Cu$^{2+}$ ions occupy just one crystallographically independent site and have a Jahn-Teller distorted coordination polyhedra (Cu1F$_6$) in the form of an octahedron stretched along the axial F2-Cu1-F2 direction of the (4+2) type, in which the lengths of four short bonds are in the range d(Cu1-F1) = 1.897−1. 903 Å, while those of two long ones - d(Cu1-F2) = 2.33-2.36 Å.

According to our calculations (figure 6, supplementary table 1), the parameters of magnetic couplings in these three antiferromagnetics – Cs$_2$Cu$_3$BF$_{12}$ (B – Sn, Zr and Ti) – are in ideal with the criteria of spin-liquid emergence. Let us examine it in detail. As herbertsmithite (ZnCu$_3$(OH)$_6$Cl$_2$), they comprise structurally perfect kagome antiferromagnets with the strong AFM nearest neighbor $J1$ ($J1$ =-0.1021 − -0.1101 Å$^{-1}$, d(Cu1-Cu1) = 3.551, − 3.583 Å) couplings in the kagome plane. From our data, the strengths of the AFM $J1$ couplings in fluorides exceed those in herbertsmithite 1.52–1.64-fold. According to other estimates, the AFM $J1$ couplings strengths are equal to 240 and 182.4 K in $Cs_2Cu_3SnF_{12}$ [80] and $ZnCu_3(OH)_6Cl_2$ [25], respectively (1.32-fold).

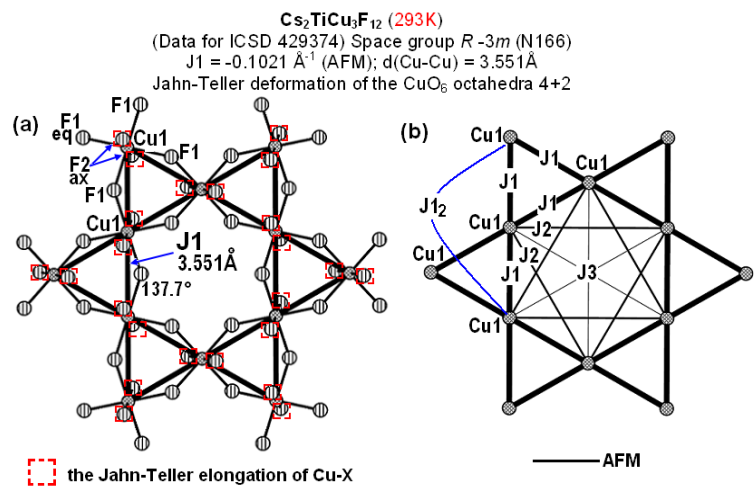

**Figure 6.** Kagome layer CuF$_6$ octahedra (a) and $Jn$ couplings in the kagome lattice (b) in Cs$_2$TiCu$_3$F$_{12}$ (projected perpendicular to the *c* axis).

All other couplings (except one – $J3(J1_2)$, as in herbertsmithite) are weak or absent and cannot compete with the dominant AFM $J1$ couplings. The values of the $Jn/J1$ ratios ($J2/J1$ = 0.017, $J3(J_d)/J1$ = 0.004–0.007, $J3(J1_2)/J1$ = 0.26–0.29, $J_{interplane}/J1$ = 0–0.007) of couplings strengths in fluoride differ insignificantly from respective values in herbertsmithite ($J2/J1$ = 0.016, $J3(J_d)/J1$ = –0.03, $J3(J1_2)/J1$ = 0.45, $J_{interplane}/J1$ = 0.002–0.09). In theory, for each of these fluorides ($Cs_2Cu_3BF_{12}$), just like for Zn-herbertsmithite, there exists the possibility to eliminate the anisotropy of the crystallographically equivalent $J3(J_d)$ and $J3(J1_2)$ couplings through shifts of two intermediate oxygen ions located in the critical position "$a$" ($\Delta a$ = 0.038–0.047 Å) slightly deeper inside the local space of the $J3(J1_2)$ coupling.

However, the above is virtually impossible. Upon the temperature decrease, these compounds undergo structural phase transitions [70, 72, 80-82] from the high-temperature rhombohedral phase with an ideal kagome lattice to the low-temperature monoclinic phases with distorted kagome lattices, in which magnetic couplings are spatially nonuniform. In [81, 82], Ono et al. assume it to be one of the main reasons of the transition into the magnetically ordered state upon further temperature decrease. Let us examine it in detail.

*3.3.3.1. Low-temperature phase of $Cs_2SnCu_3F_{12}$.* The structural phase transition of $Cs_2SnCu_3F_{12}$ [70, 80] to the low-temperature monoclinic phase with the space group $P2_1/n$ (N14) occurs at T$_c$ = 185 K and results in nonequivalence of three Cu-Cu bonds in small triangles of the kagome lattice and, accordingly, in nonequivalence of three strong AFM nearest-neighbor couplings $J1^1$ ($J1^1$ = -0.1079 Å$^{-1}$, d(Cu1-Cu2) = 3.576 Å), $J1^2$ ($J1^2$ = -0.1008 Å$^{-1}$, d(Cu2-Cu2) = 3.551 Å, $J1^2/J1^1$ = 0.93) and $J1^3$ ($J1^3$ = -0.0991 Å$^{-1}$, d(Cu1-Cu2) = 3.537 Å, $J1^3/J1^1$ = 0.92) (figure 7, supplementary table 2). However, the strengths of these AFM $J1^1$, $J1^2$, and $J1^3$ couplings differ insignificantly (maximum by 8 %). The ratios of the strengths of the AFM couplings $J2^n/J1^n$ between the values in large ($J2^n$) and small ($J1^n$) triangles of the kagome plane are in the range 0.15−0.19, which constitutes a negligible difference from the ratio $J2/J1 = 0.16$ in herbertsmithite. One should mention that in the space group $P2_1/n$ (N14) the diagonal $J3(J_d)$ couplings ($J3(J_d)$ = -0.0002 – -0.0006 Å$^{-1}$) and the next-nearest AFM $J3(J1_2^n)$ couplings ($J3(J1_2^n)$ = -0.065 – -0.0318 Å$^{-1}$) in the chains along the sides of small triangles are nonequivalent in both crystallographic and magnetic terms. Interplane couplings are virtually absent. Transition to the magnetically ordered state emerges below $T_N$ = 20.2 K [80].

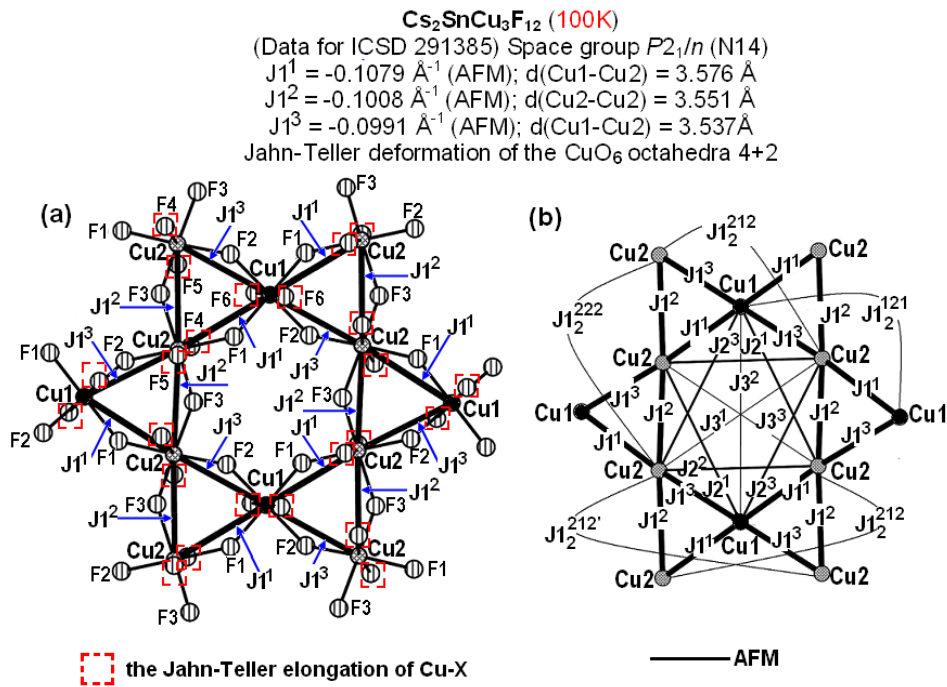

**Figure 7.** $Cs_2SnCu_3F_{12}$ in monoclinic space group $P2_1/n$ (N14) at 100 K: kagome layer $CuF_6$ octahedra (a) and $Jn$ couplings in the kagome lattice (b) in projected perpendicular to the $c$ axis.

*3.3.3.2. Low-temperature phase of Cs$_2$ZrCu$_3$F$_{12}$.* The crystal structure of Cs$_2$ZrCu$_3$F$_{12}$ transforms from rhombohedral ($R\bar{3}m$ - N166) [72] to monoclinic ($P2_1/m$ - N11) [83] near 225 K (figure 8(a), (b) and (c), supplementary table 2). In this case, the kagome plane is distorted to even larger degree than at the phase transition in Cs$_2$SnCu$_3$F$_{12}$ – it becomes slightly corrugated (figure 8c). Four nonequivalent strong AFM couplings emerge in the kagome plane: $J1^1$ ($J1^1$ = -0.1066 Å$^{-1}$, d(Cu1-Cu3) = 3.657 Å), $J1^2$ ($J1^2$ = -0.1062 Å$^{-1}$, d(Cu3-Cu3) = 3.606 Å, $J1^2/J1^1$ = 1.0), $J1^3$ ($J1^3$ = -0.0971 Å$^{-1}$, d(Cu1-Cu2) = 3.603 Å, $J1^3/J1^1$ = 0.91), and $J1^4$ ($J1^4$ = -0.0653 Å$^{-1}$, d(Cu2-Cu2) = 3.606 Å, $J1^4/J1^1$ = 0.61). These couplings are directed along the sides of two nonequivalent small triangles: $J1^1$-$J1^1$-$J1^2$ in the Cu1Cu3Cu3 triangle and $J1^3$-$J1^3$-$J1^4$ in the Cu1Cu2Cu2 triangle. If in the first triangle the strengths of the $J1^1$ and $J1^2$ couplings are virtually equal, in the second triangle the strength of the $J1^3$ coupling is 1.5-fold larger than that of the $J1^4$ coupling. The remaining couplings correspond to the characteristics of the structural-magnetic model of herbertsmithite. Three nonequivalent AFM $J2^1$($J2^1$ = -0.0098 Å$^{-1}$, d(Cu1-Cu2) = 6.244 Å), $J2^2$($J2^2$ = -0.0168 Å$^{-1}$, d(Cu1-Cu3) = 6.275 Å), and $J2^3$($J2^3$ = -0.0160 Å$^{-1}$, d(Cu2-Cu3) = 6.260 Å) couplings are relatively weak and do not compete with the AFM nearest neighbor $J1^n$ couplings ($J2^n/J1^1$ = 0.09−0.160). The diagonal ($J3^1(J_d)$ = 0 and $J3^2(J_d)$ = 0.0016 Å$^{-1}$ FM) and interplane ($J4$–$J10$ = 0–0.0016 Å$^{-1}$) couplings are virtually absent. Five nonequivalent AFM $J3(J1_2^n)$ ($J3(J1_2^n)$ = -0.0273 – -0.0307 Å$^{-1}$, d(Cu-Cu) = 7.206 - 7.314) couplings, except one ($J3(J1_2^{213})$), are relatively strong ($J3(J1_2^n)/J1^n$ = 0.28 – 0.42) and compete with the $J1^n$ couplings.

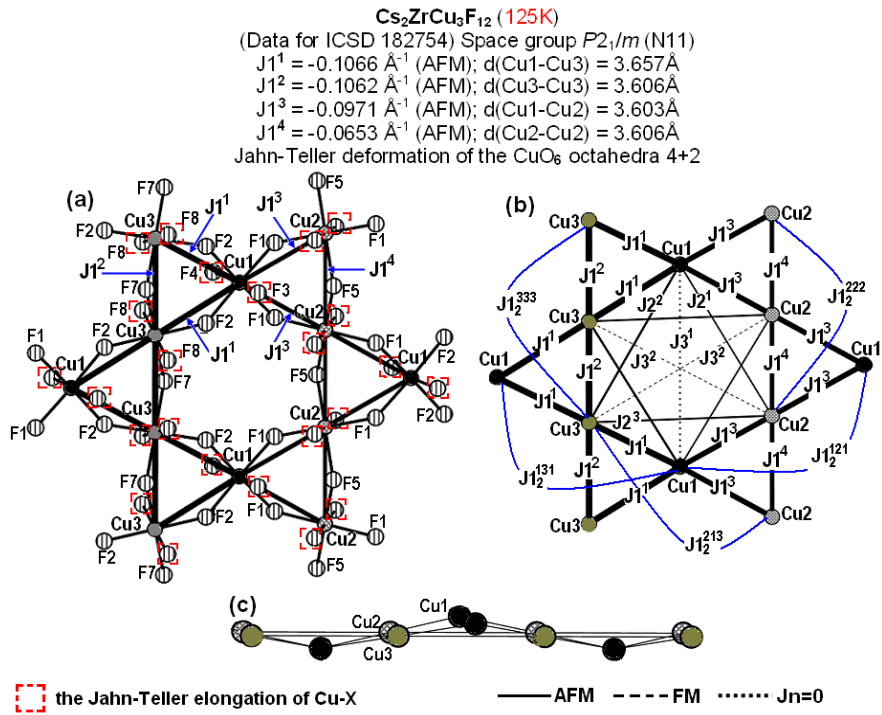

**Figure 8.** Cs$_2$ZrCu$_3$F$_{12}$ in monoclinic space group $P2_1/m$ (N11) at 125K: kagome layer CuF$_6$ octahedra (a), $Jn$ couplings in the kagome lattice (b) and corrugated the kagome plane (c).

Note that in the space group $P2_1/m$ (N11) the crystallographically equivalent diagonal $J3^2(J_d)$ ($J3^2(J_d)$ = 0.0016 Å$^{-1}$ FM, d(Cu2-Cu3) = 7.225 Å) and the next-nearest AFM $J3(J1_2^{213})$ ($J3(J1_2^{213})$ = -0.0156 Å$^{-1}$ AFM, d(Cu2-Cu3) = 7.225 Å) couplings in the chain along the sides of small triangles are not magnetically equivalent. At a temperature of 23.5 K [81, 83], there is a transition of $Cs_2Cu_3ZrF_{12}$ to the magnetically ordered state.

*3.3.3.3. Low-temperature phases of $Cs_2TiCu_3F_{12}$.* Upon cooling, $Cs_2TiCu_3F_{12}$ [72], just like other members of this family, undergoes a structural-phase transition with the symmetry reduction. The nature of this phase transition is different for single crystal and polycrystalline samples. As a result, two polymorphous forms emerge, whereas the monocrystalline preserves the trigonal/rhombohedral symmetry with the transition to the space group R -3 (N148) [72] and the powder form becomes monoclinic $P2_1/n$ (N14) [72]. In both cases, the long-range antiferromagnetic order is present in the range 16–20 K. In the low-temperature monoclinic phase $P2_1/n$, the parameters of magnetic couplings in $Cs_2TiCu_3F_{12}$ and $Cs_2SnCu_3F_{12}$ are virtually identical (figures 9 and 7, supplementary table 2). The insignificant decrease of the values of three strong AFM nearest-neighbor couplings in $Cs_2TiCu_3F_{12}$ ($J1^1$(Ti)/$J1^1$(Sn) = 0.92; $J1^2$(Ti)/$J1^2$(Sn) = 0.94; $J1^3$(Ti)/$J1^3$(Sn) = 0.93) is caused by the decrease of the Cu-Cu distances in the kagome lattice because of the smaller radius of $Ti^{4+}$ (0.605 Å) than that of $Sn^{4+}$ (0.69 Å). As a result, the necessity to preserve the Cu-F bond lengths induced the removal of intermediate F$^-$ ions from the kagome plane and corresponding decrease of the Cu-F-Cu angles in $Cs_2TiCu_3F_{12}$, which was responsible for the above effect.

The structural phase transition of the monocrystalline form of $Cs_2TiCu_3F_{12}$ from the centrosymmetric space group R -3m (№166) into another centrosymmetric space group R -3 (No. 148) [72] yields a significantly more complex magnetic structure (figure 10, Supplementary table 2). Four nonequivalent strong AFM couplings emerge in the kagome plane: $J1^1$ ($J1^1$ = -0.1056 Å$^{-1}$, d(Cu1-Cu1) = 3.545 Å), $J1^2$ ($J1^2$ = -0.1040 Å$^{-1}$, d(Cu1-Cu2) = 3.566 Å, $J1^2/J1^1$ = 0.98), $J1^3$ ($J1^3$ = -0.0975 Å$^{-1}$, d(Cu1-Cu2) = 3.512 Å, $J1^3/J1^1$ = 0.92), and $J1^4$ ($J1^4$ = -0.0795 Å$^{-1}$, d(Cu2-Cu2) = 3.550 Å, $J1^4/J1^1$ = 0.75). Three strong AFM nearest-neighbor couplings ($J1^1$-$J1^2$-$J1^3$) are located along the sides of the first triangle (Cu2Cu1Cu1), whereas three weaker AFM couplings ($J1^4$) are located along the sides of the second triangle (Cu2Cu2Cu2).

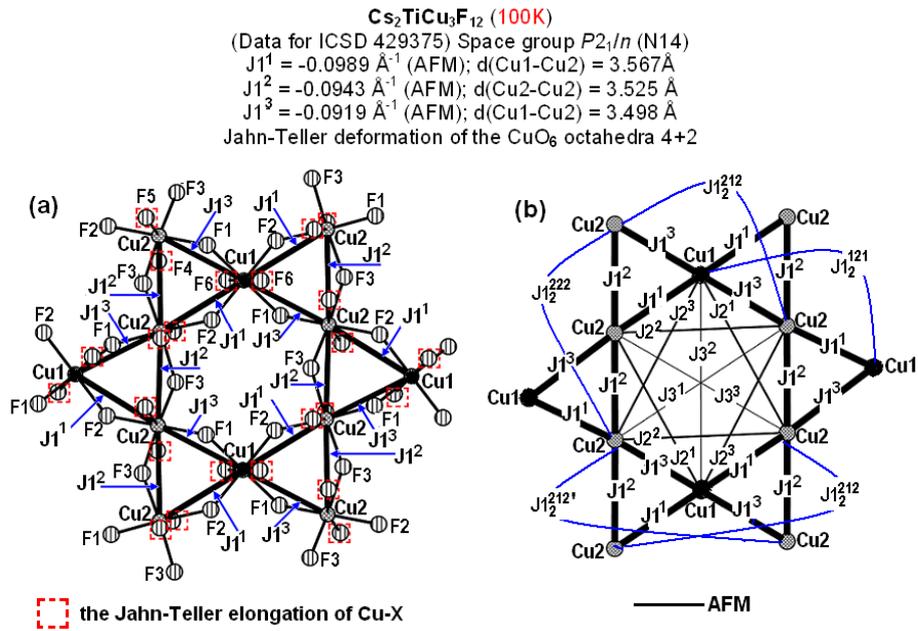

**Figure 9.** $Cs_2TiCu_3F_{12}$ in the monoclinic $P2_1/n$ (N14) space group at 100 K: kagome layer CuF$_6$ octahedra (a), $Jn$ couplings in the kagome lattice (b).

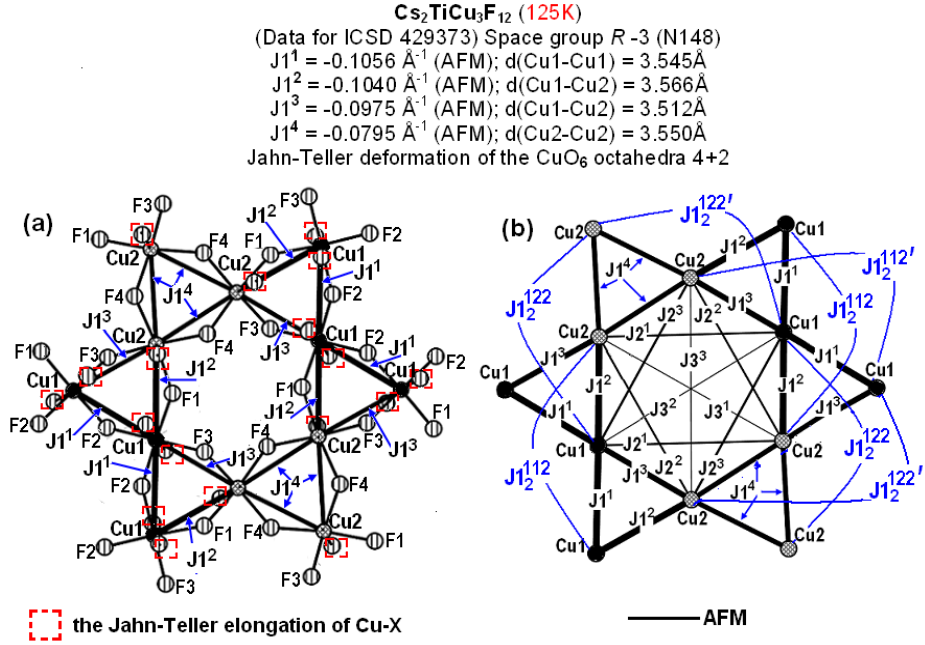

**Figure 10.** $Cs_2TiCu_3F_{12}$ in trigonal/rhombohedral space group $R$ -3 (N148) at 125 K: kagome layer $CuF_6$ octahedra (a), $Jn$ couplings in the kagome lattice (b).

The remaining couplings correspond to the characteristics of the structural-magnetic model of herbertsmithite: the ratios of strengths of the AFM couplings $J2^n/J1^n$ are in the range 0.16 − 0.19; the diagonal couplings are very weak ($J3(J_d)$ = 0.005 – 0.009 Å$^{-1}$). Four nonequivalent next-to-nearest-neighbour $J3(J1_2^n)$ couplings ($J3(J1_2^n)$ =-0.0257 – -0.0271 Å$^{-1}$, d(Cu1-Cu2) = 7.054 – 7.112 Å) in linear chains along the triangles sides are relatively strong ($J3(J1_2^n)/ J1^n$ = 0.26 – 0.32) and can compete with the $J1^n$ couplings. There is preserved a magnetic isolation of the frustrated kagome planes. The interplane $J4$–$J13$ couplings at distances d(Cu-Cu) = 6.905 – 7.910 Å are absent, except two very weak FM $J8$ ($J8$ = 0.0006 Å$^{-1}$, d(Cu1-Cu2) = 7.655Å) and FM $J13$ ($J13$ = 0.0003 Å$^{-1}$, d(Cu1-Cu2) = 7.910 Å) ones. All the crystallographically nonequivalent couplings are magnetically nonequivalent as well.

*3.3.3.4. $Rb_2SnCu_3F_{12}$*. Unlike the discussed above members of the $A_2BCu_3F_{12}$ family, $Rb_2SnCu_3F_{12}$ [84-85] (figure 11, supplementary table 2) is characterized with a strongly distorted triclinic structure (R -3 (N148)) even at room temperature. The parameters of magnetic couplings in $Rb_2SnCu_3F_{12}$ are virtually identical to respective values for $Cs_2TiCu_3F_{12}$ in the monocrystalline form after transition to the space group R -3 upon cooling. However, one should take into account that the markings of copper ions (Cu1 and Cu2) are mutually replaced in these compounds structures. The problem of stabilization of an ideal kagome geometry at low temperatures to preserve the magnetically frustrated ground state in $A_2BCu_3F_{12}$ fluorides was discussed extensively [70-72, 80-87]. According to Downie et al. [72], to ensure the stability of the quantum spin-liquid state in this system is possible through fine-tuning of different crystal chemistry factors. In a majority of works, to realize such a scenario, it was suggested to introduce larger atoms to the A position, but the performed experimental and theoretical studies have not yet corroborated such a suggestion.

In our opinion, the reason of such an easiness of distortion of the kagome plane in $A_2BCu_3F_{12}$ fluorides at low temperatures consists in weaker binding of the copper kagome lattice by anions than in herbertsmithite $ZnCu_3(OH)_6Cl_2$ and similar compounds. Unlike fluorides, copper ions in the kagome lattice of herbertsmithite (figure 12(a)) are linked to each other through octahedra edges formed by oxygen ions located in the equatorial plane and chlorine ions located in axial vertices of $Cu^{2+}$ octahedra. In fluorides (figure 12 (b)), binding between copper ions is realized only through vertices – fluorine anions located in the octahedron equatorial plane. Fluoride ions located in octahedra axial vertices do not participate in binding. A stronger polymerization in herbertsmithite is caused by the necessity of saturation of the copper ions coordination capacity. The point is, the number of anions per Cu$^{2+}$ cation in ZnCu$_3$(OH)$_6$Cl$_2$ (8/3) is 1.5-fold smaller than in A$_2$BCu$_3$F$_{12}$ (12/3). In other words, the entire copper kagome lattice is bound by an extra -Cu-Cl$_{ax}$-Cu- bond in herbertsmithite. There is no such a bond in fluorides, as the F$_{ax}$ anions are end ones.

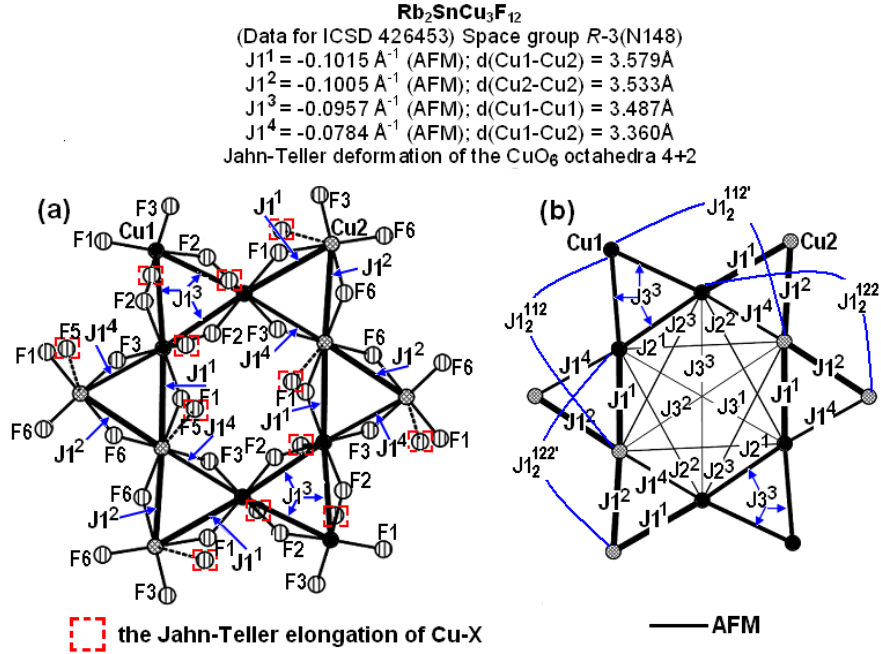

**Figure 11.** Rb$_2$SnCu$_3$F$_{12}$: kagome layer CuF$_6$ octahedra (a), $Jn$ couplings in the kagome lattice (b).

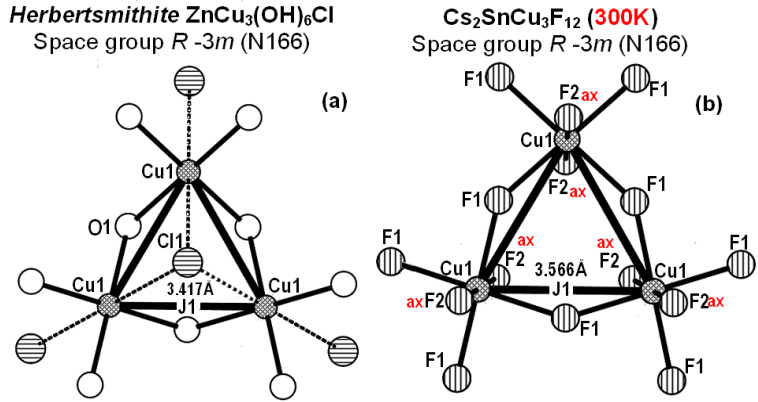

**Figure 12.** Bond between copper ions in the kagome plane through the O-Cl edge in herbertsmithite ZnCu$_3$(OH)$_6$Cl$_2$ (a) and through the vertex (F ion) in A$_2$BCu$_3$F$_{12}$ fluorides (b).

To sum up, according to our calculations (supplementary table 2), the parameters of magnetic couplings in these antiferromagnetics (Cs$_2$Cu$_3$BF$_{12}$ (B – Sn, Zr and Ti)) crystallizing in the rhombohedral system (*R-3m* (N166)) ideally fit the criteria of the spin-liquid emergence. However, structural distortions upon the temperature decrease yielding nonequivalence of the AFM $J1^n$ nearest-neighbor couplings and unattainability of equalization of the strengths of the $J3(J1_2^n)$ and $J3(J_d)$ couplings finally result in the transition to the magnetically ordered state and, therefore, block the spin-liquid formation.

*3.3.4. Antiferromagnet edwardsite $Cd_2Cu_3(SO_4)_2(OH)_6·4H_2O$ with deformed kagome lattice.* Edwardsite $Cd_2Cu_3(SO_4)_2(OH)_6·4H_2O$ [88] (figure 13, supplementary table 3) crystallizes in the monoclinic centrosymmetric space group $P2_1/c$ (N14) and contains 4 symmetrically distinct $Cu^{2+}$ sites. The Jahn–Teller distortion of Cu1, Cu2, Cu3, and Cu4 copper octahedra are of the (4+2) type, i.e., four short and two long Cu-O bonds, just like for herbertsmithite. The lengths of four short bonds in the octahedra equatorial plane are in the range $d(Cu-O_{eq})$ = 1.95 − 1.99 Å, while those of long axial bonds – in the range $d(Cu-O_{ax})$ = 2.36–2.84 Å. Therefore, the contributions to the formation of magnetic interactions are provided only by equatorial oxygen ions. One should emphasize that these intermediate oxygen ions are located outside the triangle as one per each side, just like in herbertsmithite.

According to our calculations (figure 13, supplementary table 3), the kagome lattice of edwardsite contains 6 strong nonequivalent AFM nearest-neighbor couplings: $J1^1$ ($J1^1$ = -0.0626 Å$^{-1}$, d(Cu2-Cu4) = 3.323 Å), $J1^2$ ($J1^2$ = -0.0517 Å$^{-1}$, d(Cu1-Cu4) = 3.217 Å), $J1^3$ ($J1^3$ = -0.0494 Å$^{-1}$, d(Cu1-Cu2) = 3.0220 Å, $J1^4$ ($J1^4$ = -0.0621 Å$^{-1}$, d(Cu1-Cu3) = 3.364 Å), $J1^5$ ($J1^5$ = -0.0567 Å$^{-1}$, d(Cu1-Cu4) = 3.233 Å), and $J1^6$ ($J1^6$ = -0.0538 Å$^{-1}$, d(Cu3-Cu4) = 3.198 Å), differing insignificantly from each other ($J1^n/J1^1$ = 0.79 – 1). These strong nonequivalent AFM nearest-neighbor couplings form two types of frustrated small triangles in the kagome plane: $J1^1$ - $J1^2$ - $J1^3$ and $J1^4$ - $J1^5$ - $J1^6$.

The strengths of the next in length six nonequivalent AFM couplings ($J2^1$ – $J2^6$) along the sides of two large triangles in the kagome plane are in the range -0.015 Å$^{-1}$ – -0.017 Å$^{-1}$. The value of the ratio $J2^n/J1^n$ = 0.24 – 0.26 exceeds that in herbertsmithite ($J2/J1$ = 0.16), which indicates to possible competition between $J2^n$ and $J1^n$ couplings. Moreover, one of them is unstable (AFM $J2^3$). An insignificant shift of the O13 ion along the Cu4-Cu3 bond line to its center by just 0.008 Å from the critical position *l'/l* would result in the increase of the $J2^3$ value up to -0.0615 Å$^{-1}$ (3.8-fold).

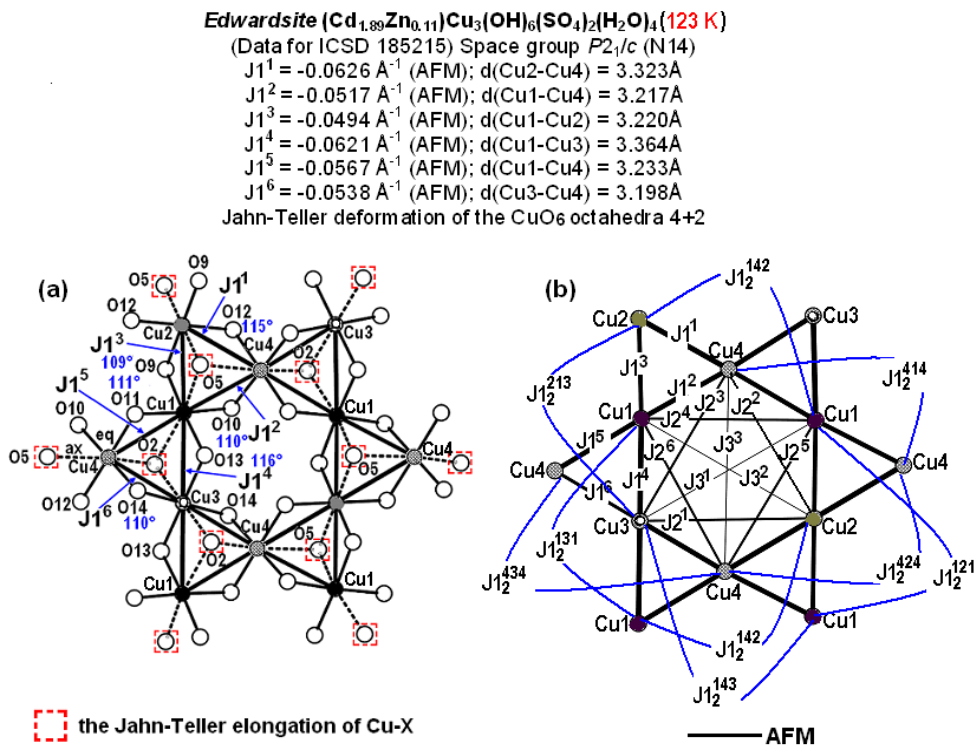

**Figure 13.** Kagome layer $CuO_6$ octahedra (a) and Jn couplings in the kagome lattice (b) in edwardsite $Cd_2Cu_3(SO_4)_2(OH)_6·4H_2O$.

Three diagonal AFM $J3^1(J_d)$ couplings ($J3^1(J_d)$ = -0.0016 Å$^{-1}$, d(Cu1-Cu3) = 6.482 Å, $J3^1(J_d)/J1^1$ = 0.03), $J3^2(J_d)$ ($J3^2(J_d)$ = -0.0014, d(Cu1-Cu2) = 6.484 Å, $J3^2(J_d)/J1^1$ = 0.02) and AFM $J3^3(J_d)$ ($J3^3(J_d)$ = -0.0022 Å$^{-1}$, d(Cu4-Cu4) = 6.567 Å, $J3^3(J_d)/J1^1$ = 0.03) are very weak. Let us emphasize that, in case of this space group ($P2_1/c$ N14), there is no crystallographic equivalence between the $J3^3(J_d)$ and $J3(J1_2^n)$ couplings, as in a majority of crystal structures discusses above. The strength of eight nonequivalent AFM $J3(J1_2^n)$ next-nearest couplings ($J3(J1_2^n)$ = -0.0193 – -0.0325 Å$^{-1}$; d(Cu-Cu) = 6.411 – 6.728 Å) in the chains along the sides of small triangles significantly (9–15-fold) exceeds that of diagonal AFM $J3^n(J_d)$ couplings. There exists a competition between the AFM $J3^n(J1_2^n)$ and AFM nearest-neighbor $J1^n$ couplings ($J3^n(J1_2^n)/J1^n$ = 0.31 – 0.64). The kagome planes are at large distances from each other. All the interplane Cu-Cu distances ($J4 – J11$) in the range from 10.030 to 10.427 Å are ferromagnetic and very weak ($J^n$ = 0 – 0.0009 Å$^{-1}$; $J^n/J1^1$ = 0 – -0.014).

To sum up, not all crystal chemistry criteria for the emergence of quantum spin liquid are valid for this compound. First, there exists the competition of the AFM nearest-neighbor couplings ($J1^n$) not only between each other, but also with other couplings ($J2^n$ and $J3^n$ ($J1_2^n$)). Second, there are nonstoichiometry in $(Cd_{1.89}Zn_{0.11})Cu_3(SO_4)_2(OH)_6 \cdot 4(H_2O)_4$ and concern that Cu positions in the kagome plane are partially occupied by Zn or Cd.

According to Ishikawa et al. [89], who studied the magnetic properties of the stoichiometric sample, edwardsite $Cd_2Cu_3(SO_4)_2(OH)_6 \cdot 4H_2O$ showed an antiferromagnetic order accompanied by a small ferromagnetic moment below 4.3 K. The weak ferromagnetism is likely due to spin canting caused by sizable Dzyaloshinsky–Moriya interactions, which may stabilize the long-range magnetic order instead of a spin-liquid state expected for the kagome antiferromagnet.

### 3.4. Antiferromagnets with oxocentered OCu$_3$ triangles

Here we describe magnetics based on AFM frustrated kagome lattices, in which the way of formation of dominant antiferromagnetic couplings along the sides of small triangles differs from that in herbertsmithite. Here, these nearest-neighbor $J1$ couplings are formed mainly due to one intermediate oxygen ion located in the center of the OCu$_3$ triangle above or below its plane (figures 2(c) and (d)). In this case, when triangles of the kagome plane are centered by oxygen ions, there emerges the same effect as in the AFM spin-frustrated layers of corner-sharing OCu$_4$ tetrahedra on the kagome lattice in volcanic minerals $Cu_5O_2(VO_4)_2(CuCl)$ [45], $NaCu_5O_2(SeO_3)_2Cl_3$ [45], and $K_2Cu_5Cl_8(OH)_4 \cdot 2H_2O$ [45]. The oxygen ions centering the OCu$_4$ tetrahedra in averievite and ilinskite provide the main contribution to the formation of AFM interactions along the tetrahedra edges. Making an analogy, we can consider this group of compounds as AFM spin-frustrated layers of corner-sharing OCu$_3$ triangles on the kagome lattice.

There is a possibility of an additional entering to the interaction space from intermediate oxygen ions (figures 2(e) and (f)) that are located not in the triangle center, but approximately in the middle of its sides, as in $ZnCu_3(OH)_6Cl_2$. However, these additional oxygen ions make insignificant contributions to the interaction AFM or FM components.

*3.4.1. Structurally-perfect kagome antiferromagnets engelhauptite $KCu_3(V_2O_7)(OH)_2Cl$ and beta-vesignieite $BaCu_3V_2O_8(OH)_2$ and deformed kagome antiferromagnet alpha-$BaCu_3V_2O_8(OH)_2$ with (2 + 4)-compressed octahedra.* In the crystal structures of the minerals engelhauptite $KCu_3(V_2O_7)(OH)_2Cl$ [90] (figure 14(a) and (b), supplementary table 4) and vesignieite beta-$BaCu_3V_2O_8(OH)_2$ [46] (figure 14 (c) and (d), Supplementary table 4), an ideal kagome lattice of Cu$^{2+}$ ions is realized, just like in herbertsmithite ($ZnCu_3(OH)_6Cl_2$). Moreover, *beta*-vesignieite ($BaCu_3V_2O_8(OH)_2$) [46] crystallizes in the same centrosymmetric trigonal space group *R*-3*m* (N166) as herbertsmithite, whereas engelhauptite $KCu_3(V_2O_7)(OH)_2Cl$ crystallizes in another centrosymmetric hexagonal space group *P*6$_3$/*mmc* (№194) [90].

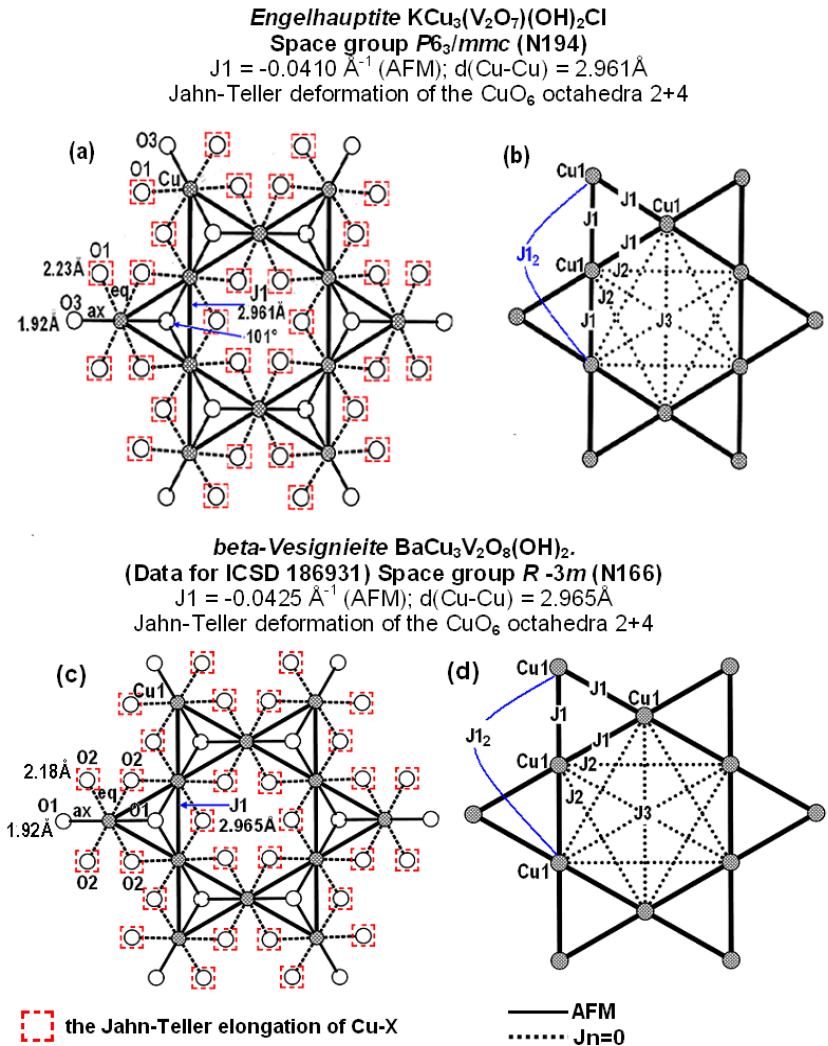

**Figure 14.** Kagome layer $CuO_6$ octahedra (a and c) and $J$n couplings in the kagome lattice (b and d) in engelhauptite $KCu_3(V_2O_7)(OH)_2Cl$ and *beta*-vesignieite ($BaCu_3V_2O_8(OH)_2$).

However, the similarity of the crystal structures of these compounds to that of herbertsmithite is limited to the above aspects. According to the structural data [23, 81], there are two important differences from the crystal structure of herbertsmithite. One of them consists in the fact that just one common intermediate oxygen ion located in the center of the triangle above or below its plane according to a definite order participates in the emergence of AFM couplings along three sides of each small triangle in the kagome plane (figure 14 (a) and (c)). At the same time, in herbertsmithite, the formation of couplings along each of three sides of small triangles is associated with one individual intermediate oxygen ion. Another, not less important difference consists in the fact that in $KCu_3(V_2O_7)(OH)_2Cl$ and beta-$BaCu_3V_2O_8(OH)_2$ the Jahn–Teller effect is manifested geometrically in compression of (2+4) octahedra ($Cu^{2+}$), rather than in (4+2) stretching, as in $ZnCu_3(OH)_6Cl_2$. In both of these compounds, magnetic $Cu^{2+}$ ions each occupy one crystallographically independent site, and, in their octahedra compressed along the axial $O_{ax}$-Cu1-$O_{ax}$ direction, the bond lengths are $d(Cu1-O3_{ax})$ = 1.918 Åx2 and $d(Cu1-O1_{eq})$ = 2.228Åx4 in $KCu_3(V_2O_7)(OH)2Cl$ and $d(Cu1-O1_{ax})$ = 1.916Åx2 and $d(Cu1-O2_{eq})$ = 2.183Åx4 in *beta*-$BaCu_3V_2O_8(OH)_2$.

In the *alpha*-$BaCu_3V_2O_8(OH)_2$ modification [91, 92], the symmetry was represented not by trigonal, but by monclinic $C2/m$ space group with an almost ideal kagome lattice. However, here, the [2+4] type of distortion of copper octahedra was preserved. In the monoclinic modification of vesignieite, $Cu^{2+}$ ions occupy two crystallographically independent positions – Cu1 ($d(Cu1-O1_{ax})$ = 1.913 Åx2 and $d(Cu1-O2_{eq})$ = 2.183 Åx4) and Cu2 ($d(Cu2-O1_{ax})$ = 1.905 Åx2 and $d(Cu2-O2_{eq})$ = 2.175Åx2, $d(Cu2-O3_{eq})$ = 2.184 Åx2).

Since two opposite Cu-Oax bonds are shorter than other four Cu-Oeq bonds in engelhauptite and both modifications of vesignieite, one can conclude that the $z^2$ orbital is occupied by an unpaired electron. In herbertsmithite, oppositely, the $x^2 - y^2$ orbital is occupied, since the two Cu–Oax bonds are longer than the others.

There always exist many contradictory opinions regarding the validity of one or another *JT* configuration [90-95]. For example, Pekov et al. [90] believe that the distortion of the [2+4] copper octahedron they found in engelhauptite ($KCu_3(V_2O_7)(OH)_2Cl$) comprises an overlapping of two [4+2] distorted geometries in different orientations and an artifact of the model of the intermediate structure obtained in the course of analysis of the diffraction structure. They also believe that studies of crystals of better quality would enable one to reveal the [4+2] copper coordination in engelhauptite. The same conclusions were made by Boldrin et al. [93, 94] in the course of correcting the structure of vesignieite $BaCu_3V_2O_8(OH)_2$ and demonstrating the dynamic *JT* effect on the $Cu^{2+}$ kagome sublattice. According to the powder neutron diffraction studies [93], the dynamic *JT* effect in the vesignieite structure yields the noncentrosymmetric space group $P3_12_1$ (No. 152), in which tiny diverse triangle distortions correspond to differences in lengths of three Cu-Cu bonds by less than 1 %. Besides, splitting of the sites of O11, O12, and O13 (figure 16) yields the difference in Cu1-O and Cu2-O bond lengths explained by the fact that the $Cu1O_6$ octahedra are in a static *JT* state and $Cu2O_6$ octahedra – in a dynamic *JT* state [93]. Possibly, the problem solution could be contributed by the methods including the analysis of electron density distribution determined by single-crystal X-ray diffraction in combination with spectroscopic measurements implemented by Udovenko and Laptash [96].

We calculated the parameters of magnetic couplings in engelhauptite $KCu_3(V_2O_7)(OH)_2Cl$ [90] and three models of vesignieite $BaCu_3V_2O_8(OH)_2$ presented in different space groups: I – *R-3m* [46], II – *C 2/m* [91], and III – $P3_121$ (Supplementary table 4). The parameters of magnetic couplings in $KCu_3(V_2O_7)(OH)_2Cl$ (figure 14(a) and (b)) and $BaCu_3V_2O_8(OH)_2$–I (figure 14(c) and (d)) calculated using these compounds coordinates in the *P6₃/mmc* [90] and *R-3m* [46] are very close to each other. Both these compounds are characterized with ideal kagome lattices with virtually equal triangles sides and identical ways of formation of magnetic couplings in the kagome plane. The *J*1 couplings along three triangle sides are formed due to one intermediate oxygen ion located in the triangle center above or below its plane according to a definite order (figure 15).

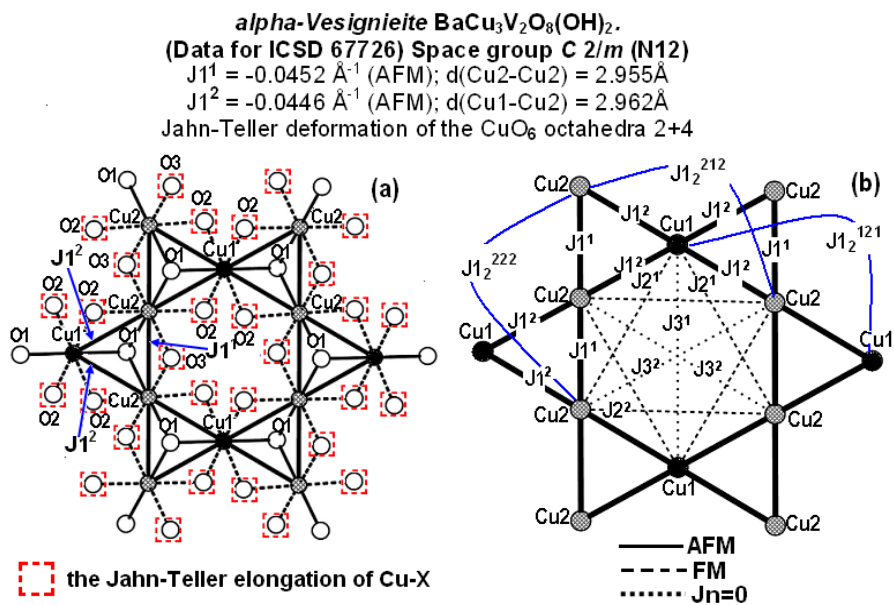

**Figure 15.** Kagome layer $CuO_6$ octahedra (a) and *J*n couplings in the kagome lattice (b) in edwardsite *alpha*-$BaCu_3V_2O_8(OH)_2$.

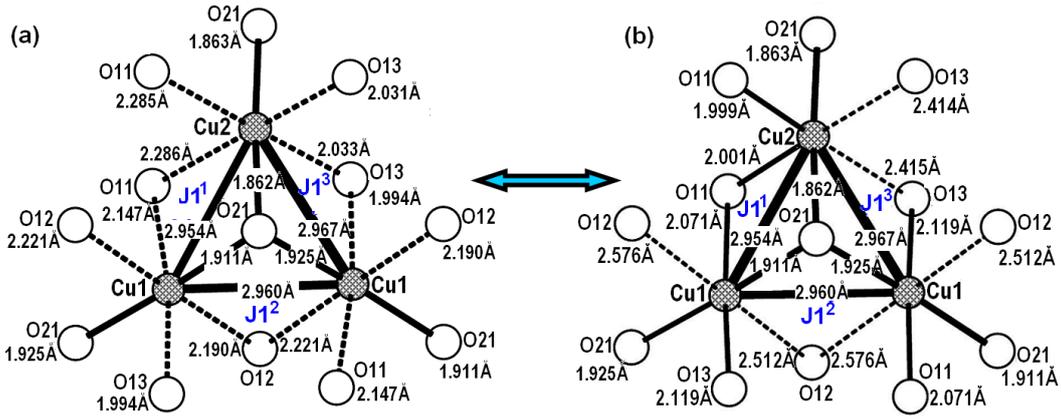

**Figure 16.** Jahn–Teller distortion of copper octahedral and $Cu^{2+}$ kagome sublattice in the trigonal $P3_121$ structure of vesignieite, $BaCu_3V_2O_8(OH)_2$.

In $KCu_3(V_2O_7)(OH)_2Cl$ and $BaCu_3V_2O_8(OH)_2$ (models I and II) (Supplementary table 4), just like in herbertsmithite ($ZnCu_3(OH)_6Cl_2$), the AFM nearest-neighbor $J1$ couplings ($J1$ =-0.0410 − -0.0425 Å$^{-1}$, d(Cu1-Cu1) = 2.961 − 2.965 Å) in the kagome plane are dominant. However, the strengths of the AFM $J1$ couplings in these compounds are approximately 1.6-fold weaker than in herbertsmithite. Other couplings in the ($J2 = 0$, $J3(J_d) = 0$) plane, except one AFM $J3(J1_2)$ coupling, just like in herbertsmithite, cannot compete with the dominant $J1$ couplings, since they are absent. The strong AFM $J3(J1_2)$ coupling ($J3(J1_2)$ = -0.0360 Å$^{-1}$, $J3(J1_2)/J1$ = 0.85 - 0.88, d(Cu1-Cu1) = 5.922 − 5.930 Å) is formed due to a large contribution of the intermediate copper ion and a small AFM contribution from two intermediate oxygen ions ($2j(O3)$ in engelhauptite and $2j(O1)$ in vesignieite (I)). Unlike herbertsmithite, it is impossible to achieve the equality of the parameters of crystallographically equivalent $J3(J_d)$ and $J3(J1_2)$ couplings, since there are no extra oxygen ions, which could make a substantial FM contribution and eliminate such a nonequivalence.

In the trigonal $P3_121$ vesignieite, $BaCu_3V_2O_8(OH)$ (model III), in which the sites of O11, O12, and O13 are split, we calculated the parameters of magnetic $J_{ij}$ couplings for both variants ('$a$' and '$b$') (figure 16). As in other models of vesignieite, the strong AFM $J1^1$, $J1^2$, and $J1^3$ couplings along three sides of the triangle are formed due to one intermediate O21 oxygen ion located in the triangle center above or below its plane. Since the sites of Cu1, Cu2, Cu3, and O21 are not split, the AFM $J1^1$ ($J1^1$ = 0.0521 Å$^{-1}$, d(Cu1-Cu2) = 2.954 Å), $J1^2$ ($J1^2$ = 0.0412 Å$^{-1}$, d(Cu1-Cu1) = 2.960 Å), and $J1^3$ ($J1^3$ = 0.0507 Å$^{-1}$, d(Cu1-Cu2) = 2.967 Å) couplings for the variants '$a$' and '$b$' are identical. The couplings in the plane, $J2^n$ ($J2^1$, $J2^2$, $J2^3$, and $J2^4$; d(Cu-Cu) = 5.035-5.215 Å), are weak ($J2^n/J1^1$ is in the ranges 0 – -0.08 and 0 – 0.09 for the variants '$a$' and '$b$', respectively). All the diagonal $J3^n(J_d)$ (d(Cu-Cu) = 5.918 Å) couplings are even weaker than $J2^n$. The values of the ratios of the couplings strengths $J3^1(J_d^{Cu1Cu1})/J1^1$ and $J3^2(J_d^{Cu2Cu2})/J1^1$ for the variant '$a$' are equal to -0.004 and -0.004, whereas for the variant '$b$' – to -0.05 and 0.02', respectively. The crystallographically equivalent to the diagonal $J3^n(J_d)$ couplings, the next-to-nearest-neighbour $J3(J1^n_2)$ couplings in linear chains along the triangle sides are, in opposite, strong AFM ones ($J1_2^{Cu1Cu1Cu1}/J1^1$= 0.59, $J1_2^{Cu1Cu2Cu1}/J1^1$= 0.61 and $J1_2^{Cu2Cu1Cu2}/J1^1$= 0.60). These couplings compete with the strong AFM $J1^1$, $J1^2$, and $J1^2$ nearest-neighbour couplings, since the $J1^n_2/J1^1$ ratio exceeds significantly the critical value as 1⁄6 in [97] and 0.2411 in [52, 55, 98].

We believe that this very drastic strength nonequivalence of the crystallographically equivalent $J3(J1^n_2)$ and $J3^n(J_d)$ magnetic interactions serves as a reason of the magnetic ordering in vesignieite. In this case, under effect of the strong $J3(J1^n_2)$ interactions, there could emerge geometrically non-frustrated AFM spin networks, which could, upon the temperature decrease, suppress the frustration of the nearest-neighbour $J1^n$ couplings in small triangles of the kagome plane and bring it to the state of magnetic ordering (figure 4).

To sum up, the long-range nonequivalent $J3$ ($J1^n_2$ and $J_d$) couplings enhance the effect of DM interactions in suppressing of frustration.. Boldrin et al. [93, 94] pay the main attention to this third-neighbor exchange as well. They believe that this structure is stabilized by a dominant antiferromagnetic third-neighbor exchange *J*3 with minor first- or second-neighbor exchanges. The study of the magnetic properties of BaCu$_3$V$_2$O$_8$(OH)$_2$ [99] demonstrates that this mineral undergoes a transition to the magnetically ordered state upon cooling below 9 K, probably, under effect of the DM interaction that it is large enough.

*3.4.2. KCu$_3$(OH)$_2$(AsO$_4$) (HAsO4) and volborthite Cu$_3$(V$_2$O$_7$)(OH)$_2$(H$_2$O)$_2$ in space group C2/m with deformed kagome lattice and two types – (2+4) and (4+2) – of JT distorted octahedral copper coordination.* KCu$_3$(OH)$_2$(AsO$_4$) (HAsO4) [100] and volborthite, Cu$_3$(V$_2$O$_7$)(OH)$_2$(H$_2$O)$_2$ [34], crystallize in the centrosymmetric monoclinic space group *C*2/*m* (N12) and contain two symmetrically distinct Cu$^{2+}$ sites. In spite of different ionic and molecular compositions filling voids inside kagome layers and between them, the parameters of unit cells and distortions of octahedra in KCu$_3$(OH)$_2$(AsO$_4$) (HAsO4) and volborthite Cu$_3$(V$_2$O$_7$)(OH)$_2$(H$_2$O) are very similar. In both structures, the Cu(l) octahedron has the (2+4) distortion type, whereas the Cu2 octahedron – the (4+2) distortion type. The kagome sublattice of copper ions is formed by just one type of isosceles Cu1Cu2Cu2 triangles. According to our calculations (figure 17(b) and (d),

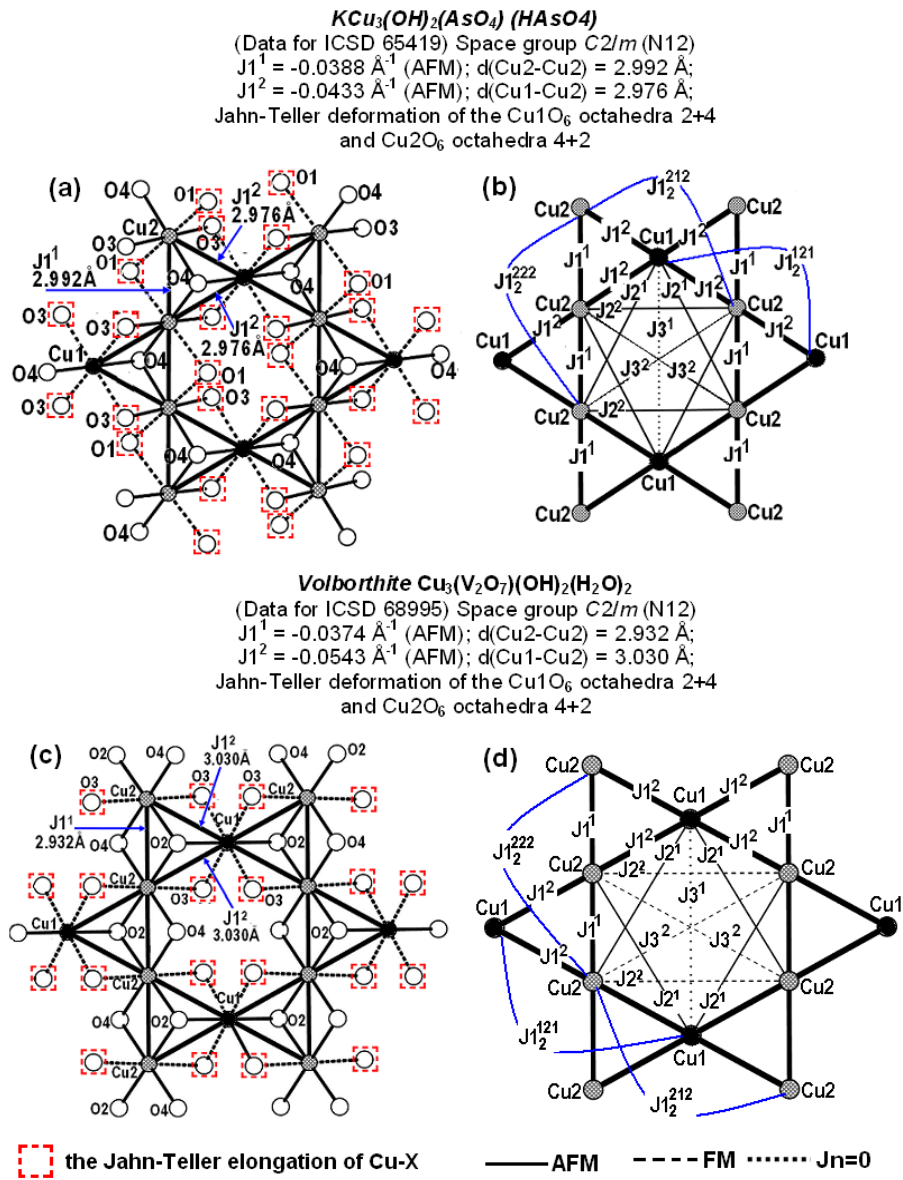

Figure 17. Kagome layer CuO$_6$ octahedra (a and c) and *J*n couplings in the kagome lattice (b and d) in KCu$_3$(OH)$_2$(AsO$_4$) (HAsO4) and volborthite Cu$_3$(V$_2$O$_7$)(OH)$_2$(H$_2$O)$_2$.

supplementary table 4), in both magnetics, two strong nonequivalent AFM nearest-neighbor $J1^1$ and $J1^2$ couplings along the sides of the $Cu_3$ triangle are formed due to one intermediate oxygen ion located in the triangle center above or below its plane. In $KCu_3(OH)_2(AsO_4)$ $(HAsO4)$, the intermediate O4 ion participates in the formation of the $J1^1$ couplings ($J1^1$ = -0.0388 Å$^{-1}$, d(Cu2-Cu2) = 2.992 Å) and $J1^2$ ($J1^2$ = -0.0433 Å$^{-1}$, d(C1-Cu2) = 2.976 Å).

In $Cu_3(V_2O_7)(OH)_2(H_2O)$, the strong AFM $J1^1$ ($J1^1$ = -0.0374 Å$^{-1}$, d(Cu2-Cu2) = 2.932 Å) and $J1^2$ ($J1^2$ = -0.0543 Å$^{-1}$, d(C1-Cu2) = 3.030 Å) couplings emerge under effect of just one O2 ion. However, the space of the $J1^1$ coupling of volborthite contains, aside from the O2 ion (j(O2) = -0.0445 Å$^{-1}$ AFM), the extra O3 ion (j(O4) = 0.0071 Å$^{-1}$ FM), which makes a small FM contribution slightly decreasing the strength of this AFM interaction. The above AFM $J1^1$ and $J1^2$ couplings in triangles are frustrated, since the ratio of their strengths approaches 1 ($J1^1/J1^2$ is equal to 0.90 and 0.69 in $KCu_3(OH)_2(AsO_4)$ $(HAsO4)$ and $Cu_3(V_2O_7)(OH)_2(H_2O)$, respectively). In both compounds, the AFM $J2^1$ and $J2^2$ couplings in the kagome plane are weak ($J2^n/J1^n$ does not exceed 0.1) and do not compete with the AFM nearest neighbor $J1^1$ and $J1^2$ couplings, so that they cannot serve as a reason of the magnetic ordering in these compounds. A drastic strength nonequivalence of the crystallographic equivalent $J3(J1_2^{212})$ and $J3^2(J_d)$ ($J3^2(J_d)/J3(J1_2^{212})$ = -0.03) magnetic interactions upon the temperature decrease. In both compounds, all the $J3^1(J_d)$ diagonal couplings are eliminated, whereas the $J3^2(J_d)$ couplings are very weak AFM ones ($J3^2(J_d)/J1^2$ -0.01) in $KCu_3(OH)_2(AsO_4)$ $(HAsO4)$ and FM ones ($J3^2(J_d)/J1^2$ - -0.02) in $(Cu_3(V_2O_7)(OH)_2(H_2O)$. In opposite, all three nonequivalent $J3$ couplings ($J1_2^{222}$, $J1_2^{212}$, and $J1_2^{121}$) in both compounds are the strong AFM ones ($J3(J1_2^n)/J1^n$ = 0.66 – 0.97). All the interplane $J4 - J7$ couplings in the range from 7.116 Å to 8.030 Å are ferromagnetic (except one) and very weak ($Jn/J1^2$ varies from -0.04 to 0) for both compounds. Just one virtually disappearing $J6$ coupling ($J6$ = -0.0001 Å$^{-1}$, d(Cu2-Cu2) = 7.787 Å) in $Cu_3(V_2O_7)(OH)_2(H_2O)$ is antiferromagnetic. Besides, the space of interplane interactions of this compound contains the oxygen ions from water molecules located between kagome planes and not included into the copper coordination. If one takes them into account at calculations of the magnetic coupling parameters, the $J4 - J7$ value would increase dramatically.

*3.4.3. Volborthite $Cu_3(V_2O_7)(OH)_2(H_2O)_2$ in space groups C2/c (N15) and Ia (N9) with deformed kagome lattice and one type of (4+2)-JT distorted octahedral copper coordination.* Around the room temperature, there exists another polymorph – volborthite $Cu_3(V_2O_7)(OH)_2(H_2O)_2$ [35] (figure 18(a) and (b)) crystallizing in another monoclinic centrosymmetric space group C2/c (N15) and containing three symmetrically distinct $Cu^{2+}$ sites. The Jahn–Teller distortion of the Cu1, Cu2, and Cu3 copper octahedra is of the (4+2) type, i.e., four short- and two long-range Cu-O bonds. In this case, the $d_{x2-y2}$ orbital is occupied by an unpaired electron. Taking the above into account, we included the contributions only from equatorial oxygen ions into calculations of the parameters of magnetic couplings. The oxygen ions occupying the octahedra axial vertices do not participate in the coupling.

According to our calculations (figure 18(a) and (b), Supplementary table 5), in this polymorph of volborthite, the kagome lattice contains just one type of small triangles with the nonequivalent AFM exchange $J1^1$ couplings ($J1^1$ = -0.0390 Å$^{-1}$, d(Cu2-Cu3) = 2.935 Å, $J1^1/J1^2$ = 0.85), $J1^2$ ($J1^2$ = -0.0461 Å$^{-1}$, d(Cu1-Cu2) = 3.032 Å), and $J1^3$ ($J1^3$ = -0.0440 Å$^{-1}$, d(Cu1-Cu3) = 3.032 Å, $J1^3/J1^2$ = 0.95). As in the first polymorph, these strong AFM nearest-neighbor couplings along the sides of the small $Cu_3$ triangle are formed due to one intermediate O2 oxygen ion located in the triangle center above or below its plane (figure 18 (a) and (b)). Two nonequivalent AFM $J2^2$ couplings ($J2^2$ = -0.0081 Å$^{-1}$, d(Cu1-Cu3) = 5.141 Å, $J2^2/J1^2$ = 0.18) and FM $J2^3$ ($J2^3$ = 0.0044 Å$^{-1}$, d(Cu2-Cu3) = 5.306 Å, $J2^3/J1^2$ = 0.09) in the kagome plane are weak and do not compete with the AFM nearest-neighbor $J1^n$ couplings. However, the strength of the AFM $J2^1$ coupling (d(Cu1-Cu2) = 5.141 Å) could drastically change from -0.0060 ($J2^1/J1^2$ = 0.13) to -0.0290 Å$^{-1}$ ($J2^1/J1^2$ = 0.63) depending on the shift along the Cu1-Cu2 bond line of the intermediate O3 ion in the critical position "c" ($l'_n/l_n \approx 2.0$) [11, 12]. This will introduce extra fluctuations into the magnetic state. Such an instability of $J2^n$ couplings exists in three more low-symmetry compounds crystallizing in the $P12_1/m1$ N11, $C2/c$ N15, and *Ia* N9 space groups considered below.

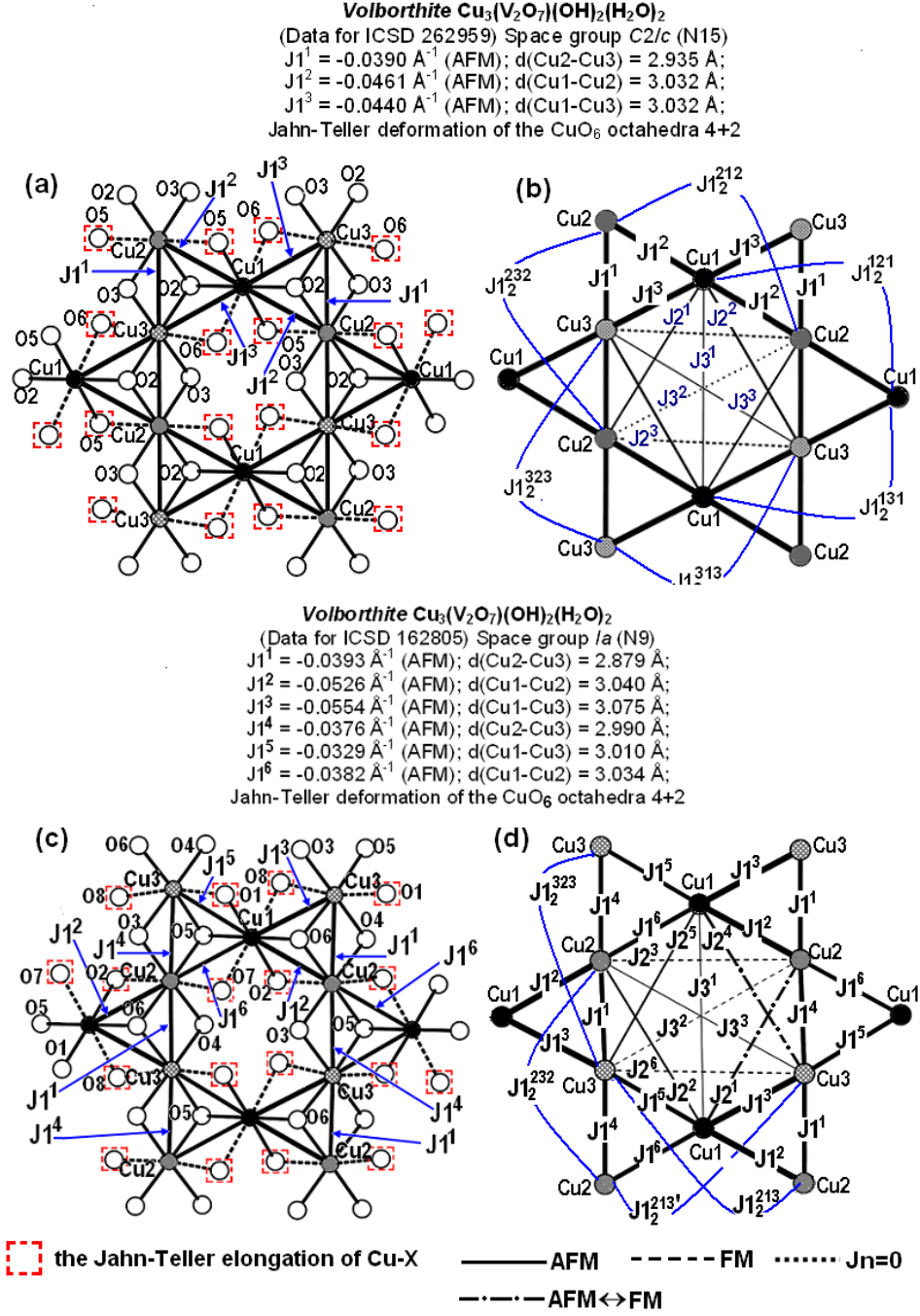

**Figure 18.** Kagome layer $CuO_6$ octahedra (a and c) and $Jn$ couplings in the kagome lattice (b and d) in volborthite $Cu_3(V_2O_7)(OH)_2(H_2O)_2$ in space groups C2/c (N15) and *Ia* (N9).

The diagonal AFM $J3^1(J_d)$ couplings ($J3^1(J_d)$ = -0.0002 Å$^{-1}$, d(Cu1-Cu1) = 5.871 Å, $J3^1(J_d)/J1^2$ = 0.004), $J3^2(J_d)$ ($J3^2(J_d)$ = 0, d(Cu2-Cu2) = 6.064 Å), and AFM $J3^3(J_d)$ ($J3^3(J_d)$ = -0.0006 Å$^{-1}$, d(Cu3-Cu3) = 6.064 Å, $J3^3(J_d)/J1^2$ = 0.01) are virtually absent. All the next-nearest AFM $J3^n(J1_2^{232}$ - d(Cu2-Cu2) = 5.871 Å; $J1_2^{323}$ - d(Cu3-Cu3) = 5.871 Å; $J1_2^{212}$ - d(Cu2-Cu2) = 6.064 Å; $J1_2^{121}$ - d(Cu1-Cu1) = 6.064 Å; $J1_2^{131}$ - d(Cu1-Cu1) = 6.064 Å; and $J1_2^{313}$ = d(Cu3-Cu3) = 6.064 Å) in the chains along the sides of small triangles are relatively strong ($J3^n(J1_2^n)/J1^n$ = 0.75 – 0.93). They compete with the AFM nearest-neighbor $J1^n$ couplings. Let us emphasize that in this space group two pairs of crystallographically equivalent $J3^2(J_d)$ and $J3^2(J1_2^{212})$ (d(Cu2-Cu2) = 6.064 Å, $J3^2(J_d)/J1_2^{212}$ = 0) and $J3^3(J_d)$ and $J3^3(J1_2^{313})$ (d(Cu3-Cu3) = 6.064 Å, $J3^3(J_d)/J3^3(J1_2^{313})$ = 0.009) couplings are magnetically nonequivalent.

All the interplane $J4 – J7$ couplings in the range from 7.209 to 8.032 Å are ferromagnetic (except $J5'$) and very weak ($Jn/J1^2$ varies from -0.07 to 0). The $J5'$ coupling ($J5' = -0.0011$ Å$^{-1}$, d(Cu1-Cu3) = 7.7603 Å) is antiferromagnetic and weak ($J5'/J1^2 = 0.02$). However, as in the former compound, the local spaces of interplane interactions contain O4 oxygen ions from water molecules that are located between kagome planes and not included into the copper coordination. If one takes them into account at calculations of the magnetic coupling parameters, the $J4 – J7$ value would increase dramatically.

The existence of two polymorphs of volborthite $Cu_3(V_2O_7)(OH)_2(H_2O)_2$ in the work by Yoshida et al. [36] was considered as a unique structural transition in single crystals of the spin-1/2 quasi-kagome antiferromagnet, at which an unpaired electron is "switched" from one *d* orbital to another. This is not a usual orbital transition of the order-disorder type, but, rather, an orbital "switching" that was not observed earlier. However, it was predicted by Bersucker [29, 31]. Our calculations confirm that the structural transition induced by the orbital "switching" results in changes in the parameters of magnetic interactions.

The above changes are expressed even more clearly at reduction of the symmetry of volborthite $Cu_3(V_2O_7)(OH)_2(H_2O)_2$ until the noncentrosymmetric monoclinic space group *Ia* N9 [101] (supplementary Note 7). In this case, the number of symmetrically distinct $Cu^{2+}$ sites remains to be equal to 3, and the Jahn-Teller deformation of all $CuO_6$ octahedra is of the 4+2 type, so that the spin-carrying orbital on $Cu^{2+}$ sites is the $d_{x2-y2}$ orbital. The number of nonequivalent nearest-neighbor $J1^n$ couplings would increase up to 6.

According to our calculations (figure 18(c) and (d), Supplementary table 5), in this noncentrosymmetric sample of volborthite, the kagome lattice contains two types of small triangles with the nonequivalent AFM exchange. In the first Cu1Cu2Cu3 triangle, the AFM nearest-neighbor $J1^1$ ($J1^1 = -0.0393$ Å$^{-1}$, d(Cu2-Cu3) = 2.879 Å, $J1^1/J1^3 = 0.71$), $J1^2$ ($J1^2 = -0.0526$ Å$^{-1}$, d(Cu1-Cu2) = 3.040 Å, $J1^2/J1^3 = 0.95$), and $J1^3$ ($J1^3 = -0.0554$ Å$^{-1}$, d(Cu1-Cu3) = 3.075 Å) couplings are formed due to the intermediate O6 oxygen ion. In the second Cu1Cu2Cu3 triangle, the AFM nearest-neighbor $J1^4$ ($J1^4 = -0.0376$ Å$^{-1}$, d(Cu2-Cu3) = 2.990 Å, $J1^4/J1^6 = 0.98$), $J1^5$ ($J1^5 = -0.0329$ Å$^{-1}$, d(Cu1-Cu3) = 3.010 Å, $J1^5/J1^6 = 0.86$), and $J1^6$ ($J1^6 = -0.0382$ Å$^{-1}$, d(Cu1-Cu2) = 3.034 Å) couplings are formed due to the intermediate O5 oxygen ion. Both of these ions are located in the centers of respective triangles above or below their planes, whereas the AFM contributions from the O6 ion exceed in strength those from the O5 ion.

The magnetic parameters of the second in length AFM $J2^1$ (d(Cu1-Cu2) = 5.148 Å) and AFM $J2^4$ (d(Cu1-Cu3) = 5.221 Å) couplings are unstable, since the intermediate O3 ion making the main contribution to their formation is located in the critical position «c» (*l'/l*) [11, 12]. In this case, the insignificant displacement of the O3 ion dramatic increase of the strength of the AFM $J2^1$ coupling from -0.0060 Å$^{-1}$ ($J2^1/J1^3 = 0.11$) to -0.0277 Å$^{-1}$ ($J2^1/J1^3 = 0.50$) and the strength of the AFM $J2^4$ coupling from -0.0080 Å$^{-1}$ ($J2^4/J1^2 = 0.15$) to -0.0312 Å$^{-1}$ ($J2^4/J1^2 = 0.60$). Such an increase of the strengths of the $J2^1$ and $J2^4$ couplings would enable them to compete with the AFM nearest-neighbor $J1^n$ couplings. Two other $J2^2$, $J2^3$, $J2^5$ and $J2^6$ couplings are very weak and cannot compete with the strong AFM nearest-neighbor $J1^n$ couplings. The diagonal AFM $J3^1(J_d)$, FM $J3^2(J_d)$ and AFM $J3^3(J_d)$ couplings are weak to a degree that they are virtually absent. In the low-symmetry space group *Ia* N9, there is no complete crystallographic equivalence between the $J3^n(J_d)$ and $J3(J1_2^n)$ couplings. Nevertheless, as in a majority of the examined above crystal structures, the strengths of the $J3(J1_2^n)$ couplings in the chains along the sides of small triangles exceeds significantly that of the diagonal AFM $J3^n(J_d)$ couplings. Nevertheless, as in a majority of the examined above crystal structures, the strengths of the $J3(J1_2^n)$ couplings in the chains along the sides of small triangles exceeds significantly that of the diagonal AFM $J3^n(J_d)$ couplings. The entire interplane Cu-Cu $J4 – J14$ couplings within the range from 7.175 Å to 8.036 Å are very weak. However, as in the former samples, the local spaces of the interplane interactions contain the O10 and O11 oxygen ions from water molecules that are not included to the copper coordination and could be removed upon the sample heating.

As was shown in [99, 102-105], the distortion of the kagome lattice and the octahedron about $Cu^{2+}$ resulted in the nonequivalent AFM nearest-neighbor $J1^n$ couplings and served as a reason of the absence of the magnetic ordering down to very low temperatures (1K).

Ishikawa et al. [106] synthesized the mineral engelhauptite KCu$_3$(V$_2$O$_7$)(OH)$_2$Cl with the space group P12$_1$/m1 (N11) and demonstrated its similarity with volborthite Cu$_3$(V$_2$O$_7$)(OH)$_2$(H$_2$O)$_2$ in space groups C2/c (N15) with respect to both crystal structure and magnetic properties.

Unlike the natural mineral engelhauptite KCu$_3$(V$_2$O$_7$)(OH)$_2$Cl with the hexagonal structure, ideal kagome lattice, and compressed (type 2+4) copper (Cu$^{2+}$) octahedra, the synthetic crystal has a monoclinic symmetry, distorted kagome lattice, and elongated (4+2) copper octahedral, just like in volborthite. We calculated the parameters of magnetic couplings in the synthetic crystal of engelhauptite using the structural data of [106] (supplementary table 5) and confirm this conclusion.

To sum up, in the case of low-symmetry distorted kagome stricture, we pose the emergence of the DM interaction because of the integral emergence of anisotropy in some individual systems of the kagome plane (hexagonal and triangular), rather than not only in the case of magnetic nonequivalence of crystallographically equivalent interactions. Nonequivalence of the magnetic parameters couplings in kagome systems could induce and really induces the magnetic ordering. We observed this phenomenon for Cr$_{1/3}$NbS$_2$ [16]. Here, we posed a more general phenomenon, when entire distorted blocks participated in the process. In one of recent works [107], similar situation was examined, when in the case of Cu$_3$Nb$_2$O$_8$ the magnetic structure disrupted the inversion symmetry, whereas the crystal structure remained centrosymmetric.

*3.4.4. Bayldonite PbCu$_3$(AsO4)$_2$(OH)$_2$ with deformed kagome lattice and (4+2)-JT distorted octahedral copper coordination.* Bayldonite (PbCu$_3$(AsO4)$_2$(OH)$_2$) [47] (figure 19, Supplementary table 5, Supplementary Note 8) crystallizes in the monoclinic centrosymmetric space group C2/c (N15), just like one of the polymorphs of volborthite Cu$_3$(V$_2$O$_7$)(OH)$_2$(H$_2$O)$_2$, and has much in common with it. The above compounds have similar unit cell parameters, whereas the Jahn–Teller distortion of Cu1, Cu2, and Cu3 copper octahedra is of the (4+2) type and identical way of formation of the dominant AFM nearest-neighbor couplings $J1^n$ due to one intermediate oxygen ion located in the triangles centers above or below their planes. The lengths of four short bonds in the equatorial octahedra plane are in the range d(Cu-O$_{eq}$) = 1.88−2.09 Å, while those of two long axial bonds – d(Cu-O$_{ax}$) = 2.27–2.45 Å (figure 19(a)), so that the contribution to the formation of magnetic interaction is provided only by equatorial oxygen ions.

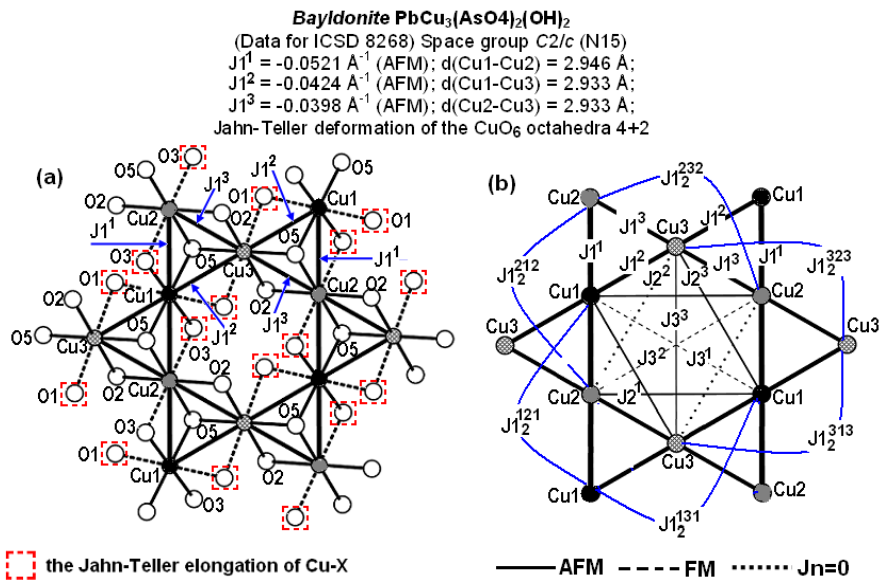

**Figure 19.** Kagome layer CuO$_6$ octahedra and $J$n couplings in the kagome lattice in bayldonite, PbCu$_3$(AsO4)$_2$(OH)$_2$.

According to our calculations (figure 19, Supplementary table 5), the kagome lattice of bayldonite, just like that of volborthite (figure 18(a) and (b)), contains small triangles of just one type. They are characterized with strong nonequivalent AFM nearest-neighbor $J1^1$ couplings ($J1^1$ = -0.0521 Å$^{-1}$, d(Cu1-Cu2) = 2.946 Å), $J1^2$ ($J1^2$ = -0.0424 Å$^{-1}$, d(Cu1-Cu3) = 2.933 Å, $J1^2/J1^1$ = 0.81), and $J1^3$ ($J1^3$ = -0.0398 Å$^{-1}$, d(Cu2-Cu3) = 2.933 Å, $J1^3/J1^2$ = 0.76).

Besides, there exists some analogy with the $J2^n$ couplings as well. Two nonequivalent $J2^2$ ($J2^2$ = 0 Å, d(Cu2-Cu3) = 5.095 Å) and AFM $J2^3$ ($J2^3$ = -0.0066 Å$^{-1}$, d(Cu1-Cu3) = 5.095 Å, $J2^3/J1^1$ = 0.13) couplings in the kagome plane do not compete with the AFM nearest-neighbor $J1^n$ couplings, as one is eliminated, while another is weak, whereas the strength of the AFM $J2^1$ coupling (d(Cu1-Cu2) = 5.074 Å) could dramatically change from -0.0053 Å$^{-1}$ ($J2^1/J1^2$ = 0.10) to -0.0264 Å$^{-1}$ ($J2^1/J1^2$ = 0.51), depending on the shift of the intermediate O2 ion located in the critical position «c». Three diagonal FM $J3^1(J_d)$, FM $J3^2(J_d)$ and AFM $J3^3(J_d)$ are very weak. All the five nonequivalent next-nearest AFM $J3(J1_2^n)$ couplings in the chains along the sides of small triangles are sufficiently strong. They compete with the AFM nearest-neighbor $J1^n$ couplings ($J3^n(J1_2^n)/J1^n$ = 0.75 – 0.93). The crystallographically equivalent $J3^1(J_d)$ and $J3(J1_2^{131})$ couplings between the Cu1-Cu1 ions at a distance of 5.867 Å are magnetically nonequivalent ($J3^1(J_d)/J3(J1_2^{131})$ = -0.04). The magnetic nonequivalence of crystallographically equivalent $J3^2(J_d)$ and $J3(J1_2^{232})$ ($J3^2(J_d)/J3(J1_2^{232})$ = -0.02) couplings is also observed between the Cu2-Cu2 ions at a distance of 5.867 Å.

All the interplane $J4 - J7$ couplings in the range from 6.950 to 7.632 Å are ferromagnetic (except AFM $J4$) and weak. However, the local spaces of interplane interactions contain O4 oxygen ions from AsO$_4$-groups located between kagome planes and not included into the copper coordination and Pb$^{2+}$ ions. If one takes them into account at calculations of the magnetic coupling parameters, the $J4 - J7$ value would increase dramatically.

The magnetic properties of this compound have not yet been studied; however, based on the similarity of structural-magnetic models of bayldonite, PbCu$_3$(AsO4)$_2$(OH)$_2$, and polymorph of volborthite, Cu$_3$(V$_2$O$_7$)(OH)$_2$(H$_2$O)$_2$, crystallizing in the same space group C2/c N15, they must be similar.

*3.4.5. Pb$_2$Cu$_3$O$_2$(NO$_3$)$_2$(SeO$_3$)$_2$ with deformed kagome lattice and the (4+0) type of JT copper coordination distortion.* Pb$_2$Cu$_3$O$_2$(NO$_3$)$_2$(SeO$_3$)$_2$ is a frustrated antiferromagnet with deformed kagome lattice and the (4+0) type of *JT* distortion of copper coordination. The available literature describes only its crystal structure [108], which was determined with an insufficient accuracy (due to low quality of crystals) by means of X-ray single-crystal diffraction (the refinement converged to the residual factor ($R$) values $R$ = 0.07). The layered kagome compound Pb$_2$Cu$_3$O$_2$(NO$_3$)$_2$(SeO$_3$)$_2$ [108] (figure 20(a) and (b), Supplementary table 5) crystallizes in a noncentrosymmetric orthorhombic space group *Cmc*2$_1$ No. 36 ($a$ = 5.884, $b$ = 12.186, c = 19.371Å, $α$ = $β$ = $γ$ = 90º, Z = 4) and contains 2 symmetrically distinct Cu2$^+$ sites. The Jahn–Teller distortion of Cu1 and Cu2 copper octahedra is of the (4+0) type, i.e., two crystallographically different Cu atoms are characterized with approximately flat square coordination with average Cu – O distances of 1.94 and 1.96 Å. The O-Cu-O angles between adjacent O atoms vary from 84.6 to 95.7°, whereas the O-Cu-O angles between the diametrally located O atoms exceed 177.2°. Other oxygen ions are located not closer than at 3.00 Å from copper ions. In Pb$_2$Cu$_3$O$_2$(NO$_3$)$_2$(SeO$_3$)$_2$, the kagome plane is formed by two types of isosceles triangles: O2Cu1Cu2Cu2 and O1Cu1Cu2Cu2. The first triangle is centered by the O2 ion, whereas the second one – by the O1 ion. These O1 and O2 ions deviate insignificantly from both the plane and their own triangle centers (figure 20(c)).

According to our calculations (figure 20(a) and (b), Supplementary table 5), only O1 and O2 centering the OCu$_3$ triangles are included into the space of magnetic couplings and participate in the formation of strong, but not equal in strength AFM nearest-neighbor $J1^n$ couplings. Along the sides of the first triangle under effect of the centered intermediate O2 oxygen atom, there emerge the competing

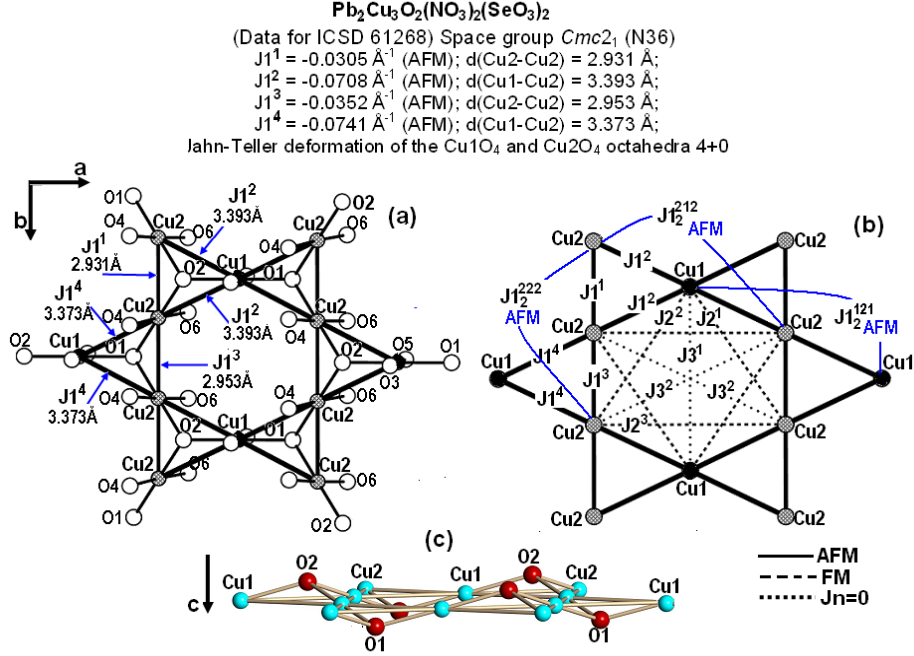

**Figure 20.** Kagome layer $CuO_6$ octahedra (a) and $Jn$ couplings in the kagome lattice (b) in $Pb_2Cu_3O_2(NO_3)_2(SeO_3)_2$, oxocentered $OCu_3$ triangle (c).

AFM $J1^1$ ($J1^1$ = -0.0305 Å$^{-1}$, d(Cu2-Cu2) = 2.931 Å, $J1^1/J1^2$ = 0.43) and two AFM $J1^2$ ($J1^2$ = -0.0708 Å$^{-1}$, d(Cu1-Cu2) = 3.393 Å) couplings that are 2.32-fold stronger than the $J1^1$ ones. Similar situation is observed for the second triangle ($J1^3$ - $J1^4$ - $J1^4$) as well, in which, under effect of the O1 ion, the competing AFM $J1^3$ ($J1^3$ = -0.0352 Å$^{-1}$, d(Cu2-Cu2) = 2.953 Å, $J1^3/J1^4$ = 0.47) and two AFM $J1^4$ ($J1^4$ = -0.0741 Å$^{-1}$, d(Cu1-Cu2) = 3.373 Å) couplings that are 2.10-fold stronger than the $J1^3$ ones are formed. Three nonequivalent $J2^1$, $J2^2$, and $J2^3$ couplings along the sides of large Cu1-Cu2-Cu2 triangles in the kagome lattice are ferromagnetic and weak ($J2^n$ = 0.0020 - 0.0024 Å$^{-1}$, $J2^n/1^4$ = 0.03), so that they cannot compete with the strong AFM $J1^n$ nearest-neighbor couplings. Three diagonal $J3^1(J_d)$ (d(Cu1-Cu1) = 5.884 Å) and two $J3^2(J_d)$ (d(Cu1-Cu2) = 6.766 Å) couplings are eliminated ($J3^n(J_d)$ = 0).

The strengths of two nonequivalent AFM $J3(J1_2^n)$ ($J3(J1_2^{222})$ = -0.0357 Å$^{-1}$, d(Cu2-Cu2) = 5.884 Å, $J3(J1_2^{222})/J1^4$ = 0.48; $J3(J1_2^{121})$ = -0.0306 Å$^{-1}$, d(Cu1-Cu1) = 6.766 Å, $J3(J1_2^{121})/J1^4$ = 0.41) next-nearest couplings in the chains along the sides of small triangles are sufficiently high, unlike those of diagonal couplings. These AFM $J3(J1_2^{222})$ and $J3(J1_2^{121})$ couplings are capable to compete with the AFM nearest-neighbor $J1^n$ ones. The third $J3(J1_2^{212})$ coupling ($J3(J1_2^{212})$ = -0.0112 Å$^{-1}$, d(Cu2-Cu2) = 6.766 Å, $J3(J1_2^{212})/J1^4$ = 0.15) is significantly weaker and has the crystallographically, but not magnetically, equivalent $J3^2(J_d)$ one ($J3^2(J_d)/J3(J1_2^{212})$ = 0). In theory, the AFM $J3(J1_2^n)$ couplings can be significantly reduced down to the value, so that they would not be able to compete with the AFM nearest-neighbor $J1^n$ couplings. For this purpose, it is necessary to just slightly shift he oxygen ions located in the critical position "a" (O6 − Δa = 0.014 Å in $J3(J1_2^{222})$; O4 − Δa = 0.023 Å and O6 − Δa = 0.078 Å in $J3(J1_2^{121})$; Δa = 0.051 Å in $J3(J1_2^{212})$) a bit deeper inside the local space of the $J3(J1_2^n)$ coupling.

The kagome planes are located at very large distances from each other. All the interplane $J4 - J8$ couplings in the range from 9.794 to 10.644 Å, except the FM $J6$ ($J6/J1^4$ = -0.13), are weak AFM ones ($Jn/J1^4$ varies from -0.06 to 0.12).

To sum up, the asymmetry of the magnetic structure (strength inequality of four AFM $J1^n$ couplings and drastic differences in the parameters of spacially similar $J3^n(J_d)$ and $J3(J1_2^n)$ couplings), in addition to the noncentrosymmetry character of the space group $Cmc2_1$, could serve as a reason for the magnetic ordering of $Pb_2Cu_3O_2(NO_3)_2(SeO_3)_2$ under effect of the *DM* forces.

One should mention that, from the crystal chemistry point of view, the compounds $PbCu_3(AsO4)_2(OH)_2$ and $Pb_2Cu_3O_2(NO_3)_2(SeO_3)_2$ could be of a great interest as objects with a complex of potential magnetic or electric properties. They contain $Pb^{2+}$ and $Se^{4+}$ ions having a stereochemically active lone electron pair. Such an electronic formation creates nonrigid parts in the structure and makes it unstable through possible easy changes in the shape and position under effect of temperature, pressure, introduction of vacancies, or ions substitution. As we demonstrated on the example of using $BiFeO_3$ [109], the metal–insulator and paraelectric–ferroelectric phase transitions result directly from changes in the degree of stereochemical activity of the lone pair – its sequential increase along with the decrease of temperature. The emergence of the magnetic ordering upon changes in the characteristics of magnetic couplings comprises a secondary effect of changes in the stereochemical activity of the lone pair inducing shifts of the intermediate ions in the local space between magnetic ions.

## 4. Conclusions

We have built up the structural-magnetic models of 25 spin-1/2 frustrated kagome antiferromagnets in order to establish crystal chemistry parameters of both existence and destruction of the spin liquid on the kagome lattice. We have demonstrated that strong frustration of dominating AFM nearest-neighbor $J1$ couplings in kagome plane triangles resists under effect of just the antisymmetric anisotropic exchange interaction (Dzyaloshinskii-Moriya), which is present in all the kagome antiferromagnetics, including herbertsmithite ($ZnCu_3(OH)_6Cl_2$), since magnetic ions are not linked to the center of inversion in the kagome lattice. Additional forces are required for the magnetic ordering. We have established that such forces include dual (anisotropic) $J3$ couplings ($J3(J1_2)$ next-to-nearest-neighbor couplings in linear chains along the triangles edges and $J3(J_d)$ along hexagon diagonals). The crystallographically equivalent $J3(J1_2)$ and $J3(J_d)$ couplings differ dramatically with respect to the strength of magnetic interactions ($J3(J1_2) \gg J3(J_d)$) and reduce the symmetry of the magnetic lattice relatively to the symmetry of its crystal structure.

It became possible to suppress such a duality (anisotropy) of the $J3$ couplings only through a lucky accident in herbertsmithite ($ZnCu_3(OH)_6Cl_2$) – by insignificant shifts of the intermediate oxygen ions, which could make the crystallographically equivalent magnetic interactions to become magnetic equivalents as well. The latter allows preserving the frustration of magnetic interactions upon the temperature decrease down to 0 K and, as a result, transforms herbertsmithite into a spin liquid.

Duality is a fundamental phenomenon in nature producing equilibrium for contradictions. In physics, the term "duality" has a rather broad concept [110–113]. In this case, duality – magnetic nonequivalence (weakness and strength) of crystallographically equivalent interactions – serves as a reason for creating the forces destructing the spin liquid and resulting in the magnetic ordering of frustrated AFM systems.

Based on the performed studies, we have concluded that centrosymmetric frustrated AFM systems, in which the duality (anisotropy) is eliminated, and crystallographically identical couplings are magnetically identical as well, must be searched as candidates for the spin liquid.

**Data availability statement**
All data that support the findings of this study are included within the article (and any supplementary files).

**Additional information**
Supplementary information is available in the online version of the paper.

**Acknowledgment**
The work was financially supported within the frames of the State Order of the Institute of Chemistry FEBRAS, project No. 0205-2021-0001.

**References**

[1] Balents L 2010 *Nature* **464** 199
[2] Savary L and Balents L 2017 *Rep. Prog. Phys*. **80** 16502
[3] Norman M R 2016 *Rev. Mod. Phys*. **88** 41002
[4] Gong S S, Zhu W, Balents L and Sheng D N 2015 *Phys. Rev*. B **91** 75112
[5] Tchernyshyov O 2004 *J. Phys.: Condens. Matter.* **16** S709
[6] Mondal K and Kadolkar C 2017 *Phys. Rev*. B **95** 134404
[7] Smirnov A I 2006 *Physics-Uspekhi* **49** 649
[8] Inosov D S 2018 *Adv. Phys*. **67** 149
[9] Braithwaite R S W, Mereiter K, Paar W H and Clark A M 2004 *Mineral. Mag.* **68** 527
[10] Shores M P, Nytko E A, Bartlett B M and Nocera D G 2005 *J. Am. Chem. Soc.* **127** 13462
[11] Volkova L M and Polyshchuk S A 2005 *J. Supercond.* **18** 583
[12] Volkova L M and Marinin D V 2009 *J. Phys.: Condens. Matter* **21** 015903
[13] Volkova L M 2009 *J Struct. Chem.* **50**, *(Suppl.)* S49
[14] Dzyaloshinsky I 1958 *J. Phys. Chem. Solids* **4** 241
[15] Moriya T 1960 *Phys. Rev*. **120** 91
[16] Volkova L M and Marinin D V 2014 *J. App. Phys.* **116**, 133901
[17] Kramers H A 1934 *Physica* **1** 182
[18] Goodenough J B 1955 *Phys. Rev.* 100 564
[19] Kanamori J 1959 *J. Phys. Chem. Solids* **10** 87
[20] Anderson P W 1963 *Solid State Physics* vol 14, ed F Seitz and D Turnbull (New York: Academic) pp 99–214
[21] Vonsovsky S V 1971 *Magnetism* (Moscow: Nauka)
[22] Bert F, Nakamae S, Ladieu F, Hôte D L, Bonville P, Duc F, Trombe J-C and Mendels P 2007 *Phys. Rev*. B **76** 132411
[23] Helton J S, Matan K, Shores M P, Nytko E A, Bartlett B M, Yoshida Y, Takano Y, Suslov A, Qiu Y, Chung J-H, Nocera D G and Lee Y S 2007 *Phys. Rev. Lett.* **98** 107204
[24] Misguich G and Sindzingre P 2007 *Eur. Phys. J.* B **59** 305
[25] Jeschke H O, Salvat-Pujol F and Valenti R 2013 *Phys. Rev*. B **88** 75106
[26] Arh T, Gomilšek M, Prelovšek P, Pregelj M, Klanjšek M, Ozarowski A, Clark S J, Lancaster T, Sun W, Mi J-X and Zorko A 2020 *Phys. Rev. Lett.***125** 27203
[27] Shannon R D 1976 *Acta Crystallogr*. A **32** 751
[28] Jahn H A and Teller 1937 *Proc. R. Soc. London* Ser. A **161** 220
[29] Bersuker I B, Vekhter B J and Ogurtsov I Ya 1975 *Sov. Phys. Usp.* **18** 569
[30] Kugel' K I and Khomskii D I 1982 *Sov. Phys. Usp.* **25** 231
[31] Bersuker I B 1986 *Russ. Chem. Rev*. **55** 581
[32] Streltsov S V and Khomskii D I 2017 *Phys. Usp.* **60** 1121
[33] Kugel' K I and Khomskii D I 1973 *Zh. Eksp. Teor. Fiz*. **64** 1429
[34] Lafontaine M A, Le Bail A and Ferey G J 1990 *Solid State Chem.* **85** 220
[35] Ishikawa H, Yamaura J, Okamoto Y, Yoshida H, Nilsen G J and Hiroi Z 2012 *Acta Crystallogr*. C **68** i41
[36] Yoshida H, Yamaura J, Isobe M, Okamoto Y, Nilsen G J and Hiroi Z 2012 *Nat. Commun.* **3** 860
[37] Towler M D, Dovesi R and Saunders V R 1995 *Phys. Rev*. B **52** 10150
[38] Yamada Y and Kato N 1994 *J. Phys. Soc. Japan* **63** 289
[39] Paolasini L, Caciuffo R, Sollier A, Ghigna P and Altarelli M 2002 *Phys. Rev. Lett.* **88** 106403
[40] Hutchings M T, Samuelson E J, Shirane G and Hirakawa K 1969 *Phys. Rev.* **188** 919
[41] Oleś A M, Horsch P, Feiner L F and Khaliullin G 2006 *Phys. Rev. Lett.* **96** 147205
[42] Satija S K, Axe J D, Shirane G, Yoshizawa H and Hirakawa K 1980 Phys. Rev. B. **21**, 2001
[43] Tanaka K, Konishi M and F Marume 1979 *Acta Cryst.* B **35** 1303
[44] Volkova L M and Marinin D V 2018 *Phys. Chem. Miner.* **45** 655
[45] Volkova L M and Marinin D V 2018 *J. Phys.: Condens. Matter* **30** 425801
[46] Yoshida H, Michiue Y, Takayama-Muromachi E and Isobe M 2012 *J. Mater. Chemi.* **22** 18793

[47] Ghose S and Wan C 1979 *Acta Crystallogr.* B **35** 819
[48] Kolley F, Depenbrock S, McCulloch I P, Schollwock U and Alba V 2015 *Phys. Rev.* B **91** 104418
[49] Jiang H C, Wang Z and Balents L 2012 *Nat. Phys.* **8** 901–5
[50] Iqbal Y, Poilblanc D and Becca F 2015 *Phys. Rev.* B **91** 020402
[51] Liao H J, Xie Y, Chen J, Liu Z Y, Xie H D, Huang R Z, Normand B and Xiang T 2017 *Phys. Rev. Lett.* **118** 137202
[52] Okamoto K and Nomura K 1992 *Phys. Lett.* A **169,** 433.
[53] Bonca J, Rodriguez J P, Ferrer J and Bedell K S 1994 *Phys. Rev.* B **50** 3415
[54] White S and Affleck I 1996 *Phys. Rev.* B **54** 9862.
[55] Eggert S 1996 *Phys. Rev.* B **54** R9612
[56] Gong S S, Zhu W and Sheng D N 2014 *Sci. Rep.* **4** 6317
[57] Wietek A, Sterdyniak A and Läuchli A M 2015 *Phys. Rev.* B **92** 125122
[58] He Y-C and Chen Y 2015 *Phys. Rev. Lett.* **114** 037201
[59] Zhu W, Gong S S and Sheng D N 2015 *Phys. Rev.* B **92** 14424
[60] Götze O and Richter J 2016 *EPL* **114** 67004
[61] Zorko A, Nellutla S, van Tol J, Brunel L C, Bert F, Duc F, Trombe J-C, de Vries M A, Harrison A and Mendels P 2008 *Phys. Rev. Lett.* **101** 026405
[62] Zorko **A**, Herak M, Gomilšek M, van Tol J, Velázquez M, Khuntia P, Bert F and Mendels P 2017 *Phys. Rev. Lett.* **118** 017202
[63] Olariu A, Mendels P, Bert F, Duc F, Trombe J C, de Vries M A and Harrison A 2008 *Phys. Rev. Lett.* **100** 087202
Cépas O, Fong C M, Leung P W and Lhuillier C 2008 *Phys. Rev.* B **78,** 140405.
[64] Colman R H, Sinclair A and Wills A S 2011 *Chem. Mater.* **23** 1811
[65] Kermarrec E, Mendels P, Bert F, Colman R H, Wills A S, Strobel P, Bonville P, Hillier A. and Amato A 2011 *Phys. Rev.* B.**84**.100401
[66] Sun W, Huang Y X, Nokhrin S, Pan Y and Mi J X 2016 *J. Mater. Chem.* C **4** 8772
[67] Malcherek T and Schlüter J 2007 *Acta Crystallogr.* B **63** 157
[68] Wei Y, Feng Z, d. Cruz C, Yi W, Meng Z Y, Mei J-W, Shi Y and Li S 2019 *Phys. Rev.* B **100** 155129
[69] Oswald H R 1969 *Helvetica Chimica Acta* **52** 2369
[70] Downie L J, Black C, Ardashnikova E I, Tang C C, Vasiliev A N, Golovanov A N, Berdonosov P S, Dolgikh V A and Lightfoot P 2014 *CrystEngComm* **16** 7419
[71] Mueller M and Mueller B G 1995 *Z. anorg. allg. Chem.* **621** 993
**[72]** Downie L J, Ardashnikova E I, Tang C C, Vasiliev A N, Berdonosov P S, Dolgikh V A, de Vries M A and Lightfoot P 2015 *Crystals* **5** 226
**[73]** Barthélemy Q, Puphal P, Zoch K, Krellner C, Luetkens H, Baines C, Sheptyakov D, Kermarrec E, Mendels P and Bert F 2019 *Phys. Rev. Materials* **3** 074401
**[74]** Zorko A, Pregelj M, Klanjšek M, Gomilšek M, Jaglicic Z, Lord J S, Verezhak J A T, Shang T, Sun W and Mi J-X 2019 *Phys. Rev.* B **99** 214441
**[75]** Zorko A, Pregelj M, Gomilšek M, Klanjšek M, Zaharko O, Sun W and Mi J-X 2019 *Phys. Rev.* B **100** 144420
**[76]** Colman R H, Sinclair A and Wills A S 2010 *Chem. Mater.* **22** 5774
[77] Boldrin D, Fåk B, Enderle M, Bieri S, Ollivier J, Rols S, Manuel P and Wills A S 2015 *Phys. Rev.* B **91**220408
[78] Janson O, Richter J and Rosner H 2008 *Phys. Rev. Lett.* **101** 106403
[79] Okuma R, Yajima T, Nishio-Hamane D, Okubo T and Hiroi Z 2017 *Phys. Rev.* B **95** 094427
[80] Matan K, Ono T, Gitgeatpong G, de Roos K, Miao P, Torii S, Kamiyama T, Miyata A, Matsuo A, Kindo K, Takeyama S, Nambu Y, Piyawongwatthana P, Sato T J and Tanaka H 2019 *Phys. Rev.* B **99** 224404
[81] Ono T, Morita K, Yano M, Tanaka H, Fujii K, Uekusa H, Narumi Y and Kindo K 2009 *Phys. Rev.* B **79** 174407
[82] Ono T, Matan K, Nambu Y, Sato T J, Katayama K, Hirata S and Tanaka H J. 2014 *Phys. Soc. Jpn.* **83** 043701
[83] Reisinger S A, Tang C C, Thompson S P, Morrison F D and Lightfoot P 2011 *Chem. Mater.* **23** 4234
[84] Downie L J, Thompson S P, Tang C C, Parsons S and Lightfoot P T 2013 *CrystEngComm* **15** 7426
[85] Morita K, Yano M, Ono T, Tanaka H, Fujii K and Uekusa H 2008 *J. Phys. Soc. Jpn.* **77** 043707
[86] Matan K, Ono T, Fukumoto Y, Sato T J, Yamaura J, Yano M, Morita K. and Tanaka H 2010 *Nature Phys.* **6** 865
[87] Matan K, Nambu Y, Zhao Y, Sato T J, Fukumoto Y, Ono T, Tanaka H, Broholm C, Podlesnyak A and Ehlers G 2014 *Phys. Rev.* B 89, 024414
[88] Elliott P, Brugger J and Caradoc-Davies T 2010 *Mineralogical Magazine* **74** 39
[89] Ishikawa H, Okamoto Y and Hiroi Z 2013 *J. Phys. Soc. Jpn.* **82** 063710
[90] Pekov I V, Siidra O I, Chukanov N V, Yapaskurt V O, Britvin S N, Krivovichev S V, Schüller W and Ternes B 2015 *Mineral. Petrol.* **109** 705
[91] Ma Zhesheng, He Ruilin and Zhu. Xiaoling 1991 *Acta Geologica Sinica* **4** 145
[92] Okamoto Y, Yoshida H and Hiroi Z 2009 *J. Phys. Soc. Jap.* **78** 033701
[93] Boldrin D, Knight K and Wills A S 2016 *J. Mater. Chem.* C **4** 10315

[94] Boldrin D, Fåk B, Canévet E, Ollivier J, Walker H C, Manuel P, Khalyavin D D and Wills A S 2018 *Phys. Rev. Lett.* **121** 107203.
[95] Burns P C and Hawthorne F C 1996 *Canadian Mineralogist* **34** 1089
[96] Udovenko A A and Laptash N M 2019 *Acta Cryst.* B **75** 1164
[97] Haldane F D M 1982 *Phys. Rev.* B **25** 4925
[98] Eggert S and Affleck I 1992 *Phys. Rev.* B **46** 10866
[99] Yoshida M, Okamoto Y, Takigawa M and Hiroi Z 2013 *J. Phys. Soc. Jap.* **82** 013702
[100] Efenberger H 1989 *Z. Kristallographie* **188** 43
[101] Kashaev A A, Rozhdenstvenskaya I V, Bannova I I, Sapozhnikov A N, Glebova O D 2008 *J. Struct. Chem.* (USSR) **49** 708
[102] Nilsen G J, Coomer F C, de Vries M A, Stewart J R, Deen P P, Harrison A and Rønnow H M 2011 *Phys. Rev.* B **84** 172401
[103] Watanabe D, Sugii K, Shimozawa M, Suzuki Y, Yajima T, Ishikawa H, Hiroi Z, Shibauchi T, Matsuda Y and Yamashita M 2016 *Proc. Natl. Acad. Sci. U.S.A.* **113** 8653.
[104] Bert F, Bono D, Mendels P, Ladieu F, Duc F, Trombe J-C and Millet P 2005 *Phys. Rev. Lett.* **95** 087203
[105] Yoshida M, Takigawa M, Yoshida H, Okamoto Y and Hiroi Z 2009 *Phys. Rev. Lett.* **103** 077207
[106] Ishikawa H, Nishio-Hamane D, Miyake A, Tokunaga M, Matsuo A, Kindo K and Hiroi Z 2019 *Phys. Rev. Materials* **3** 064414
[107] Giles-Donovan N, Qureshi N, Johnson R D, Zhang L Y, Cheong S-W, Cochran S and Stock C 2020 *Phys. Rev.* B **102** 024414
[108] Effenberger H 1986 *Monatsh. Chem.* **117** 1099
[109] Volkova L M and Marinin D V 2011 *J. Supercond* Nov Magn **24** 2161
[110] Lv J-P, Deng Y, Jacobsen J L, Salas J and Sokal A D 2018 *Phys. Rev.* E **97** 040104(R)
[111] Fruchart M, Zhou Y and Vitelli V 2020 *Nature* **577** 636
[112] Otto H H 2020 *J Modern Physics* **11** 98
[113] Gorsky A S 2005 *Usp. Fiz. Nauk* 175 1145-62
Gorsky A S 205 *Phys. Usp.* 48 1093-108

— *Supplementary Material* —
*Crystal Chemistry Criteria of the Existence of Spin Liquids on the Kagome Lattice*

*L M Volkova and D V Marinin*

**Supplementary Note 1: Figure 1 and Table 1.**

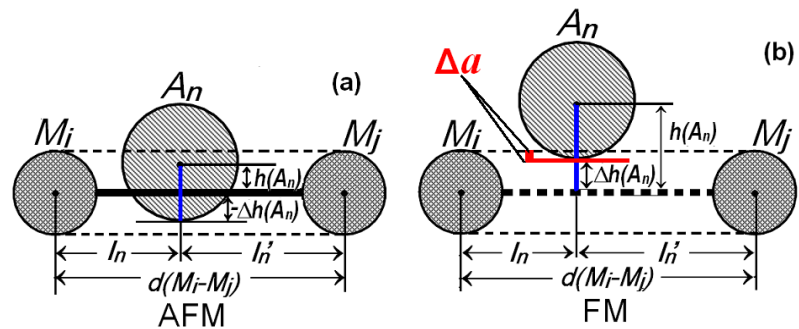

**Figure 1.** A schematic representation of the intermediate A$n$ ion arrangement in the local space between magnetic ions $M_i$ and $M_j$ in cases where the A$n$ ion initiates the emerging of the antiferromagnetic (**a**) and ferromagnetic (**b**) interactions. $h(A_n)$, $l_n$, $l_n'$, and $d(M_i–M_j)$ are the parameters determining the sign and strength of magnetic interactions $(J_n)$, critical positions "$a$" ($\Delta a \sim 0.1$ Å ($\Delta a = (r_M + r_{An}) - h_{An}$).

**Table 1.** Crystallographic characteristics and parameters of magnetic couplings ($J_n$) calculated on the basis of structural data and respective distances between magnetic $Cu^{2+}$ ions in kagome antiferromagnets with (4 + 2) - elongated octahedral Cu2 + - analogs of herbertsmithite

| Crystallographic and magnetic parameters | Herbertsmithite $ZnCu_3(OH)_6Cl_2$ [9] (Data for ICSD – 425834) Space group $R$ -3$mH$ (N166) $a = b = 6.834$, $c = 14.075$ Å $\alpha = \beta = 90°$, $\gamma = 120°$, Z =3 Method[a] - XDS (296 K); $R$-value[b] = 0.0118 *(no ordering)* | Herbertsmithite (Mg) $\gamma$-$Cu_{3.038}Mg_{0.962}(OH)_6Cl_2$ [66] (Data for ICSD – 182038) Space group $R$ -3$mH$ (N166) $a = b = 6.8389$, $c = 14.0212$Å $\alpha = \beta = 90°$, $\gamma = 120°$, Z =3 Method[a] - XDP (295 K); $R$-value[b] = No *(ordering down to 1.8 K)* | $YCu_3(OH)_6Cl_2$ [68] (No in ICSD) Space group $P$-3$m$1 (N164) $a = b = 6.7490$, $c$=5.6244Å $\alpha = \beta = 90°$, $\gamma = 120°$, Z =1 Method[a] - XDS (295 K); $R$-value[b] = 0.026 *(ordering at 15 K)* | Haydeeite $\alpha$-$MgCu_3(OH)_6Cl_2$ [69] (Data for ICSD – 240663) Space group $P$-3$m$1 (N164) $a = b = 6.2733$, $c$ =5.7472Å $\alpha = \beta = 90°$, $\gamma = 120°$, Z =1 Method[a] - XDS (293 K); $R$-value[b] = 0.0225 *(ordering at 4.2 K [69, 70])* | $MgCu_3(OH)_6Br_2$ [70] Space group $P$-3$m$1 (N164) $a = b = 6.2931$, $c = 6.1064$Å $\alpha = \beta = 90°$, $\gamma = 120°$, Z =1 Method[a] - XDS (10 K); $R$-value[b] = 0.0225 *(ordering at 5.4 K)* | $CdCu_3(OH)_6(NO_3)_2H_2O$ [71] (Data for ICSD – 16962) Space group $P$-3$m$1 (N164) $a = b = 6.522$, $c = 7.012$ Å $\alpha = \beta = 90°$, $\gamma = 120°$, Z =1 Method[a] - XDS (296 K); $R$-value[b] = 0.111 *(order below $T_N \sim 4$)* |
|---|---|---|---|---|---|---|
| d(Cu-X) (Å) | Cu-O = 1.984 x4 -Cl = 2.779 x2 | Cu-O = 1.993 x4 -Cl = 2.765 x2 | Cu-O = 1.977 x4 -Cl = 2.811 x2 | Cu-O = 1.977 x4 -Cl = 2.765 x2 | Cu-O = 1.975 x4 -Br = 2.860 x2 | Cu-O1 = 2.036 x4 -O2 = 2.425 x2 |
| **Kagome plane** | | | | | | |
| Bond | Cu1-Cu1 | Cu1-Cu1 | Cu1-Cu1 | Cu1-Cu1 | Cu1-Cu1 | Cu1-Cu1 |
| d(Cu-Cu) (Å) | 3.417 | 3.419 | 3.375 | 3.137 | 3.147 | 3.261 |
| $J1$[c] (Å$^{-1}$) | -0.0670 (AFM) | -0.0641 (AFM) | -0.0649 (AFM) | -0.0399 (AFM) | -0.0416 (AFM) | -0.0340 (AFM) |
| $j(O1)$[d] (Å$^{-1}$) | -0.0670 | -0.0641 | -0.0649 | -0.0399 | -0.0416 | -0.0340 |
| ($\Delta h(O1)$[e] Å, $l_n'/l_n$[f], CuO1Cu[g]) | (-0.391, 1.0, 118.9°) | (-0.375, 1.0, 118.1°) | (-0.369, 1.0, 117.2°) | (-0.196, 1.0, 105.0°) | (-0.206, 1.0, 105.6°) | (-0.181, 1.0, 106.4°) |
| Bond | Cu1-Cu1 | Cu1-Cu1 | Cu1-Cu1 | Cu1-Cu1 | Cu1-Cu1 | Cu1-Cu1 |
| d(Cu-Cu) (Å) | 5.918 | 5.923 | 5.845 | 5.433 | 5.450 | 5.648 |

| $J2^{(c)}$ (Å$^{-1}$) | -0.0108 (AFM) | -0.0104 (AFM) | -0.0160 (AFM) | -0.0162 (AFM) | -0.0174 (AFM) | -0.0130 (AFM) |
|---|---|---|---|---|---|---|
| $j$(O1)$^d$ (Å$^{-1}$) | -0.0054x2 | -0.0052x2 | -0.0080x2 | -0.0081x2 | -0.0087x2 | -0.0065x2 |
| ($\Delta h$(O1)$^e$ Å, $l_n$'/$l_n^f$, CuO1Cu$^g$) | (-0.451, 2.4, 138.6°) | (-0.434, 2.4, 138.0°) | (-0.610, 2.2, 145.4°) | (-0.500, 2.1, 139.2°) | (-0.536, 2.1, 140.8°) | (-0.442, 2.1, 137.9°) |
| $J2/J1^x$ | 0.16 | 0.16 | 0.25 | 0.41 | 0.42 | 0.38 |
| Bond | Cu1-Cu1 | Cu1-Cu1 | Cu1-Cu1 | Cu1-Cu1 | Cu1-Cu1 | Cu1-Cu1 |
| d(Cu-Cu) (Å) | 6.834 | 6.839 | 6.749 | 6.273 | 6.293 | 6.522 |
| $J3(J_d)^{(c)}$ (Å$^{-1}$) | 0.0018 (FM) | 0.0020 (FM) | -0.0012 (AFM) | -0.0028 (AFM) | -0.0036 (AFM) | 0 |
| $j$(O1)$^d$ (Å$^{-1}$) | 0.00045x4 | 0.0005x4 | $j$(O1): -0.0003x4 | $j$(O1): -0.0007x4 | $j$(O1): -0.0009x4 | $j$(O1): 0x4 |
| ($\Delta h$(O1)$^e$ Å, $l_n$'/$l_n^f$, CuO1Cu$^g$) | (0.091, 4.2, 116.2°) | (0.102, 4.2, 115.9°) | (-0.0058, 3.6, 123.0°) | (-0.009, 3.2, 123.2°) | (-0.0118, 3.2, 124.6°) | (-0.003, 3.4, 121.2°) |
| $J3(J_d)/J1$ | 0 | -0.03 | -0.03 | 0.02 | 0.07 | 0.09 |
| Bond | Cu1-Cu1 | Cu1-Cu1 | Cu1-Cu1 | Cu1-Cu1 | Cu1-Cu1 | Cu1-Cu1 |
| d(Cu-Cu) (Å) | 6.834 | 6.839 | 6.749 | 6.273 | 6.293 | 6.522 |
| $J3(J1_2)^{(c)}$ (Å$^{-1}$) | -0.0300(AFM) ↔ 0.0178(FM) | -0.0298(AFM)↔0.0188(FM) | -0.0304AFM↔0.0196 (FM) | -0.0324 (AFM) | -0.0322 (AFM) | -0.0296 (AFM) |
| $j$(Cu1)$^d$ (Å$^{-1}$) | -0.0244 | -0.0244 | -0.0250 | -0.0290 | -0.0288 | -0.0268 |
| ($\Delta h$(O1)$^e$ Å, $l_n$'/$l_n^f$, CuCuCu$^g$) | (-0.570, 1.0, 180°) | (-0.570, 1.0, 180°) | (-0.570, 1.0, 180°) | (-0.570, 1.0, 180°) | (-0.570, 1.0, 180°) | (-0.570, 1.0, 180°) |
| $j$(O1)$^d$ (Å$^{-1}$) | -0.0028x2 | -0.0027x2 | -0.0027x2 | -0.0017x2 | -0.0017x2 | -0.0014x2 |
| ($\Delta h$(O1)$^e$ Å, $l_n$'/$l_n^f$, CuO1Cu$^g$) | (-0.391, 3.0, 138.3°) | (-0.375, 3.0, 137.7°) | (-0.369, 3.0, 137.1°) | (-0.196, 3.0, 128.1°) | (-0.206, 3.0, 128.6°) | (-0.181, 3.0, 129.2°) |
| $j$(O1)$^d$ (Å$^{-1}$) | 0.0239x2 | 0.0243x2 | 0.0250x2 | — | — | — |
| ($\Delta h$(O1)$^e$ Å, $l_n$'/$l_n^f$, CuO1Cu$^g$) | (0.544$^h$, 1.26, 120.2°) | (0.553$^h$, 1.26, 120.6°) | (0.563$^h$, 1.15, 119.4°) | — | — | — |
| $\Delta a^h$ (Å) | $\Delta a$ = 0.026 | $\Delta a$ = 0.017 | $\Delta a$ = 0.007 | | | |
| $J3(J1_2)/J1$ | 0.45 ↔ -0.27 | 0.46 ↔ -0.29 | 0.47 ↔ -0.30 | 0.81 | 0.77 | 0.87 |
| *Interplane couplings* | | | | | | |
| Bond | Cu1-Cu1 | Cu1-Cu1 | Cu1-Cu1 | Cu1-Cu1 | Cu1-Cu1 | Cu1-Cu1 |
| d(Cu-Cu) (Å) | 5.090 | 5.075 | 5.624 | 5.747 | 6.106 | 7.012 |
| $J4^{(c)}$ (Å$^{-1}$) | -0.0020 AFM | -0.0024 AFM | 0 | 0 | 0 | $J4$ = 0.0052 (FM) |
| $j$(O1)$^d$ (Å$^{-1}$) | -0.0010x2 | $j$(O1): -0.0012x2 | | | | $j$(O1): 0.0013x4 |
| ($\Delta h$(O1)$^e$ Å, $l_n$'/$l_n^f$, CuO1Cu$^g$) | (-0.067, 2.5, 117.6°) | (-0.074, 2.4, 118.3°) | | | | (0.402, 6.4, 101.2°) |
| $J4/J1$ | 0.03 | 0.04 | 0 | 0 | 0 | $J4/J1$ = -0.15 |
| Bond | Cu1-Cu1 | Cu1-Cu1 | Cu1-Cu1 | Cu1-Cu1 | Cu1-Cu1 | Cu1-Cu1 |
| d(Cu-Cu) (Å) | 6.130 | 6.119 | 6.559 | 6.547 | 6.869 | 7.733 |
| $J5^{(c)}$ (Å$^{-1}$) | 0.0032 (FM) | 0.0034 (FM) | $J5$ = -0.0009 (AFM) | $J5$ = -0.0011 (AFM) | $J5$ = -0.0006 (AFM) | $J5$ = -0.0003 (AFM) |
| $j$(O1)$^d$ (Å$^{-1}$) | 0.0016x2 | 0.0017x2 | -0.0009x1 | -0.0011 | -0.0006 | -0.0003 |
| ($\Delta h$(O1)$^e$ Å, $l_n$'/$l_n^f$, CuO1Cu$^g$) | (0.315, 5.1, 101.7°) | (0.315, 5.1, 101.9°) | (-0.129, 3.3, 125.9°) | (-0.157, 3.3, 127.1°) | (-0.099, 3.6, 125.2°) | (-0.075, 4.0, 127.3°) |
| $J5/J1$ | -0.05 | -0.05 | 0.01 | 0.03 | 0.01 | $J5/J1$ = 0.01 |
| Bond | Cu1-Cu1 | Cu1-Cu1 | Cu1-Cu1 | Cu1-Cu1 | Cu1-Cu1 | Cu1-Cu1 |
| d(Cu-Cu) (Å) | 6.130 | 6.119 | 8.111 | 7.909 | 8.185 | 9.004 |
| $J5'^{(c)}$ (Å$^{-1}$) | -0.0012 (AFM) | -0.0012 (AFM) | $J6$ = -0.0058 (AFM) | $J6$ = 0.0010 (FM) | $J6$ = 0.0008 (FM) | $J6$ = 0.0008 (FM) |
| $j$(O1)$^d$ (Å$^{-1}$) | -0.0003x4 | -0.0003x4 | -0.0029x2 | 0.0005x2 | $j$(O1): 0.0004x2 | $j$(O1): 0.0004x2 |
| ($\Delta h$(O1)$^e$ Å, $l_n$'/$l_n^f$, CuO1Cu$^g$) | (-0.037, 3.3, 120.4°) | (-0.040, 3.2, 120.7°) | (0.0649, 3.4, 150.9°) | (0.0169, 5.6, 114.3°) | (0.0174, 5.9, 114.5°) | (0.0231, 6.4 115.0°) |
| $J5'/J1$ | 0.02 | 0.02 | 0.04 | -0.02 | $J6/J1$ = -0.02 | $J6/J1$ = -0.02 |
| Bond | Cu1-Cu1 | Cu1-Cu1 | Cu1-Cu1 | Cu1-Cu1 | Cu1-Cu1 | Cu1-Cu1 |
| d(Cu-Cu) (Å) | 7.018 | 7.009 | 8.111 | 7.909 | 8.185 | 9.004 |
| $J6^{(c)}$ (Å$^{-1}$) | -0.0025 (AFM) | -0.0026 (AFM) | $J6'$ = 0.0012 (FM) | $J6'$ = -0.0072 (AFM) | $J6'$ = -0.0056 (AFM) | $J6'$ = -0.0038 (AFM) |
| $j$(O1)$^d$ (Å$^{-1}$) | -0.0025 | -0.0026 | 0.0006x2 | -0.0036x2 | $j$(O1): -0.0028x2 | $j$(O1): -0.0019x2 |
| ($\Delta h$(O1)$^e$ Å, $l_n$'/$l_n^f$, CuO1Cu$^g$) | (-0.382, 3.1, 138.3°) | (-0.388, 3.1, 138.7°) | (0.288, 6.9, 108.0°) | (-0.730, 3.3, 153.9°) | (-0.0645, 3.5, 150.8°) | (-0.0581, 3.8, 149.7°) |
| $J6/J1$ | 0.04 | 0.04 | -0.02 | 0.18 | $J6/J1$ = 0.13 | $J6/J1$ = 0.11 |

| Bond | Cu1-Cu1 | Cu1-Cu1 | Cu1-Cu1 | Cu1-Cu1 | Cu1-Cu1 | Cu1-Cu1 |
|---|---|---|---|---|---|---|
| d(Cu-Cu) (Å) | 8.521 | 8.516 | 8.785 | 8.508 | 8.769 | 9.576 |
| $J7$ [c] (Å$^{-1}$) | 0.0079 (FM) | 0.0068 (FM) | $J7$ = -0.0014 (AFM) | $J7$ = -0.0026 (AFM) | $J7$ = -0.0028 (AFM) | $J7$ = -0.0024 (AFM) |
| $j$(O1)[d] (Å$^{-1}$) | 0.0079 | 0.0073 | $j$(O1): -0.0007x2 | $j$(O1): -0.0013x2 | $j$(O1): -0.0014x2 | $j$(O1): -0.0012x2 |
| ($\Delta h$(O1)[e] Å, $l_n$'/$l_n$ [f], CuO1Cu[g]) | (0.282, 1.2, 136.7°) | (0.263, 1.2, 137.2°) | (-0.0236, 4.5, 134.6°) | (-0.0385, 4.0, 140.6°) | (-0.0445, 4.1, 143.4°) | (-0.0456, 4.3, 145.4°) |
| $J7/J1$ | -0.12 | -0.11 | 0.02 | -0.07 | -0.07 | $J7/J1$ = -0.07 |
| Bond | Cu1-Cu1 | Cu1-Cu1 | Cu1-Cu1 | Cu1-Cu1 | Cu1-Cu1 | Cu1-Cu1 |
| d(Cu-Cu) (Å) | 9.181 | 9.177 | 8.785 | 8.508 | 8.769 | 9.576 |
| $J8$ [c] (Å$^{-1}$) | -0.0062 (AFM) | -0.0062 (AFM) | $J7$' = -0.0036 (AFM) | $J7$' = 0 | $J7$' = 0 | $J7$' = -0.0014 (AFM) |
| $j$(O1)[d] (Å$^{-1}$) | 0.0031x2 | 0.0031x2 | $j$(O1): -0.0018x2 | | | $j$(O1): -0.0007x2 |
| ($\Delta h$(O1)[e] Å, $l_n$'/$l_n$ [f], CuO1Cu[g]) | (-0.0974, 3.7, 164.2°) | (-0.0980, 3.7, 164.5°) | (-0.0537, 4.0, 147.1°) | | | (-0.282, 4.6, 138.6°) |
| $J8/J1$ | 0.09 | 0.1 | 0.06 | 0 | 0 | $J7'/J1$ = -0.04 |

[a]XDS – X-ray diffraction from single crystal; XDP - X-ray diffraction (powder)
[b]The refinement converged to the residual factor (R) values.
[c]$Jn$ in Å$^{-1}$ ($Jn$ (meV) = $Jn$ (Å$^{-1}$)×K, where scaling factors K$_{middle}$ = 74) – the magnetic couplings ($Jn<0$ - AFM, $Jn>0$ – FM).
[d]$j$(X) – contributions of the intermediate X ion into the AFM ($j$(X) <0) and FM ($j$(X)>0) components of the $Jn$ coupling
[e]$\Delta h$(X) – the degree of overlapping of the local space between magnetic ions by the intermediate ion X.
[f]$l_n$'/$l_n$ is the asymmetry of position of the intermediate X ion relatively to the middle of the Cu$_i$–Cu$_j$ bond line.
[g]Cu$_i$XCu$_j$ bonding angle
[h]$\Delta a \sim 0.1$ Å ($\Delta a = (r_M + r_{An})$ - $h_{An}$) - critical position of intermediate A$n$ ion.

**Supplementary Note 2:**
**Table 2.** Crystallographic characteristics and parameters of magnetic couplings (*J*n) calculated on the basis of structural data and respective distances between magnetic Cu$^{2+}$ ions in kagome antiferromagnets mixed metal fluorides Cs$_2$SnCu$_3$F$_{12}$, Cs$_2$ZrCu$_3$F$_{12}$ and Cs$_2$TiCu$_3$F$_{12}$

| Crystallographic and magnetic parameters | Cs$_2$SnCu$_3$F$_{12}$[72] (Data for ICSD – 291388) Space group $R$ -3$mH$ (N166) $a = b$ = 7.1315, $c$ = 20.3609Å $\alpha = \beta = 90º, \gamma = 120º$, Z =3 Method[a] - NDP (300 K) $R$-value[b] – no given | Cs$_2$ZrCu$_3$F$_{12}$[73] (Data for ICSD – 79114) Space group $R$ -3$mH$ (N166) $a=b$ = 7.1661, $c$=20.4640 Å $\alpha = \beta = 90º, \gamma = 120º$, Z =3 Method[a] - XDS (293 K); $R$-value[b] – 0.02 | Cs$_2$TiCu$_3$F$_{12}$[74] (Data for ICSD – 429374) Space group $R$-3$mH$ (N166) $a = b$ = 7.1014, $c$=19.955Å $\alpha = \beta = 90º, \gamma = 120º$, Z =3 Method[a] - XDS (293 K); $R$-value[b] = 0.0326 |
|---|---|---|---|
| d(Cu-X) (Å) | Cu-F1 = 1.898 x4 -F2 = 2.356 x2 | Cu-F1 = 1.897 x4 -F2 = 2.340 x2 | Cu-F1 = 1.903x4 -F2 = 2.333x2 |
| *Kagome plane* | | | |
| Bond | Cu1-Cu1 | Cu1-Cu1 | Cu1-Cu1 |
| d(Cu-Cu) (Å) | 3.566 | 3.583 | 3.551 |
| *J*1[c] (Å$^{-1}$) | -0.1067 (AFM) | -0.1101 (AFM) | -0.1021 (AFM) |
| *j*(F1)[d] (Å$^{-1}$) | -0.1067 | -0.1101 | -0.1021 |
| ($\Delta h$(F1)[e] Å, $l_n'/l_n$[f], CuF1Cu[g]) | (-0.652, 1.0, 139.8°) | (-0.707, 1.0, 141.6°) | (-0.644, 1.0, 137.7°) |
| Bond | Cu1-Cu1 | Cu1-Cu1 | Cu1-Cu1 |
| d(Cu-Cu) (Å) | 6.176 | 6.206 | 6.150 |
| *J*2[c] (Å$^{-1}$) | -0.0180 (AFM) | -0.0184 (AFM) | -0.0174 (AFM) |
| *j*(F1)[d] (Å$^{-1}$) | -0.0090x2 | -0.0092x2 | -0.0087x2 |
| ($\Delta h$(F1)[e] Å, $l_n'/l_n$[f], CuF1Cu[g]) | (-0.816 2.4, 157.5°) | (-0.844 2.4, 158.8°) | (-0.780, 2.4, 156.0°) |
| *J*2/*J*1 | 0.17 | 0.17 | 0.17 |
| Bond | Cu1-Cu1 | Cu1-Cu1 | Cu1-Cu1 |
| d(Cu-Cu) (Å) | 7.132 | 7.166 | 7.101 |
| *J*3(*J*$_d$)[c] (Å$^{-1}$) | -0.0004 (AFM) | -0.0008(AFM) | -0.0004(AFM) |
| *j*(F1)[d] (Å$^{-1}$) | -0.0001x4 | -0.0002x4 | -0.0001x4 |
| ($\Delta h$(F1)[e] Å, $l_n'/l_n$[f], CuF1Cu[g]) | (-0.020, 4.1, 124.0°) | (-0.033, 4.2, 124.2°) | (-0.020, 4.1, 123.6°) |
| *J*3(*J*$_d$)/*J*1 | 0.004 | 0.007 | 0.004 |
| Bond | Cu1-Cu1 | Cu1-Cu1 | Cu1-Cu1 |
| d(Cu-Cu) (Å) | 7.132 | 7.166 | 7.101 |
| *J*3(*J*1$_2$)[c] (Å$^{-1}$) | -0.0312(AFM) ↔ 0.0112(FM) | -0.0314 (AFM) ↔ 0.0104 (FM) | -0.0312 (AFM) ↔ 0.0120 (FM) |
| *j*(Cu)[d] (Å$^{-1}$) | -0.0224 | -0.0222 | -0.0226 |
| ($\Delta h$(Cu)[e] Å, $l_n'/l_n$[f], CuCuCu[g]) | (-0.570, 1.0, 180°) | (-0.570, 1.0, 180°) | (-0.570, 1.0, 180°) |
| *j*(F1)[d] (Å$^{-1}$) | -0.0044x2 | -0.0046x2 | -0.0043x2 |
| ($\Delta h$(F1)[e] Å, $l_n'/l_n$[f], CuFCu[g]) | (-0.678, 3.0, 153.0°) | (-0.707, 3.0, 154.2°) | (-0.644, 3.0, 151.5°) |
| *j*(F1)[d] (Å$^{-1}$) | 0.0212x2 | 0.0209x2 | 0.0216x2 |
| ($\Delta h$(F1)[e] Å, $l_n'/l_n$[f], CuF1Cu[g]) | (0.526[h], 1.25, 124.6°) | (0.523[h], 1.26, 124.8°) | (0.532[h], 1.25, 124.2°) |
| $\Delta a$[h] (Å) | $\Delta a$ = 0.044 | $\Delta a$ = 0.047 | $\Delta a$ = 0.038 |
| *J*3(*J*1$_2$)/*J*1 | 0.29 ↔ -0.10 | 0.29 ↔ -0.09 | 0.31 ↔ -0.12 |
| *Interplane couplings* | | | |
| Bond | Cu1-Cu1 | Cu1-Cu1 | Cu1-Cu1 |
| d(Cu-Cu) (Å) | 7.092 | 7.128 | 6.960 |
| *J*4[c] (Å$^{-1}$) | 0 | 0 | 0 |
| Bond | Cu1-Cu1 | Cu1-Cu1 | Cu1-Cu1 |
| d(Cu-Cu) (Å) | 7.938 | 7.978 | 7.814 |
| *J*5[c] (Å$^{-1}$) | 0 | 0 | 0 |
| Bond | Cu1-Cu1 | Cu1-Cu1 | Cu1-Cu1 |
| d(Cu-Cu) (Å) | 7..938 | 7..978 | 7.814 |
| *J*5'[c] (Å$^{-1}$) | 0.0008 (FM) | 0.0008 (FM) | 0.0004 (FM) |
| *j*(F1)[d] (Å$^{-1}$) | 0.0004x2 | 0.0004x2 | 0.0002x2 |
| ($\Delta h$(F1)[e] Å, $l_n'/l_n$[f], CuF1Cu[g]) | (0.122, 5.5, 117.9°) | (0.157, 5.8, 116.0°) | (0.077, 5.1, 120.2°) |
| *J*5'/*J*1 | -0.007 | -0.007 | -0.002 |
| Bond | Cu1-Cu1 | Cu1-Cu1 | Cu1-Cu1 |
| d(Cu-Cu) (Å) | 8.702 | 8.746 | 8.583 |
| *J*6[c] (Å$^{-1}$) | 0.0007 (FM) | 0.0002 (FM) | 0.0003 (FM) |
| *j*(F1)[d] (Å$^{-1}$) | 0.0005 | 0.0002 | -0.0001 |
| ($\Delta h$(F1)[e] Å, $l_n'/l_n$[f], CuF1Cu[g]) | (0.312, 8.1, 100.8°) | (0.104, 6.0, 120.1°) | (-0.039, 5.4, 123.7°) |
| *j*(F1)[d] (Å$^{-1}$) | 0.0002 | - | 0.0004 |
| ($\Delta h$(F1)[e] Å, $l_n'/l_n$[f], CuF1Cu[g]) | (0.072, 5.8, 121.7°) | - | (0.217, 7.5., 109.8°) |
| *J*6/*J*1 | -0.007 | -0.002 | -0.004 |

[a]XDS – X-ray diffraction from single crystal; XDP - X-ray diffraction (powder)
[b]The refinement converged to the residual factor (*R*) values.
[c]*J*n in Å$^{-1}$ (*J*n (meV) = *J*n (Å$^{-1}$)×K, where scaling factors K$_{middle}$ = 74) – the magnetic couplings (*J*n<0 - AFM, *J*n>0 – FM).
[d]*j*(X) – contributions of the intermediate X ion into the AFM (*j*(X) <0) and FM (*j*(X)>0) components of the *J*n coupling
[e]$\Delta h$(X) – the degree of overlapping of the local space between magnetic ions by the intermediate ion X.
[f]$l_n'/l_n$ is the asymmetry of position of the intermediate X ion relatively to the middle of the Cu$_i$–Cu$_j$ bond line.
[g]Cu$_i$XCu$_j$ bonding angle
[h]$\Delta a \sim 0.1$ Å ($\Delta a = (r_M + r_{An}) - h_{An}$) - critical position of intermediate A*n* ion.

**Supplementary Note 3:**
**Table 3.** Crystallographic characteristics and parameters of magnetic couplings ($J_n$) calculated on the basis of structural data and respective distances between magnetic $Cu^{2+}$ ions in the monoclinic and rhombohedral phases of mixed metal fluorides and edwardsite $Cd_2Cu_3(SO_4)_2(OH)_6 \cdot 4H_2O$ with deformed kagome lattice

| Crystallographic and magnetic parameters | $Cs_2SnCu_3F_{12}$ [72] (Data for ICSD – 291385) Space group $P2_1/n$ (N14)) $a = 7.8966, b = 7.0975, c = 10.5876$ Å $α = 90º, β = 97.8, γ = 90º, Z =2$ Method[a] - NDP (100 K); No $R$-value[b] | $Cs_2TiCu_3F_{12}$ [74] (Data for ICSD – 429375) Space group $P2_1/n$ (N14)) $a = 7.7578, b = 7.0432, c = 10.4344$Å $α = 90º, β = 96.9, γ = 90º, Z =2$ Method[a] - SRP (100 K); No $R$-value[b] | $Cs_2TiCu_3F_{12}$ [74] (Data for ICSD – 429373) Space group $R$ -$3H$ (N148) $a = b = 14.163, c = 19.865$ Å $α =90º, β = 90º, γ = 120º, Z =12$ Method[a] - XDS (125 K); $R$-value[b] = 0.0438 | $Rb_2SnCu_3F_{12}$ [86] (Data for ICSD – 426453) Space group $R$ -$3H$ (N148) $a = b = 13.905, c = 20.333$ Å $α =90º, β = 90º, γ = 120º, Z =12$ Method[a] - XDS (300 K); $R$-value[b] = 0.0287 | $Cs_2ZrCu_3F_{12}$ [85] (Data for ICSD – 182754) Space group $P2_1/m$ (N11) $a = 7.798, b = 7.212, c = 10.351$ Å $α = 90º, β = 93.97, γ = 90º, Z =2$ Method[a] - XDS (125 K); $R$-value[b] – 0.0489 | Edwardsite $Cd_2Cu_3(OH)_6(SO_4)_2(H_2O)_4$[90] (Data for ICSD – 185315) Space group $P2_1/c$ N14 $a$=10.863, $b$=13.129, $c$=11.169 Å $α = 90º, β = 113.04, γ = 90º, Z$ =4 Method[a] – XDS (123K); $R$-value[b] = 0.0321 | |
|---|---|---|---|---|---|---|---|
| d(Cu-X) (Å) | Cu1-F1 = 1.874x2 -F2 = 1.933x2 -F6 = 2.304x2 Cu2-F3 = 1.839 -F1 = 1.878 -F2 = 1.925 -F3 = 1.976 -F4 = 2.351 -F5 = 2.355 | Cu1-F2 = 1.912x2 -F1 = 1.917x2 -F6 = 2.310x2 Cu2-F3 = 1.875 -F1 = 1.903 -F2 = 1.921 -F3 = 1.952 -F4 = 2.319 -F5 = 2.325 | Cu1-F3 = 1.877 -F1 = 1.890 -F3 = 1.902 -F2 = 1.911 -F6 = 2.307 -F8 = 2.334 Cu2-F3 = 1.900 -F1 = 1.918 -F4 = 1.944 -F4 = 1.974 -F7 = 2.325 -F5 = 2.359 | Cu1-F2 = 1.897x2 -F3 = 1.900 -F1 = 1.913 -F4 = 2.390 -F7 = 2.582 -F10 = 2.594 Cu2-F6 = 1.890 -F3 = 1.899 -F6 = 1.912 -F1 = 1.916 -F5 = 2.315 -F9 = 2.360 | Cu1-F2 = 1.930x2 -F1 = 1.959x2 -F3 = 2.236 -F4 = 2.368 Cu2-F1 = 1.906x2 -F5 = 2.018x2 -F6 = 2.191x2 Cu3-F7 = 1.913x2 -F2 = 1.930x2 -F8 = 2.303x2 | Cu1-O10 = 1.950 -O11 = 1.958 -O9 = 1.972 -O13 = 1.982 -O2 = 2.539 -O5 = 2.611 Cu2-O9 = 1.978 x2 -O12 = 1.981 x2 -O5 = 2.430 x2 | Cu3-O14 = 1.947 x2 -O13 = 1.982 x2 -O2 = 2.537 x2 Cu4-O11 = 1.956 x2 -O14 = 1.962 -O10 = 1.985 -O2 = 2.362 -O5 = 2.838 |
| *Kagome plane* | *Triangle 1 - Cu1Cu2Cu2* | *Triangle 1 - Cu1Cu2Cu2* | *Triangle 1 – Cu2Cu1Cu1* | *Triangle 1 - Cu1Cu2Cu2* | *Triangle 1 - Cu1Cu3Cu3* | *Triangle1-Cu1Cu2Cu4* | *Triangle2-Cu1Cu3Cu4* |
| Bond | Cu1-Cu2 | Cu1-Cu2 | Cu1-Cu1 | Cu1-Cu2 | Cu1-Cu3 | Cu2-Cu4 | Cu1-Cu3 |
| d(Cu-Cu) (Å) | 3.576 | 3.567 | 3.545 | 3.579 | 3.657 | 3.323 | 3.364 |
| $J_n$[c] (Å$^{-1}$) | $J1^1$ = -0.1079 (AFM) | $J1^1$ = -0.0989 (AFM) | $J1^1$ = -0.1056 (AFM) | $J1^1$ = -0.1015 (AFM) | $J1^1$ = -0.1066 (AFM) | $J1^1$ = -0.0626 | $J1^4$ = -0.0621 |
| $j(X)^d$ (Å$^{-1}$) | $j$(F1): - 0.1079 | $j$(F2): -0.989 | $j$(F2): - 0.1056 | $j$(F1): - 0.1015 | $j$(F2): - 0.1066 | $j$(O12):-0.0626 | $j$(O13): -0.0621 |
| ($Δh(X)^e$ Å, $l_n'/l_n^f$, CuXCu$^g$) | (-0.689, 1.0, 140,6°) | (-0.629, 1.0, 137.1°) | (-0.664, 1.0, 138,8°) | (-0.650, 1.0, 138,4°) | (-0.713, 1.0, 142,7 °) | (-0.345, 1.0, 115.2°) | (-0.351, 1.0, 116.1°) |
| $J_n/J_{max}$ | $J1^1/J1^1$ = 1 | $J1^1/J1^1$ = 1 | $J1^1/J1^1$ = 1.0 | $J1^1/J1^1$ = 1.0 | $J1^1/J1^1$= 1 | $J1^1/J1^1$ = 1 | $J1^4/J1^1$ = 1 |
| Bond | Cu2-Cu2 | Cu2-Cu2 | Cu1-Cu2 | Cu1-Cu2 | Cu3-Cu3 | Cu1-Cu4 | Cu1-Cu4 |
| d(Cu-Cu) (Å) | 3.551 | 3.525 | 3.566 | 3.533 | 3.606 | 3.217 | 3.233 |
| $J_n$[c] (Å$^{-1}$) | $J1^2$ = -0.1008 (AFM) | $J1^2$ = -0.0943 (AFM) | $J1^2$ = -0.1040 (AFM) | $J1^2$ = -0.1005 (AFM) | $J1^2$ = -0.1062 (AFM) | $J1^2$ = -0.0517 | $J1^5$ = -0.0567 |
| $j(X)^d$ (Å$^{-1}$) | $j$(F3): - 0.1008 | $j$(F3): - 0.943 | $j$(F1): -0.1040 | $j$(F6): - 0.1005 | $j$(F7): -0.1062 | $j$(O10): -0.0517 | $j$(O11): -0.0567 |
| ($Δh(X)^e$ Å, $l_n'/l_n^f$, CuXCu$^g$) | (-0.633, 1.1, 137,1°) | (-0.585, 1.0, 134,2°) | (-0.661, 1.0, 138,9°) | (-0.627, 1.0, 136,6°) | (-0.690, 1.0, 140,9°) | (-0.267, 1.0, 109.7°) | (-0.296, 1.0, 111.4°) |
| $J_n/J_{max}$ | $J1^2/J1^1$ = 0.93 | $J1^2/J1^1$ = 0.95 | $J1^2/J1^1$ = 0.98 | $J1^2/J1^1$ = 0.98 | $J1^2/J1^1$ = 1.0 | $J1^2/J1^1$ = 0.82 | $J1^5/J1^1$ = 0.90 |
| Bond | Cu1-Cu2 | Cu1-Cu2 | Cu1-Cu2 | Cu1-Cu2 | *Triangle 2 - Cu1Cu2Cu2* | Cu1-Cu2 | Cu3-Cu4 |
| d(Cu-Cu) (Å) | 3.537 | 3.498 | 3.512 | 3.360 | Cu1-Cu2 | 3.220 | 3.198 |
| $J_n$[c] (Å$^{-1}$) | $J1^3$= -0.0991 (AFM) | $J1^3$ = -0.0919 (AFM) | $J1^3$ = -0.0975 (AFM) | $J1^4$ = -0.0784 (AFM) | 3.603 | $J1^3$ = -0.0494 | $J1^6$ = -0.0538 |
| $j(X)^d$ (Å$^{-1}$) | $j$(F2): -0.0991 | $j$(F1): - 0.919 | $j$(F3): - 0.0975 | $j$(F3): - 0.0784 | $J1^3$ = -0.0971 (AFM) | $j$(O9): -0.0494 | $j$(O14): -0.0538 |
| ($Δh(X)^e$ Å, $l_n'/l_n^f$, CuXCu$^g$) | (-0.619, 1.0, 136.2°) | (-0.563, 1.0, 132,6°) | (-0.601, 1.0, 134,9°) | (-0.443, 1.0, 124,3°) | $j$(F1): - 0.0971 | (-0.256, 1.0, 109.2°) | (-0.275, 1.0, 109.7°) |
| | $J1^3/J1^1$ = 0.92 | $J1^3/J1^1$ = 0.93 | $J1^3/J1^1$ = 0.92 | $J1^4/J1^1$ = 0.77 | (-0.630, 1.0, 137,5°) | $J1^3/J1^1$ = 0.79 | $J1^8/J1^1$ = 0.85 |
| | $J2^n$ | $J2^n$ | *Triangle 2 – Cu2Cu2Cu2* | *Triangle 2 – Cu1Cu1Cu1* | $J1^3/J1^1$ = 0.91 | — | — |
| | Cu1-Cu2 | Cu1-Cu2 | Cu2-Cu2 | Cu1-Cu1 | Cu2-Cu2 | — | — |
| | 6.023 | 6.243 | 3.550 | 3.487 | 3.606 | — | — |
| | $J2^1$ = -0.0159 (AFM) | $J2^1$ = -0.0191 (AFM) | $J1^4$ = -0.0795 (AFM) | $J1^3$ = -0.0957 (AFM) | $J1^4$ = -0.0653 (AFM) | — | — |
| | $j$(F2): -0.0069 | $j$(F3): -0.0078 | $j$(F4): - 0.0795 | $j$(F2): -0.0957 | $j$(F5): - 0.0653 | — | — |
| | (-0.593, 2.4, 147.7°) | (-0.757, 2.5, 154.9°) | (-0.501, 1.0, 129,9°) | (-0.582, 1.0, 133,6°) | (-0.424, 1.0, 126,7°) | — | — |

|   |   |   |   |   |   |   |
|---|---|---|---|---|---|---|
| *j*(F2): -0.0090 | *j*(F2): -0.0113 | J1⁴/J1¹ = 0.75 | J1³/J1¹ = 0.94 | J1⁴/J1¹ = 0.61 | — | — |
| (-0.793, 2.4, 155.8) | (-1.015, 2.3, 166.84) | ***J2ⁿ*** | ***J2ⁿ*** | ***J2ⁿ*** |  | ***J2ⁿ*** |
| J2¹/J1¹ = 0.15 | J2¹/J1¹ = 0.19 | Cu1-Cu2 | Cu1-Cu2 | Cu1-Cu2 | Cu2-Cu3 | Cu1-Cu1 |
| Cu2-Cu2 | Cu2-Cu2 | 6.003 | 5.763 | 6.244 | 5.585 | 5.586 |
| 6.163 | 6.124 | J2¹ = -0.0165 (AFM) | J2¹ = -0.0132 (AFM) | J2¹ = -0.0098 (AFM) | J2¹ = -0.0162 | J2⁴ = -0.0160 |
| J2² = -0.0177 (AFM) | J2² = -0.0167 (AFM) | *j*(F3): -0.0065 | *j*(F2): -0.0082 | *j*(F1): -0.0077 | *j*(O12): -0.0112 | *j*(O10): -0.0074 |
| *j*(F2): -0.0073 | *j*(F1): -0.0059 | (-0.574, 2.4, 145.5°) | (-0.620, 2.3, 148.0°) | (-0.711, 2.4, 153.6°) | (-0.704, 2.0, 148.9°) | (-0.513, 2.2, 140.0°) |
| (-0.689, 2.5, 151.7°) | (-0.554, 2.5, 159.0°) | *j*(F4): -0.0100 | *j*(F3): -0.0050 | *j*(F5): -0.0021 | *j*(O14): -0.0050 | *j*(O11): -0.0086 |
| *j*(F1): -0.0104 | *j*(F2): -0.0108 | (-0.799, 2.2, 156.8°) | (-0.408, 2.5, 138.3°) | (-0.221, 2.7, 133.0) | (-0.369, 2.4, 133.3°) | (-0.575, 2.4, 142.9°) |
| (-0.905, 2.3, 161.7) | (-0.920, 2.3, 162.2) | J2¹/J1¹ = 0.16 | J2¹/J1¹ = 0.13 | J2¹/J1¹ = 0.09 | J2¹/J1¹ = 0.26 | J2⁴/J1¹ = 0.26 |
| J2²/J1¹ = 0.16 | J2¹/J1¹ = 0.17 | Cu2-Cu2 | Cu1-Cu1 | Cu1-Cu3 | Cu2-Cu4 | Cu1-Cu4 |
| Cu1-Cu2 | Cu1-Cu2 | 6.131 | 6.023 | 6.275 | 5.589 | 5.698 |
| 6.276 | 5.966 | J2² = -0.0169 (AFM) | J2² = -0.0168 (AFM) | J2² = -0.0168 (AFM) | J2² = -0.0166 | J2⁵ = -0.0154 |
| J2³ = -0.0192 (AFM) | J2³ = -0.0144 (AFM) | *j*(F1): -0.0107 | *j*(F1): -0.0132 | *j*(F2): -0.0101 | *j*(O11): -0.0123 | *j*(O12): -0.0097 |
| *j*(F1): -0.0106 | *j*(F1): -0.0051 | (-0.912, 2.3, 161.8°) | (-1.047, 2.2, 167.6°) | (-0.928, 2.3, 162.7°) | (-0.777, 2.0, 152.0°) | (-0.675, 2.1, 147.6°) |
| (-0.999, 2.4, 165.6°) | (-0.455, 2.5, 141.2°) | *j*(F3): -0.0062 | *j*(F3): -0.0036 | *j*(F7): -0.0067 | *j*(O9): -0.0043 | *j*(O9): -0.0057 |
| *j*(F3): -0.0086 | *j*(F4): -0.0093 | (-0.582, 2.5, 147.1°) | (-0.354, 2.7, 136.6°) | (-0.657, 2.5 150.9°) | (-0.320, 2.4, 131.6°) | (-0.432, 2.3, 137.0°) |
| (-0.778, 2.3 156.6°) | (-0.730, 2.2 153.8°) | J2²/J1¹ = 0.16 | J2²/J1¹ = 0.16 | J2²/J1¹ = 0.16 | J2²/J1¹ = 0.26 | J2⁵/J1¹ = 0.25 |
| J2³/J1¹ = 0.18 | J2³/J1¹ = 0.15 | Cu1-Cu2 | Cu1-Cu2 | Cu2-Cu3 | Cu2-Cu3 | Cu1-Cu4 |
| ***J3ⁿ(Jₐ)*** | ***J3ⁿ(Jₐ)*** | 6.268 | 6.253 | 6.260 | 5.662 | 5.713 |
| Cu2-Cu2 | Cu2-Cu2 | J2³ = -0.0199 (AFM) | J2³ = -0.0197 (AFM) | J2³ = -0.0160 (AFM) | J2³ = -0.0162 | J2⁶ = -0.0152 |
| 7.309 | 6.858 | *j*(F1): -0.0100 | *j*(F1): -0.0114 | *j*(F1): -0.0056 | *j*(O12): -0.0112 | *j*(O13): -0.0098 |
| J3¹ = -0.0002 (AFM) | J3¹ = -0.0002 (AFM) | (-0.938, 2.4, 163.9°) | (-1.026, 2.3, 166.9°) | (-0.569, 2.6, 146.9°) | (-0.704, 2.0, 148.9° | (-0.670, 2.1, 147.2°) |
| *j*(F2): 0x2 | *j*(F2): -0.0005x2 | *j*(F4): -0.0099 | *j*(F3): -0.0083 | *j*(F2): -0.0104 | *j*(O14): -0.0050 | *j*(O14): -0.0054 |
| *j*(F3): -0.0001x2 | *j*(F3): 0.0004x2 | (-0.883, 2.2 161.0°) | (-0.792, 2.4 156.6°) | (-0.905, 2.3, 163.3) | (-0.369, 2.4, 133.3°) | (-0.414, 2.4, 136.1°) |
| J3¹/J1¹ =-0.002 | J3¹/J1¹ = 0.002 | J2³/J1¹ = 0.19 | J2³/J1¹ = 0.19 | J2³/J1¹ = 0.15 | J2³/J1¹ = 0.26 | J2⁶/J1¹ = 0.24 |
| ***J3ⁿ(Jₐ)*** | ***J3ⁿ(Jₐ)*** | ***J3ⁿ(Jₐ)*** | ***J3ⁿ(Jₐ)*** | ***J3ⁿ(Jₐ)*** | ***J3ⁿ(Jₐ)*** | ***J3ⁿ(Jₐ)*** |
| Cu1-Cu1 | Cu1-Cu1 | Cu2-Cu2 | Cu1-Cu1 | Cu1-Cu1 | Cu1-Cu3 | Cu4-Cu4 |
| 7.098 | 7.043 | 6.879 | 6.639 | 7.212 | 6.482 | 6.567 |
| J3²= -0.0004 (AFM) | J3²= -0.0002 (AFM) | J3¹ = -0.0008 (AFM) | J3¹ = -0.0004 (AFM) | J3¹= 0 | J3¹= -0.0016 | J3³ = -0.0022 |
| *j*(F1): -0.0007x2 | *j*(F1): 0.0007x2 | *j*(F1): -0.0002x2 | *j*(F1): -0.0007x2 | *j*(F1): 0.0002x2 | *j*(O11): -0.0012; | *j*(O12): -0.0016; |
| *j*(F2): 0.0005x2 | *j*(F2): -0.0008x2 | *j*(F4): -0.0002x2 | *j*(F2): 0.0005x2 | *j*(F2): -0.0002x2 | *j*(O13): -0.0010; | *j*(O11): -0.00015; |
| J3²/J1¹ = 0.004 | J3²/J1¹ = 0.002 | J3¹/J1¹ = 0.008 | J3¹/J1¹ = 0.004 | J3¹/J1¹ = 0 | *j*(O14): 0.0004; | *j*(O10): 0.0005; |
| Cu2-Cu | Cu2-Cu2 | Cu1-Cu1 | Cu2-Cu2 | Cu2-Cu3 | *j*(O9): 0.0002 | *j*(O14): 0.0004 |
| 6.911 | 7.268 | 7.084 | 6.926 | 7.225 | J3¹/J1¹ = 0.03 | J3³/J1¹ = 0.03 |
| J3³ = -0.0006 (AFM) | J3³ = -0.0004 (AFM) | J3²= -0.0005 (AFM) | J3²= -0.0008 (AFM) | J3²= 0.0016 (FM) | Cu1-Cu2 | — |
| *j*(F1): -0.0002x2 | *j*(F1): 0.0004x2 | *j*(F1): -0.0007x2 | *j*(F1): -0.0012x2 | *j*(F1): 0.0002x2 | 6.484 | — |
| *j*(F3): -0.0001x2 | *j*(F3): -0.0008x2 | *j*(F3): 0.0005x2 | *j*(F3):0.0008x2 | *j*(F2): -0.0003 | J3²= -0.0014 | — |
| J3³/J1¹ = 0.006 | J3³/J1¹ = 0.004 | J3²/J1¹ = 0.005 | J3²/J1¹ = 0.008 | *j*(F5): 0.0013 | *j*(O13): -0.0010; | — |
| ***J3(J1₂ⁿ)*** | ***J3(J1₂ⁿ)*** | Cu2-Cu2 | Cu1-Cu1 | *j*(F7): 0.0004 | *j*(O12): -0.0007; | — |
| Cu2-Cu2 | Cu2-Cu2 | 7.284 | 7.267 | J3²/J1¹ = -0.02 | *j*(O9): 0.0002; | — |
| 7.152 | 6.997 | J3³ = -0.0010 (AFM) | J3³ = 0 | — | *j*(O10): 0.0001 | — |
| J1₂²¹² = -0.0315 ↔ 0.0129 | J1₂²¹² = -0.0309 (AFM) | *j*(F3): 0.0002x2 | *j*(F2): -0.0005x2 | — | J3²/J1¹ = 0.02 | — |
| *j*(Cu1): -0.0223 | *j*(Cu1): -0.0233 | *j*(F4): -0.0007x2 | *j*(F3): 0.0005x2 | — | — | — |
| (-0.570 1.0, 180°) | (-0.570 1.0, 180°) | J3³/J1¹ = 0.009 | J3³/J1¹ = 0 | — | — | — |
| *j*(F1): -0.0046x2 | *j*(F1): -0.0038x2 | Cu1-Cu2 | ***J3(J1₂ⁿ)*** | ***J3(J1₂ⁿ)*** |  | ***J3(J1₂ⁿ)*** |
| (-0.682, 2.9, 153.7°) | (-0.563, 3.0, 147.9°) | 7.054 | Cu1-Cu2 | Cu3-Cu3 |  | Cu4-Cu4 |
| *j*(F2): 0.0222x2 | J1₂²¹²/J1¹ = 0.34 | J3(J1₂¹¹²')= -0.0270(AFM) | 6.866 | 7.212 |  | 6.646 |
| (-0.550ʰ, 1.3, 123.9°) | — | *j*(Cu1): -0.0187 | J1₂¹²² = 0.0007 | J1₂³³³ = -0.0307 (AFM) |  | J1₂⁴²⁴ = -0.0310 |

| | | | | | |
|---|---|---|---|---|---|
| | $\Delta a^h = 0.020$ | | | | |
| | $J1_2^{212}/J1^3 = 0.29$ $0.29 \leftrightarrow -0.12$ | — | (-0.465 1.0, 176.5°) | $j$(Cu2): -0.0114 | $j$(Cu3): -0.0219 | $J1_2^{424}/J1^1 = 0.50$ |
| | Cu2-Cu2 | Cu2-Cu2 | $j$(F2): -0.0046 | (-0.268 1.0, 169.9°) | (-0.570 1.0, 180°) | Cu1-Cu3 |
| | 7.098 | 7.043 | (-0.688, 3.0, 153.0°) | $j$(F1): 0.0193 | $j$(F7): -0.0044x2 | 6.411 |
| | $J1_2^{222} = -0.0076$ (AFM) | $J1_2^{222} = -0.0246$ (AFM) | $j$(F3): -0.0037 | (-0.448, 1.2, 125.0°) | (-0.690, 3.0, 153.7°) | $J1_2^{143} = -0.0266$ |
| | $j$(Cu2): -0.0175 | $j$(Cu2): -0.0166 | (-0.569, 3.1, 148.2°) | $j$(F3): -0.0022 | $J1_2^{333}/J1^2 = 0.29$ | $J1_2^{143}/J1^2 = 0.51$; $J1_2^{143}/J1^6 = 0.49$ |
| | (-0.441, 1.0, 175.8°) | (-0.411, 1.0, 174.8°) | $J1_2^{212}/J1^1 = 0.26$; $J1_2^{212}/J1^3 = 0.28$ | (-0.333, 3.2, 137.6°) | Cu1-Cu1 | Cu4-Cu4 |
| | $j$(F3): -0.0044 | $j$(F3): -0.0033 | Cu1-Cu2 | $j$(F6): -0.0050 | 7.314 | 6.437 |
| | (-0.694, 3.1, 149.1°) | (-0.514, 2.9, 145.5°) | 7.108 | (-0.666, 2.8, 152.2°) | $J1_2^{131} = -0.0301$ (AFM) | $J1_2^{414} = -0.0211$ |
| | $j$(F3): -0.0039 | $j$(F3): -0.0047 | $J1_2^{112} = -0.0271$ (AFM) | $J1_2^{122}/J1^2 = -0.007$; $J1_2^{122}/J1^4 = -0.009$ | $j$(Cu3): -0.0213 | $J1_2^{414}/J1^2 = 0.41$; $J1_2^{414}/J1^5 = 0.37$ |
| | (-0.694, 2.9, 153.0°) | (-0.660 2.8, 152.6°) | $j$(Cu1): -0.0184 | Cu1-Cu2 | (-0.570 1.0, 180°) | Cu1-Cu1 |
| | $j$(F3): -0.0039 | $J1_2^2/J1^2 = 0.26$ | (-0.466 1.0, 176.6°) | 7.080 | $j$(F2): -0.0044x2 | 6.439 |
| | $j$(F2): 0.0182 | — | $j$(F1): -0.0045 | $J1_2^{122'} = -0.0033$ (AFM) | (-0.713, 3.0, 154.9°) | $J1_2^{121} = -0.0317$ |
| | (0.45, 1.2, 126.3°) | — | (-0.667, 2.9, 152.7°) | $j$(Cu2): -0.0091 | $J1_2^{131}/J1^1 = 0.28$ | $J1_2^{121}/J1^3 = 0.64$; |
| | $J1_2^{222}/J1^2 = 0.08$ | — | $j$(F2): -0.0042 | (-0.229 1.0, 169.0°) | Cu2-Cu3 | Cu1-Cu1 |
| | Cu2-Cu2 | Cu2-Cu2 | (-0.637 3.0, 151.3°) | $j$(F1): -0.0041 | 7.225 | 6.728 |
| | 7.074 | 7.134 | $J1_2^{112}/J1^1 = 0.26$ $J1_2^{112}/J1^2 = 0.26$ | (-0.615, 3.0, 150.4°) | $J1_2^{213} = -0.0156$ (AFM) | $J1_2^{131} = -0.0304$ |
| | $J1_2^{212'} = 0.0040$ FM | $J1_2^{212'} = -0.0306 \leftrightarrow 0.0122$ | Cu1-Cu2 | $j$(F6): -0.0036 | $j$(Cu1): -0.0081 | $J1_2^{131}/J1^4 = 0.49$ |
| | $j$(Cu1): -0.0228 | $j$(Cu2): -0.0224 | 7.112 | (-0.558 3.1, 147.7°) | (-0.212, 1.0,.168.7°) | Cu1-Cu2 |
| | (-0.570, 1.0, 180°) | (-0.570, 1.0, 180°) | $J1_2^{122'} = -0.0257$ (AFM) | $j$(F6): 0.0135 | $j$(F1): -0.0030 | 6.555 |
| | $j$(F2): -0.0045x2 | $j$(F2): -0.0041 | $j$(Cu2): -0.0181 | (0.331 1.2, 129.3°) | (-0.509, 3.2, 146.0°) | $J1_2^{142} = -0.0284$ |
| | (-0.689, 3.0, 150.2°) | (-0.629, 3.0, 151.1°) | (-0.457, 1.0, 176.4°) | $J1_2^{122}/J1^1 = 0.03$; $J1_2^{122}/J1^2 = 0.03$ | $j$(F2): -0.0045 | $J1_2^{142}/J1^2 = 0.54$; $J1_2^{142}/J1^1 = 0.45$ |
| | $j$(F1): 0.0179x2 | $j$(F'): 2x0.0214 | $j$(F1): -0.0039 | Cu1-Cu2 | 0.700, 3.0, 154.3° | Cu2-Cu3 |
| | (-0.440, 1.2, 124.7°) | (0.525$^h$,1.3,124.4) ($\Delta a^h = 0.02$) | (-0.606, 3.1, 149.8°) | 7.057 | $J1_2^{213}/J1^1 = 0.15$; $J1_2^{213}/J1^3 = 0.16$ | 6.565 |
| | $J1_2^{212}/J1^3 = -0.040$ | $J1_2^{212'}/J1^3 = 0.33 \leftrightarrow -0.13$ | $j$(F4): -0.0037 | $J1_2^{112} = -0.0240$ (AFM) | Cu2-Cu2 | $J1_2^{213} = -0.0193$ |
| | Cu1-Cu1 | Cu1-Cu1 | (-0.556, 3.0, 148.3°) | $j$(Cu1): -0.0156 | 7.212 | $J1_2^{213}/J1^3 = 0.39$; $J1_2^{213}/J1^4 = 0.31$ |
| | 7.108 | 7.059 | $J1_2^{221}/J1^2 = 0.25$; $J1_2^{221}/J1^4 = 0.32$ | (-0.388, 1.0, 174.1°) | $J1_2^{222} = -0.0273$ (AFM) | Cu4-Cu4 |
| | $J1_2^{121} = -0.0065$ (AFM) | $J1_2^{212'} = -0.0055$ (AFM) | Cu1-Cu2 | $j$(F1): -0.0038 | $j$(Cu2): -0.0219 | 6.395 |
| | $j$(Cu2): -0.0173 | $j$(Cu2): -0.0167 | 7.058 | $j$(F2): -0.0046 | (-0.570 1.0, 180°) | $J1_2^{434} = -0.0325$ |
| | (-0.436 1.0, 175.7°) | (-0.415 1.0,.175°) | $J1_2^{221'} = -0.0257$ (AFM) | (-0.674, 3.0, 152.7°) | $j$(F5): -0.0027x2 | $J1_2^{434}/J1^6 = 0.60$ |
| | (-0.647, 2.9, 152.0°) | (-0.647, 2.9, 152.0°) | (-0.647, 2.9, 152.0°) | (-0.647, 2.9, 152.0°) | (-0.424, 3.0, 143.8°) | — |
| | $j$(F1): -0.0041 | $j$(F1): -0.0041 | $j$(F1): -0.0041 | $j$(F1): -0.0041 | $J1_2^{222}/J1^4 = 0.42$ | — |
| | (-0.636, 3.1, 150.9°) | (-0.636, 3.1, 150.9°) | (-0.636, 3.1, 150.9°) | (-0.636, 3.1, 150.9°) | Cu1-Cu1 | — |
| | $j$(F3): 0.0193 FM | $j$(F3): 0.0193 FM | $j$(F3): 0.0193 FM | $j$(F3): 0.0193 FM | 7.206 | — |
| | (0.479, 1.2, 125.7°) | (0.479, 1.2, 125.7°) | (0.479, 1.2, 125.7°) | (0.479, 1.2, 125.7°) | $J1_2^{121} = -0.0302$ (AFM) | — |
| | $J1_2^{121}/J1^1 = 0.06$ | $J1_2^{121}/J1^1 = 0.06$ | $J1_2^{121}/J1^1 = 0.06$ | $J1_2^{121}/J1^1 = 0.06$ | $j$(Cu2): -0.0220 | — |
| | — | — | $J1_2^{122'}/J1^2 = 0.25$; $J1_2^{122'}/J1^4 = 0.32$ | $j$(F3): -0.0037 | (-0.570 1.0, 180°) | — |
| | — | — | — | (-0.493, 3.0, 144.6°) | $j$(F1): -0.0041x2 | — |
| | — | — | — | $j$(F2): 0.0179 | (-0.630, 2.9, 151.6°) | — |
| | — | — | — | (0.415, 1.2, 125.7°) | $J1_2^{121}/J1^3 = 0.31$ | — |
| | — | — | — | $J1_2^{112}/J1^3 = 0.06$; $J1_2^{112'}/J1^4 = 0.07$ | | |

| | | | | | |
|---|---|---|---|---|---|
| ***Interplane couplings*** | – | – | – | – | – | – |
| Bond | Cu-Cu | Cu - Cu | Cu - Cu | Cu - Cu | Cu - Cu | Cu - Cu |
| d(Cu-Cu) (Å) | 7.026 – 7.897 | 6.861 – 7.758 | 6.905 – 7.807 | 6.986 – 8.023 | 6.470 – 8.040 | 10.030 – 10.427 |
| $Jn^{(c)}$ (Å$^{-1}$) | J4 – J12<br>0 - 0.0006 (FM) | J4 – J12<br>0 – 0.0016 FM) | J4 – J12<br>0 – 0.0006 (FM) | J4 – J12<br>0 - 0025 (FM) | J4 – J10<br>0 – 0.0015 (FM) | J4 - J11<br>0 - 0.009 |

[a]XDS – X-ray diffraction from single crystal; XDP - X-ray diffraction (powder)
[b]The refinement converged to the residual factor (*R*) values.
[c]$Jn$ in Å$^{-1}$ ($Jn$ (meV) = $Jn$ (Å$^{-1}$)×K, where scaling factors $K_{middle}$ = 74) – the magnetic couplings ($Jn<0$ - AFM, $Jn>0$ – FM).
[d]$j(X)$ – contributions of the intermediate X ion into the AFM ($j(X) <0$) and FM ($j(X)>0$) components of the $Jn$ coupling
[e]$\Delta h(X)$ – the degree of overlapping of the local space between magnetic ions by the intermediate ion X.
[f]$l_n'/l_n$ is the asymmetry of position of the intermediate X ion relatively to the middle of the $Cu_i$–$Cu_j$ bond line.
[g]$Cu_iXCu_j$ bonding angle
[h]$\Delta a \sim 0.1$ Å ($\Delta a = (r_M + r_{An})$ - $h_{An}$) - critical position of intermediate A*n* ion.

**Supplementary Note 4:**
**Table 4.** Crystallographic characteristics and parameters of magnetic couplings (*Jn*) calculated on the basis of structural data and respective distances between magnetic Cu$^{2+}$ ions in kagome antiferromagnets with oxocentered triangles (OCu$_3$) and (2+4) and (4+2) types of *JT* distorted octahedral copper coordination.

| Crystallographic and magnetic parameters | Engelhauptite<br>KCu$_3$(V$_2$O$_7$)(OH)$_2$Cl [92]<br>(No in ICSD)<br>Space group $P6_3/mmc$ N194<br>$a = b = 5.922$, $c = 14.513$ Å<br>$\alpha = \beta = 90°$, $\gamma = 120°$, Z =2<br>Method[a] - XDS (296 K);<br>*R*-value[b] = 0.090 | *beta-Vesignieite*<br>BaCu$_3$V$_2$O$_8$(OH)$_2$ [26]<br>(Data for ICSD – 186931)<br>Space group *R* -3*mH* (N166)<br>$a = b = 5.9295$, $c = 20.790$ Å<br>$\alpha = \beta = 90°$, $\gamma = 120°$, Z =3<br>Method[a] - XDS (293 K);<br>*R*-value[b] = 0.0217 | *alfa-Vesignieite*<br>BaCu$_3$V$_2$O$_8$(OH)$_2$ [93]<br>(Data for ICSD – 67726)<br>Space group *C* 2/*m* (N12)<br>$a = 10.270$, $b = 5.911$, $c = 7.711$ Å<br>$\alpha = 90°$, $\beta = 116.42°$, $\gamma = 90°$, Z =2<br>Method[a] – XDS (293 K);<br>*R*-value[b] = 0.051 | KCu$_3$(OH)$_2$(AsO$_4$) (HAsO4) [104]<br>(Data for ICSD – 65419)<br>Space group C2/m (N12)<br>$a =10.292$, $b = 5.983$, $c = 7.877$ Å<br>$\alpha = 90°$, $\beta = 117.86°$, $\gamma = 90°$, Z=2<br>Method[a] – XDS (296 K);<br>*R*-value[b] = 0.033 | *Volborthite*<br>Cu$_3$(V$_2$O$_7$)(OH)$_2$(H$_2$O)$_2$ [34]<br>(Data for ICSD – 68995)<br>Space group C2/m (N12)<br>$a =10.607$, $b=5.864$, $c = 7.214$<br>$\alpha = 90°$, $\beta = 94.88°$, $\gamma = 90°$, Z =2<br>Method[a] - NDP(296 K);<br>*R*-value[b] = 0.093 |
|---|---|---|---|---|---|
| d(Cu-X) (Å) | Cu1- O1 = 1.918 x2<br>- O2 = 2.228 x4 | Cu1- O1 = 1.916 x2<br>- O2 = 2.183 x4 | Cu1- O1 = 1.913 x2<br>- O2 = 2.183 x4<br>Cu2 - O1 = 1.905 x2<br>- O2 = 2.175 x2<br>- O3 = 2.184 x2 | Cu1-O4 = 1.899 x2<br>-O3 = 2.186 x4<br>Cu2-O4 = 1.934 x2<br>-O3 = 2.000 x2<br>-O1 = 2.428 x2 | Cu1-O2 = 1.905 Åx2<br>-O3 = 2.158 Åx4<br>Cu2-O2 = 1.900 Åx2<br>-O4 = 2.048 Åx2<br>-O3 = 2.379 Åx2 |
| ***Kagome plane*** | ***Triangle - Cu1Cu1Cu1*** | ***Triangle - Cu1Cu1Cu1*** | ***Triangle - Cu1Cu2Cu2*** | ***Triangle - Cu1Cu2Cu2*** | ***Triangle - Cu1Cu2Cu2*** |
| Bond | Cu1-Cu1 | Cu1-Cu1 | Cu2-Cu2 | Cu2-Cu2 | Cu2-Cu2 |
| d(Cu-Cu) (Å) | 2.961 | 2.965 | 2.955 | 2.992 | 2.932 |
| $Jn^{(c)}$ (Å$^{-1}$) | J1 = -0.0410 (AFM) | J1 = -0.0425 (AFM) | J1$^1$ = -0.0452 (AFM) | J1$^1$ = -0.0388 (AFM) | J1$^1$ = -0.0374 (AFM) |
| $j(X)^d$ (Å$^{-1}$) | j(O3): -0.0410 | j(O1): -0.0425 | j(O1): -0.0452 | j(O4): -0.0388 | j(O2): -0.0445 |
| ($\Delta h(X)^e$ Å, $l_n'/l_n^f$, CuXCu$^g$) | (-0.180, 1.0, 101.6°) | (-0.187, 1.0, 101.4°) | (-0.197, 1.0, 101.7°) | (-0.174, 1.0, 101.3°) | (-0.191, 1.0, 101.0°) |
| $Jn/Jn^{max}$ | J1/J1 = 1 | J1/J1 = 1 | J1$^1$/J1$^1$ = 1 | J1$^1$/J1$^2$ = 0.90 | j(O4): 0.0071 FM |
| | ***J2$^n$*** | ***J2$^n$*** | Cu1-Cu2 | Cu1-Cu2 | 0.031, 1.0, 91.4°) |
| | Cu1-Cu1 | Cu1-Cu1 | 2.962 | 2.976 | (J1$^1$/J1$^2$=0.69 |
| | 5.129 | 5.135 | J1$^2$ = -0.0446 (AFM) | J1$^2$ = -0.0433 (AFM) | Cu1-Cu2 |
| | J2 = 0 | J2 = 0 | j(O1): -0.0446 | j(O2): -0.0445 | 3.030 |
| | J2/J1 = 0 | J2/J1 = 0 | (-0.195, 1.0, 101.8°) | (-0.192, 1.0, 101.8°) | J1$^2$ = -0.0543 (AFM) |
| | ***J3$^n$(J$_d$)*** | ***J3$^n$(J$_d$)*** | J1$^2$/J1$^1$ = 0.99 | J1$^2$/J1$^2$ = 1 | j(O2): -0.0543 |
| | Cu1-Cu1 | Cu1-Cu1 | ***J2$^n$*** | ***J2$^n$*** | (-0.249, 1.0, 105.6°) |
| | 5.922 | 5.930 | Cu1-Cu2 | Cu1-Cu2 | J1$^2$/J1$^2$=1 |
| | J3(J$_d$) = 0 | J3(J$_d$) = 0 | 5.123 | 5.173 | ***J2$^n$*** |
| | J3(J$_d$)/J1 = 0 | J3(J$_d$)/J1 = 0 | J2$^1$ = 0.0046 FM | J2$^1$= 0.0047 FM | Cu1-Cu2 |
| | ***J3(J1$_2^n$)*** | ***J3(J1$_2^n$)*** | j(O1): 0.0023x2 | j(O4): 0.0022 x2 | 5.136 |
| | Cu1-Cu1 | Cu1-Cu1 | (0.306, 5.0, 94.8°) | (0.302, 5.1, 94.90°) | J2$^1$= -0.0044(AFM) |
| | 5.922 | 5.930 | J2$^1$/J1$^1$ = -0.10 | J2$^1$/J1$^1$ = -0.12 | j(O4): -0.0044 |
|  | $J3(J1_2) = -0.0359$ (AFM) | $J3(J1_2)_2 = -0.0360$ (AFM) | Cu2-Cu2 | Cu2-Cu2 | (-0.235, 2.0, 126.7°) |
|  | $j$(Cu1): -0.0325 | $j$(Cu1): -0.0324 | 5.135 | 5.146 | $J2^1/J1^2 = 0.08$ |
|  | (-0.570, 1.0, 180°) | (-0.570, 1.0, 180°) | $J2^2 = 0.0046$ FM | $J2^2 = -0.0030$ (AFM) | Cu2-Cu2 |
|  | $j$(O3): -0.0017x2 | $j$(O1): -0.0018x2 | $j$(O1): 0.0023x2 | $j$(O3): -0.0040x2 | 5.304 |
|  | (-0.180, 3.0, 125.2°) | (-0.187, 3.0, 125.4°) | (0.304, 5.0, 94.9°) | (-0.232, 2.125.9°) | $J2^2 = 0.0040$ (FM) |
|  | $J3(J1_2)/J1 = 0.88$ | $J3(J1_2)/J1 = 0.85$ | $J2^2/J1^1 = -0.10$ | $j$(O4): -0.0025x2 | $j$(O2): 0.0020x2 |
| ***Interplane couplings*** | ***Interplane couplings*** | ***Interplane couplings*** | $\boldsymbol{J3^n(J_d)}$ | (0.0324, 4.9, 95.0°) | (0.264, 4.8, 98.1°) |
| Bond | Cu-Cu | Cu-Cu | Cu1-Cu1 | $J2^2/J1^2 = 0.07$ | $J2^2/J1^2 = -0.07$ |
| d(Cu-Cu) (Å) | 7.257 – 8.886 | 7.138 – 8.279 | 5.911 | $\boldsymbol{J3^n(J_d)}$ | $\boldsymbol{J3^n(J_d)}$ |
| $J_n^{(c)}$ (Å$^{-1}$) | J4 – J6 | J4 – J6 | $J3^1(J_d) = 0$ | Cu1-Cu1 | Cu1-Cu1 |
|  | -0.0004 (AFM) – 0.0022 (FM) | -0.0038 (AFM) – 0 |  |  |  |
| — | — | — | Cu2-Cu2 | 5.983 | 5.864 |
| — | — | — | 5.925 | $J3^1(J_d) = 0$ | $J3^1(J_d) = 0$ |
| — | — | — | $J3^2(J_d) = 0$ | Cu2-Cu2 | Cu2-Cu2 |
| — | — | — | $\boldsymbol{J3(J1_2^n)}$ | 5.952 | 6.060 |
| — | — | — | Cu2-Cu2 | $J3^2(J_d) = -0.0004$ AFM | $J3^2(J_d) = 0.0010$ FM |
| — | — | — | 5.911 | $j$(O3): -0.0002x2 | $j$(O4): 0.0005x2 |
| — | — | — | $J3(J1^{121}_2) = -0.0364$ (AFM) | (-0.022, 3.1, 119.5°) | (0.059, 3.2, 117.05°) |
| — | — | — | $j$(Cu2): -0.0326 | $J3^2/J1^2 = 0.009$ | $J3^2/J1^2 = -0.02$ |
| — | — | — | (-0.570, 1.0, 180°) | $\boldsymbol{J3(J1_2^n)}$ | $\boldsymbol{J3(J1_2^n)}$ |
| — | — | — | $j$(O1): -0.0019x2 | Cu2-Cu2 | Cu2-Cu2 |
| — | — | — | (-0.197, 3.0, 125.7°1 | 5.983 | 5.864 |
| — | — | — | $J3(J1_2^{121}) J1^1 = 0.81$ | $J3(J1_2^{222}) = -0.0350$ (AFM) | $J3(J1_2^{222}) = -0.0364$ (AFM) |
| — | — | — | Cu1-Cu1 | $j$(Cu2): -0.0318 | $j$(Cu2): -0.0332 |
| — | — | — | 5.925 | (-0.570, 1.0, 180°) | (-0.570, 1.0, 180°) |
| — | — | — | $J3(J1_2^{212}) = -0.0364$ (AFM) | $j$(O4): -0.0016x2 | $j$(O2): -0.0019x2 |
| — | — | — | $j$(Cu1): -0.0326 | (-0.174, 3.0, 125.4°) | (-0.191, 3.0, 125.1°) |
| — | — | — | (-0.570, 1.0, 180°) | $J3(J1_2^{222})/J1 = 0.90$ | $j$(O4): 0.0003x2 |
| — | — | — | $j$(O1): -0.0019x2 | Cu2-Cu2 | (0.031, 3.0, 117,7°) |
| — | — | — | (-0.195, 3.0, 125.8°) | 5.952 | $J3(J1_2^{222})/J1^1 = 0.97$ |
| — | — | — | $J3(J1_2^{212})/J1^1 = 0.81$ | $J3(J1_2^{212}) = -0.0358$ (AFM) | Cu2-Cu2 |
| — | — | — | ***Interplane couplings*** | $j$(Cu1): -0.0322 | 6.060 |
| — | — | — | Cu-Cu | (-0.570, 1.0, 180°) | $J3(J1_2^{212}) = -0.0356$ (AFM) |
| — | — | — | 7.113 – 7.711 | $j$(O4): -0.0018x2 | $j$(Cu1): -0.0310 |
| — | — | — | J4 – J7 | (-0.192, 2.9, 115.0°) | (-0.570, 1.0, 180°) |
| — | — | — | 0 – 0.0008 (FM) |  |  |
| — | — | — | — | $J3(J1_2^{212})/J1^2 = 0.83$ | $j$(O2): -0.0023x2 |
| — | — | — | — | Cu1-Cu1 | $J3(J1_2^{212})/J1^2 = 0.66$ |
| — | — | — | — | 5.952 | Cu1-Cu1 |
| — | — | — | — | $J3(J1_2^{121}) = -0.0358$ (AFM) | 6.060 |
| — | — | — | — | $j$(Cu2): -0.0322 | $J3(J1_2^{121}) = -0.0356$ (AFM) |
| — | — | — | — | (-0.570, 1.0, 180°) | $j$(Cu2): -0.0310 |
| — | — | — | — | $j$(O4): -0.0018x2 | (-0.570, 1.0, 180°) |
| — | — | — | — | (-0.192, 3.1, 115.01°) | $j$(O2): -0.0023x2 |
| — | — | — | — | $J3(J1_2^{121})/J1^2 = 0.83$ | (-0.249, 3.0, 128.6°) |
| — | — | — | — |  | $J3(J1_2^{121})/J1^2 = 0.66$ |
| — | — | — | — | ***Interplane couplings*** | ***Interplane couplings*** |
| — | — | — | — | Cu-Cu | Cu-Cu |
| — | — | — | — | 7.116 – 7.877 | 7.214 – 8.030 |

| | – | – | – | – | J4 – J7<br>0 – 0.0008 (FM) | J4 – J7<br>-0.0001 (AFM) – 0.0022 (FM) |

[a]XDS – X-ray diffraction from single crystal; XDP - X-ray diffraction (powder)
[b]The refinement converged to the residual factor ($R$) values.
[c]$Jn$ in Å$^{-1}$ ($Jn$ (meV) = $Jn$ (Å$^{-1}$)×K, where scaling factors K$_{middle}$ = 74) – the magnetic couplings ($Jn<0$ - AFM, $Jn>0$ – FM).
[d]$j(X)$ – contributions of the intermediate X ion into the AFM ($j(X) <0$) and FM ($j(X)>0$) components of the $Jn$ coupling
[e]$\Delta h(X)$ – the degree of overlapping of the local space between magnetic ions by the intermediate ion X.
[f]$l_n'/l_n$ is the asymmetry of position of the intermediate X ion relatively to the middle of the Cu$_i$–Cu$_j$ bond line.
[g]Cu$_i$XCu$_j$ bonding angle
verged to the residual factor ($R$) values.
[c]$Jn$ in Å$^{-1}$ ($Jn$ (meV) = $Jn$ (Å$^{-1}$)×K, where scaling factors K$_{middle}$ = 74) – the magnetic couplings ($Jn<0$ - AFM, $Jn>0$ – FM).
[d]$j(X)$ – contributions of the intermediate X ion into the AFM ($j(X) <0$) and FM ($j(X)>0$) components of the $Jn$ coupling

**Supplementary Note 5:**
**Table 5.** Crystallographic characteristics and parameters of magnetic couplings ($Jn$) calculated on the basis of structural data and respective distances between magnetic Cu$^{2+}$ ions in kagome antiferromagnets Cu$_3$(V$_2$O$_7$)(OH)$_2$(H$_2$O)$_2$ (space groups $C2/c$ and $Ia$), KCu$_3$(V$_2$O$_7$)(OH)$_2$Cl, PbCu$_3$(AsO$_4$)$_2$(OH)$_2$ and Pb$_2$Cu$_3$O$_2$(NO$_3$)$_2$(SeO$_3$)$_2$ with oxocentered OCu$_3$ triangles and deformed kagome lattice

| Crystallographic and magnetic parameters | *Volborthite*<br>Cu$_3$(V$_2$O$_7$)(OH)$_2$(H$_2$O)$_2$ [35]<br>(Data for ICSD 262959)<br>Space group C2/c (N15)<br>$a$ =10.612, $b$=5.871, $c$ = 14.418<br>$\alpha$=90° $\beta$=95.03°, $\gamma$=90°, Z =2<br>Method[a] - XDS (293K);<br>$R$-value[b] = 0.0254 | *Engelhauptite*<br>KCu$_3$(V$_2$O$_7$)(OH)$_2$Cl [110]<br>-<br>Space group $P12_1/m1$ N11<br>$a$=5.862, $b$=14.566, $c$=6.077 Å<br>$\alpha$ = 90°, $\beta$ = 119.06, $\gamma$ = 90°, Z=2<br>Method[a] – XDP (293K) | *Bayldonite*<br>PbCu$_3$(AsO$_4$)$_2$(OH)$_2$ [112]<br>(Data for ICSD – 8268)<br>Space group C2/$c$ N15<br>$a$=10.147, $b$=5.892, $c$=14.081 Å<br>$\alpha$ = 90°, $\beta$ = 106.05, $\gamma$ = 90°,Z=4<br>Method[a] - XDS; (296K)<br>$R$-value[b] = 0.052 | *Volborthite*<br>Cu$_3$(V$_2$O$_7$)(OH)$_2$(H$_2$O)$_2$ [105]<br>(Data for ICSD 162805)<br>Space group $Ia$ N9<br>$a$ =10.646, $b$=5.867, $c$ = 14.432<br>$\alpha$ = 90 °,$\beta$ = 95.19°, $\gamma$ = 90°, Z=4<br>Method[a] - XDS (296K);<br>$R$-value[b] = 0.038 | | Pb$_2$Cu$_3$O$_2$(NO$_3$)$_2$(SeO$_3$)$_2$ [113]<br>(Data for ICSD – 61268)<br>Space group $Cmc2_1$ N36<br>$a$ = 5.884, $b$ = 12.186, c = 19.371Å<br>$\alpha$ = $\beta$ = $\gamma$ = 90°, Z =4<br>Method[a] - XDS; (296K)<br>$R$-value[b] = 0.07 | |
|---|---|---|---|---|---|---|---|
| **d(Cu-X) (Å)** | Cu1-O2 = 1.941 Åx2<br>-O5 = 1.993 Åx2<br>-O6 =2.352 Åx2<br>Cu2-O2 = 1.912 Åx2<br>-O3 = 2.050 Åx2<br>-O5 = 2.370 Åx2<br>Cu3-O3 = 1.923 Åx2<br>-O3 = 2.004 Åx2<br>-O6 = 2.467 Åx2 | Cu1-O4 = 1.920 x2<br>-O3 = 2.052 x2<br>-O1 = 2.501 x2<br>Cu2-O4 = 1.927 x2<br>-O1 = 2.047 x2<br>-O2 = 2.383 x2<br>Cu3-O4 = 1.935 x2<br>-O3 = 2.026 x2<br>-O1 = 2.439 x2 | Cu1-O5 = 1.890 x2<br>-O3 = 2.039 x2<br>-O1 = 2.422 x2<br>Cu2-O5 = 1.878 x2<br>-O2 = 2.086 x2<br>-O3 = 2.271 x2<br>Cu3-O5 = 1.923 x2<br>-O2 = 1.999 x2<br>-O1 = 2.453 x2 | Cu1-O6 = 1.914 Å<br>-O1 = 1.999 Å<br>-O5 = 1.972 Å<br>-O2 = 2.000 Å<br>-O7 = 2.324 Å<br>-O8 = 2.369 Å<br>Cu2-O6 = 1.906 Å<br>-O5 = 1.927 Å<br>-O4 = 1.995 Å<br>-O3 = 2.044 Å<br>-O2 = 2.390 Å<br>-O7 = 2.454 Å | Cu3-O6 = 1.911 Å<br>-O5 = 1.942 Å<br>-O4 = 1.998 Å<br>-O3 = 2.058 Å<br>-O1 = 2.322 Å<br>-O8 = 2.572 Å | Cu1-O3 = 1.864<br>-O5 = 1.939<br>-O1 = 1.968<br>-O2 = 1.992<br>Cu2-O1 = 1.932<br>-O2 = 1.939<br>-O6 = 1.959<br>-O4 = 1.995 | |
| **Kagome plane** | *Triangle 1 - Cu1Cu2Cu3* | *Triangle 1 - Cu1Cu2Cu3* | *Triangle 1 - Cu1Cu2Cu3* | *Triangle 1 Cu1Cu2Cu3* | *Triangle 2 Cu1Cu2Cu3* | *Triangle 1 Cu1Cu2Cu2* | *Triangle 2 Cu1Cu2Cu2* |
| Bond | Cu2-Cu3 | Cu1-Cu3 | Cu1-Cu2 | Cu2-Cu3 | Cu2-Cu3 | Cu2-Cu2 | Cu2-Cu2 |
| d(Cu-Cu) (Å) | 2.935 | 2.931 | 2.946 | 2.879 | 2.990 | 2.931 | 2.953 |
| $Jn$[c] (Å$^{-1}$) | $J1^1$ = -0.0390 (AFM) | $J1^1$ = -0.0305 (AFM) | $J1^1$ = -0.0521 (AFM) | $J1^1$ = -0.0393 (AFM) | $J1^4$ = -0.0376 (AFM) | $J1^1$ = -0.0305 (AFM) | $J1^3$ = -0.0352 (AFM) |
| | Cu1-Cu2 | Cu1-Cu2 | Cu1-Cu3 | Cu1-Cu2 | Cu1-Cu3 | Cu1-Cu2 | Cu1-Cu2 |
| | 3.032 | 3.028 | 2.933 | 3.040 | 3.010 | 3.393 | 3.373 |
| | $J1^2$ = -0.0461 (AFM) | $J1^2$ = -0.0466 (AFM) | $J1^2$ = -0.0424 (AFM) | $J1^2$ = -0.0526 (AFM) | $J1^5$= -0.0329 (AFM) | $J1^2$ = -0.0708 (AFM) | $J1^4$ = -0.0741 (AFM) |
| | Cu1-Cu3 | Cu2-Cu3 | Cu2-Cu3 | Cu1-Cu3 | Cu1-Cu2 | *J2* | *J2* |
| | 3.032 | 3.039 | 2.933 | 3.075 | 3.034 | Cu2-Cu2 | Cu2-Cu2 |
| | $J1^3$ = -0.0440 (AFM) | $J1^3$ = -0.0451 (AFM) | $J1^3$ = -0.0398 (AFM) | $J1^3$ = -0.0554 (AFM) | $J1^6$= -0.0382 (AFM) | 5.350 | 6.093 |
| | *J2* | *J2* | *J2* | *J2* | | $J2^1$ = 0.0022 FM | $J2^3$= 0.0020 FM |
| | Cu1-Cu2 | Cu2-Cu3 | Cu1-Cu2 | Cu1-Cu2 | Cu1-Cu3 | Cu1-Cu2 | |
| | 5.141 | 5.128 | 5.074 | 5.148 | 5.221 | 5.374 | |
| | $J2^1$= -0.0290→ -0.0060 (AFM) | $J2^1$ = -0.0052 (AFM) | $J2^1$ = -0.0264 → -0.0053 (AFM) | $J2^1$= -0.0290→-0.0060 | $J2^4$= -0.0312→-0.0080 | $J2^2$ = 0.0024 FM | |
| | Cu1-Cu3 | Cu1-Cu3 | Cu2-Cu3 | Cu1-Cu2 | Cu1-Cu3 | \multicolumn{2}{c}{$J3^n(J_d)$} | |
| | 5.141 | 5.146 | 5.095 | 5.145 | 5.056 | Cu1-Cu1 | Cu2-Cu2 |
| | $J2^2$ = -0.0081(AFM) | $J2^2$ = -0.0066 ↔ -0.0275(AFM) | $J2^2$ = 0 | $J2^2$= -0.0081 (AFM) | $J2^5$ = -0.0027 (AFM) | 5.884 | 6.766 |
| | Cu2-Cu3 | Cu1-Cu3 | Cu1-Cu3 | Cu2-Cu2 | Cu3-Cu3 | $J3^1(J_d)$= 0 | $J3^2(J_d)$= 0 |
| | 5.306 | 5.312 | 5.095 | 5.323 | 5.324 | \multicolumn{2}{c}{$J3(J1_2{}^n)$} |
| | $J2^3$ = 0.0044 (FM) | $J2^3$ = 0.0043 (FM) | $J2^3$ = -0.0066 (AFM) | $J2^3$= 0.0047 (FM) | $J2^6$ = 0.0053 (FM) | Cu2-Cu2 | Cu2-Cu2 |
| | $J3^n(J_d)$ | $J3^n(J_d)$ | $J3^n(J_d)$ | \multicolumn{2}{c}{$J3^n(J_d)$} | 5.884 | 6.766 |
| | Cu1-Cu1 | Cu2-Cu2 | Cu1-Cu1 | Cu1-Cu1 | Cu2-Cu3 | $J3(J1_2{}^{222})$ = -0.0357 (AFM) | $J3(J1_2{}^{212})$ = -0.0112(AFM) ↔ 0.0118 (FM) |
| | 5.871 | 5.862 | 5.867 | 5.867 | 6.161 | Cu1-Cu1 | |
| | $J3^1(J_d)$ = -0.0002 (AFM) | $J3^1(J_d)$ = 0.0004 (FM) | $J3^1(J_d)$ = 0.0016 (FM) | $J3^1(J_d)$ = -0.0001 | $J3^3(J_d)$ = -0.0006 | 6.766 | |
| | Cu2-Cu2 | Cu2-Cu2 | Cu2-Cu2 | Cu2-Cu3 | — | $J3(J1_2{}^{121})$ = -0.0306 | |
| | 6.064 | 6.057 | 5.867 | 5.995 | — | \multicolumn{2}{c}{*Interplane couplings*} |
| | $J3^2(J_d)$ = 0.0 | $J3^2(J_d)$ = 0.0002 (FM) | $J3^2(J_d)$ = 0.0006 (FM) | $J3^2(J_d)$ = 0.0001 (FM) | — | \multicolumn{2}{c}{Cu-Cu} |
| | Cu3-Cu3 | Cu3-Cu3 | Cu3-Cu1 | \multicolumn{2}{c}{$J3(J1_2{}^n)$} | \multicolumn{2}{c}{9.794 – 10.644} |
| | 6.064 | 6.077 | 5.892 | Cu2-Cu2 | Cu3-Cu3 | \multicolumn{2}{c}{$J4 – J12$: -0.0078 (AFM) – 0.0097 (FM)} |
| | $J3^3(J_d)$ = -0.0006 (AFM) | $J3^3(J_d)$ = 0.0004(FM) | $J3^3$ = -0.0008 (AFM) | 5.867 | 5.867 | \multicolumn{2}{c}{—} |
| | $J3(J1_2{}^n)$ | $J3(J1_2{}^n)$ | $J3(J1_2{}^n)$ | $J3(J1_2{}^{232})$ = -0.0321 | $J3(J1_2{}^{323})$ = -0.0320 | \multicolumn{2}{c}{—} |
| | Cu2-Cu2 | Cu1-Cu1 | Cu1-Cu1 | Cu2-Cu3 | Cu2-Cu3 | \multicolumn{2}{c}{—} |
| | 5.871 | 5.862 | 5.867 | 6.049 | 6.108 | \multicolumn{2}{c}{—} |
| | $J3(J1_2{}^{232})$ = -0.0363(AFM) | $J3(J1_2{}^{313})$ = -0.0357 (AFM) | $J3(J1_2{}^{131})$ = -0.0365 (AFM) | $J3(J1_2{}^{213})$ = -0.0322 | $J3(J1_2{}^{213'})$ = -0.0325 | \multicolumn{2}{c}{—} |
| | Cu3-Cu3 | Cu2-Cu2 | Cu2-Cu2 | \multicolumn{2}{c}{*Interplane couplings*} | \multicolumn{2}{c}{—} |
| | 5.871 | 6.057 | 5.867 | \multicolumn{2}{c}{Cu-Cu} | \multicolumn{2}{c}{—} |
| | $J3(J1_2{}^{323})$ = -0.0363 (AFM) | $J3(J1_2{}^{212})$ = -0.0349 (AFM) | $J3(J1_2{}^{232})$ = -0.0363 (AFM) | \multicolumn{2}{c}{7.175 – 7.789} | \multicolumn{2}{c}{—} |
| | Cu2-Cu2 | Cu1-Cu1 | Cu3-Cu3 | \multicolumn{2}{c}{$J4 – J12$} | \multicolumn{2}{c}{—} |
| | 6.064 | 6.057 | 5.867 | \multicolumn{2}{c}{0 – 0.0033 (FM)} | \multicolumn{2}{c}{—} |
| | $J3(J1_2{}^{212})$ = -0.0348 (AFM) | $J3(J1_2{}^{121})$ = -0.0349 (AFM) | $J3(J1_2{}^{313})$ = -0.0367 (AFM) | — | | \multicolumn{2}{c}{—} |
| | Cu1-Cu1 | Cu3-Cu3 | Cu3-Cu3 | — | | \multicolumn{2}{c}{—} |
| | 6.064 | 6.077 | 5.867 | — | | \multicolumn{2}{c}{—} |
| | $J3(J1_2{}^{121})$ = -0.0350 (AFM) | $J3(J1_2{}^{323})$ = -0.0347 (AFM) | $J3(J1_2{}^{323})$ = -0.0365 (AFM) | — | | \multicolumn{2}{c}{—} |
| | Cu1-Cu1 | Cu2-Cu2 | Cu1-Cu1 | — | | \multicolumn{2}{c}{—} |
| | 6.064 | 6.077 | 5.892 | — | | \multicolumn{2}{c}{—} |
| | $J3(J1_2{}^{131})$ = -0.0348 (AFM) | $J3(J1_2{}^{232})$ = -0.0314 (AFM) | $J3(J1_2{}^{121})$ = -0.0345 (AFM) | — | | \multicolumn{2}{c}{—} |
| | Cu3-Cu3 | - | Cu2-Cu2 | — | | \multicolumn{2}{c}{—} |
| | 6.064 | - | 5.892 | — | | \multicolumn{2}{c}{—} |
| | $J3(J1_2{}^{313})$ = -0.0346 (AFM) | - | $J3(J1_2{}^{212})$ = -0.0372 (AFM) | — | | \multicolumn{2}{c}{—} |
| | *Interplane couplings* | *Interplane couplings* | *Interplane couplings* | — | | \multicolumn{2}{c}{—} |
| | Cu-Cu | Cu-Cu | Cu-Cu | — | | \multicolumn{2}{c}{—} |
| | 7.209 – 8.032 | 7.283 – 9.472 | 6.950 – 7.632 | — | | \multicolumn{2}{c}{—} |
| | $J4 – J7$ -0.0011 (AFM) – 0.0032 (FM) | $J4 – J9$ -0.0021 (AFM) – 0.0020 (FM) | $J4 – J7$ -0.0011 (AFM) – 0.0047 (FM) | — | | \multicolumn{2}{c}{—} |

[a]XDS – X-ray diffraction from single crystal; XDP - X-ray diffraction (powder)
[b]The refinement converged to the residual factor (R) values.
[c]$Jn$ in Å$^{-1}$ ($Jn$ (meV) = $Jn$ (Å$^{-1}$)×K, where scaling factors K$_{middle}$ = 74) – the magnetic couplings ($Jn<0$ - AFM, $Jn>0$ – FM).

**Supplementary Note 6: Figure 5(c).**

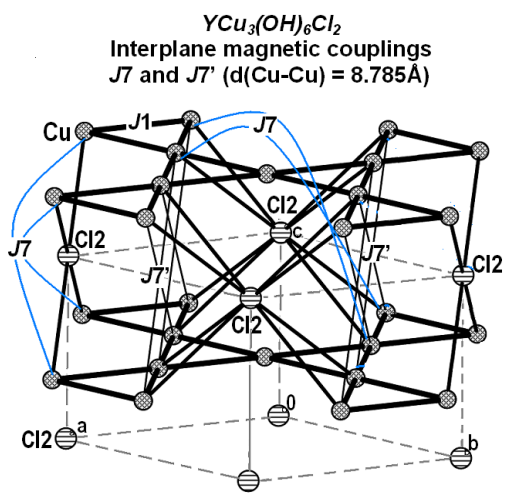

Figure 5(c). The Interplane $J7$ and $J7'$ magnetic couplings in $YCu_3(OH)_6Cl$

**Supplementary Note 7:**

*Volborthite $Cu_3(V_2O_7)(OH)_2(H_2O)_2$ in space groups C2/c (N15) and Ia (N9) with deformed kagome lattice and one type of (4+2)-JT distorted octahedral copper coordination.*

The changes of magnetic parameters are expressed even more clearly at reduction of the symmetry of volborthite $Cu_3(V_2O_7)(OH)_2(H_2O)_2$ until the noncentrosymmetric monoclinic space group *Ia* N9 [105]. In this case, the number of symmetrically distinct $Cu^{2+}$ sites remains to be equal to 3, and the Jahn-Teller deformation of all $CuO_6$ octahedra is of the 4+2 type, so that the spin-carrying orbital on $Cu^{2+}$ sites is the $d_{x2-y2}$ orbital. The number of nonequivalent nearest-neighbor $J1^n$ couplings would increase up to 6.

According to oir calculations (figure 18(c) and (d), Supplementary table 5), in this noncentrosymmetric sample of volborthite, the kagome lattice contains two types of small triangles with the nonequivalent AFM exchange. In the first Cu1Cu2Cu3 triangle, the AFM nearest-neighbor $J1^1$ ($J1^1$ = -0.0393 Å$^{-1}$, d(Cu2-Cu3) = 2.879 Å, $J1^1/J1^3$ = 0.71), $J1^2$ ($J1^2$ = -0.0526 Å$^{-1}$, d(Cu1-Cu2) = 3.040 Å, $J1^2/J1^3$ = 0.95), and $J1^3$ ($J1^3$ = -0.0554 Å$^{-1}$, d(Cu1-Cu3) = 3.075 Å) couplings are formed due to the intermediate O6 oxygen ion. In the second Cu1Cu2Cu3 triangle, the AFM nearest-neighbor $J1^4$ ($J1^4$ = -0.0376 Å$^{-1}$, d(Cu2-Cu3) = 2.990 Å, $J1^4/J1^6$ = 0.98), $J1^5$ ($J1^5$ = -0.0329 Å$^{-1}$, d(Cu1-Cu3) = 3.010 Å, $J1^5/J1^6$ = 0.86), and $J1^6$ ($J1^6$ = -0.0382 Å$^{-1}$, d(Cu1-Cu2) = 3.034 Å) couplings are formed due to the intermediate O5 oxygen ion. Both of these ions are located in the centers of respective triangles above or below their planes, whereas the AFM contributions from the O6 ion exceed in strength those from the O5 ion.

The magnetic parameters of the second in length AFM $J2^1$ (d(Cu1-Cu2) = 5.148 Å) and AFM $J2^4$ (d(Cu1-Cu3) = 5.221 Å) couplings are unstable, since the intermediate O3 ion making the main contribution to their formation is located in the critical position «c» (*l'/l*) [11, 12]. In this case, the insignificant displacement (by 0.005 Å) of the O3 ion to the center in parallel to the Cu1–Cu2 bond line results in a dramatic increase of the strength of the AFM $J2^1$ coupling from -0.0060 Å$^{-1}$ ($J2^1/J1^3$ = 0.11) to -0.0277 Å$^{-1}$ ($J2^1/J1^3$ = 0.50). At shifting of the O3 ion to the Cu1–Cu3 bond line center (by 0.017 Å), there would occur the increase of the strength of the AFM $J2^4$ coupling from -0.0080 Å$^{-1}$ ($J2^4/J1^2$ = 0.15) to -0.0312 Å$^{-1}$ ($J2^4/J1^2$ = 0.60). Such an increase of the strengths of the $J2^1$ and $J2^4$ couplings would enable them to compete with the AFM nearest-neighbor $J1^n$ couplings. Two other $J2^2$ ($J2^2$ = -0.0081 Å$^{-1}$, d(Cu1-Cu2) = 5.145 Å, $J2^2/J1^1$ = 0.21) and FM $J2^3$ ($J2^3$ = 0.0047 Å$^{-1}$, d(Cu2-Cu2) = 5.323 Å, $J2^3/J1^2$ = 0.09) couplings in the first large Cu1Cu2Cu2 triangle are very weak and cannot compete with the strong AFM nearest-neighbor $J1^n$ couplings. In the second large triangle, one observes the same stutuation for the AFM $J2^5$ ($J2^5$ = -0.0027 Å$^{-1}$,

d(Cu1-Cu3) = 5.056 Å, $J2^5/J1^1$ = - 0.07) and FM $J2^6$ ($J2^6$ = 0.0053 Å$^{-1}$, d(Cu3-Cu3) = 5.324 Å, $J2^6/J1^3$ = 0.10) ones.

The diagonal AFM $J3^1(J_d)$ ($J3^1(J_d)$ = -0.0001 Å$^{-1}$, d(Cu1-Cu1) = 5.867 Å, $J3^1(J_d)/J1^3$ = 0.002), FM $J3^2(J_d)$ ($J3^2(J_d)$ = 0.0001, d(Cu2-Cu3) = 5.995 Å, $J3^2(J_d)/J1^3$ = -0.002), and AFM $J3^3(J_d)$ ($J3^3(J_d)$ = -0.0006 Å$^{-1}$, d(Cu2-Cu3) = 6.161 Å, $J3^3(J_d)/J1^3$ = 0.01) couplings are weak to a degree that they are virtually absent. In the low-symmetry space group $Ia$ N9, there is no complete crystallographic equivalence between the $J3^n(J_d)$ and $J3(J1_2^n)$ couplings. Nevertheless, as in a majority of the examined above crystal structures, the strengths of the $J3(J1_2^n)$ couplings in the chains along the sides of small triangles exceeds significantly that of the diagonal AFM $J3^n(J_d)$ couplings. All the 4 nonequivalent next-nearest AFM $J3^n$ couplings ($J1_2^{232}$ (d(Cu2-Cu2) = 5.867 Å), $J1_2^{323}$ (d(Cu3-Cu3) = 5.867 Å), $J1_2^{213}$ (d(Cu2-Cu3) = 6.049 Å), and $J1_2^{213'}$ (d(Cu2-Cu3) = 6.108 Å)) in the chains along the sides of small triangles are sufficiently strong ($J3^n(J1_2^n)/J1^n$ = 0.59 – 0.82). They compete with the AFM nearest-neighbor $J1^n$ couplings. All the interplane Cu-Cu $J4$ – $J14$ couplings within the range from 7.175 Å to 8.036 Å are ferromagnetic (except $J7$) and very weak ($Jn/J1^3$ varies from -0.06 to 0). The $J7$ coupling (d(Cu1-Cu2) = 7.589 Å) is antiferromagnetic and weak ($J7/J1^3$ = 0.005). However, as in the former samples, the local spaces of the inetrplane interactions contain the O10 and O11 oxygen ions from water molecules that are not included to the copper coordination and could be removed upon the sample heating.

**Supplementary Note 8:**

*Bayldonite PbCu$_3$(AsO4)$_2$(OH)$_2$ with deformed kagome lattice and (4+2)-JT distorted octahedral copper coordination.*

Bayldonite (PbCu$_3$(AsO4)$_2$(OH)$_2$) [112] (figure 19, Supplementary table 5) crystallizes in the monoclinic centrosymmetric space group C2/c (N15). According to our calculations (figure 19, Supplementary table 5), the kagome lattice of bayldonite(PbCu$_3$(AsO4)$_2$(OH)$_2$) [112], just like that of volborthite (figure 18(a) and (b)), contains small triangles of just one type. They are characterized with strong nonequivalent AFM nearest-neighbor $J1^1$ couplings ($J1^1$ = -0.0521 Å$^{-1}$, d(Cu1-Cu2) = 2.946 Å), $J1^2$ ($J1^2$ = -0.0424 Å$^{-1}$, d(Cu1-Cu3) = 2.933 Å, $J1^2/J1^1$ = 0.81), and $J1^3$ ($J1^3$ = -0.0398 Å$^{-1}$, d(Cu2-Cu3) = 2.933 Å, $J1^3/J1^2$ = 0.76).

Besides, there exists some analogy with the $J2^n$ couplings as well. Two nonequivalent $J2^2$ ($J2^2$ = 0 Å, d(Cu2-Cu3) = 5.095 Å) and AFM $J2^3$ ($J2^3$ = -0.0066 Å$^{-1}$, d(Cu1-Cu3) = 5.095 Å, $J2^3/J1^1$ = 0.13) couplings in the kagome plane do not compete with the AFM nearest-neighbor $J1^n$ couplings, as one is eliminated, while another is weak, whereas the strength of the AFM $J2^1$ coupling (d(Cu1-Cu2) = 5.074 Å) could dramatically change from -0.0053 Å$^{-1}$ ($J2^1/J1^2$ = 0.10) to -0.0264 Å$^{-1}$ ($J2^1/J1^2$ = 0.51), depending on the shift of the intermediate O2 ion along the Cu1-Cu2 bond line located in the critical position «c» ($l'/l$) [11, 12].

Three diagonal FM $J3^1(J_d)$ couplings ($J3^1(J_d)$ = 0.0016 Å$^{-1}$, d(Cu1-Cu1) = 5.867 Å, $J3^1(J_d)/J1^1$ = -0.03), FM $J3^2(J_d)$ ($J3^2(J_d)$ = 0.0006, d(Cu2-Cu2) = 5.867 Å, $J3^2(J_d)/J1^1$ = -0.01), and AFM $J3^3(J_d)$ ($J3^3(J_d)$ = -0.0008 Å$^{-1}$, d(Cu3-Cu1) = 5.892 Å, $J3^3(J_d)/J1^2$ = 0.01) are very weak.

All the five noneqjuivalent next-nearest AFM $J3(J1_2^n)$ couplings ($J3(J1_2^{131})$ = -0.0365 Å$^{-1}$, d(Cu1-Cu1) = 5.867 Å; $J3(J1_2^{232})$ = -0.0363 Å$^{-1}$, d(Cu2-Cu2) = 5.867 Å; $J3(J1_2^{313})$ = -0.0367 Å$^{-1}$, d(Cu3-Cu3) = 5.867 Å); $J3(J1_2^{323})$ = -0.0365 Å$^{-1}$, d(Cu3-Cu3) = 5.867 Å; $J3(J1_2^{121})$ = - 0.0345 Å$^{-1}$, d(Cu1-Cu1) = 5.892 Å, and $J3(J1_2^{212})$ = -0.0372 Å$^{-1}$, d(Cu2-Cu2) = 5.892 Å) in the chains along the sides of small triangles are sufficiently strong. They compete with the AFM nearest-neighbor $J1^n$ couplings ($J3^n(J1_2^n)/J1^n$ = 0.75 – 0.93). The crystallographically equivalent $J3^1(J_d)$ and $J3(J1_2^{131})$ couplings between the Cu1-Cu1 ions at a distance of 5.867 Å are magnetically nonequivalent ($J3^1(J_d)/J3(J1_2^{131})$ = -0.04). The magnetic nonequivalence of crystallographically equivalent $J3^2(J_d)$ and $J3(J1_2^{232})$ ($J3^2(J_d)/J3(J1_2^{232})$ = -0.02) couplings is also observed between the Cu2-Cu2 ions at a distance of 5.867 Å.

All the interplane *J*4 – *J*7 couplings in the range from 6.950 to 7.632 Å are ferromagnetic (except *J*4) and weak (*J*n/*J*1$^2$ varies from -0.09 to 0.02). The *J*4 coupling (*J*4 = -0.0011 Å$^{-1}$, d(Cu1-Cu3) = 6.950 Å) is antiferromagnetic and weak as well (*J*4/*J*1$^1$ = 0.02). However, the local spaces of interplane interactions contain O4 oxygen ions from AsO$_4$-groups located between kagome planes and not included into the copper coordination and Pb$^{2+}$ ions. If one takes them into account at calculations of the magnetic coupling parameters, the *J*4 – *J*7 value would increase dramatically.